\documentclass[aip,jcp,reprint,amsmath,amssymb,floatfix]{revtex4-1}

\usepackage{graphicx,color,times,bm,upgreek,relsize}
\usepackage[colorlinks=true,urlcolor=blue,linkcolor=blue,
  citecolor=blue]{hyperref}
\usepackage[ignoreunlbld,norefs,nocites]{refcheck}

\newcommand{\normord}[1]{\xcentcolon\mathrel{#1}\xcentcolon}
\newcommand{\xcentcolon}{\mathrel{\vbox{\hbox{$:$}\kern.2ex}}}

\begin{document}

\title{Variational and Diffusion Quantum Monte Carlo Calculations with
  the CASINO Code}

\author{R.\ J.\ Needs} \affiliation{TCM Group, Cavendish Laboratory,
  University of Cambridge, 19 J.\ J.\ Thomson Avenue, Cambridge CB3 0HE,
  United Kingdom}

\author{M.\ D.\ Towler} \affiliation{University College London, London
  WC1E 6BT, United Kingdom}

\author{N.\ D.\ Drummond} \affiliation{Department of Physics, Lancaster
  University, Lancaster LA1 4YB, United Kingdom}

\author{P.\ L\'opez\ R\'ios} \affiliation{Max Planck Institute for
  Solid State Research, Heisenbergstra{\ss}e 1, 70569 Stuttgart,
  Germany}

\author{J.\ R.\ Trail} \affiliation{TCM Group, Cavendish Laboratory,
  University of Cambridge, 19 J.\ J.\ Thomson Avenue, Cambridge CB3 0HE,
  United Kingdom}

\begin{abstract}
We present an overview of the variational and diffusion quantum Monte
Carlo methods as implemented in the \textsc{casino} program.
We particularly focus on developments made in the last decade,
describing state-of-the-art quantum Monte Carlo algorithms and
software and discussing their strengths and their weaknesses.
We review a range of recent applications of \textsc{casino}.
\end{abstract}

\maketitle

\tableofcontents

\section{Introduction}

For nearly a century it has been accepted that a large part of
chemistry, materials science, and condensed matter physics could be
quantitatively explained if only it were possible to solve the
nonrelativistic many-electron Schr\"odinger equation for molecules,
surfaces, and bulk materials. \cite{Dirac_1929}
It is straightforward to write down the many-electron time-independent
Schr\"odinger equation, which in Hartree atomic units ($\hbar = m_{\rm
  e} = 4\pi\epsilon_0 = |e| = 1$) reads
\begin{equation}
  \label{eq:SE}
  \hat H \Psi \equiv \left( -\frac 1 2 \nabla^2 + \sum_{i>j} \frac 1
       {r_{ij}} - \sum_I \sum_i \frac{Z_I}{r_{iI}} \right) \Psi = E
       \Psi,
\end{equation}
where $\Psi({\bf r}_1,\ldots,{\bf r}_N)$ is the spatial wave function
for $N$ electrons, $-\frac 1 2 \nabla^2 \equiv -\frac 1 2 \sum_{i=1}^N
\nabla_i^2$ is the total kinetic-energy operator, $r_{ij}=|{\bf
r}_i-{\bf r}_j|$ is the distance between two electrons, $r_{iI}=|{\bf
r}_i-{\bf R}_I|$ is the distance between electron $i$ and nucleus $I$
of atomic number $Z_I$ at ${\bf R}_I$, and $E$ is an energy
eigenvalue.
The fermionic nature of the electrons imposes the important
requirement that the spatial wave function must be antisymmetric under
the exchange of same-spin electrons.
Unfortunately, solving the Schr\"odinger equation precisely for
anything but the smallest system sizes is a grand-challenge problem
due to the interactions between the electrons.

Over the last century a huge range of different techniques for
approximately solving Eq.\ (\ref{eq:SE}) has been developed.
An isolated hydrogen atom with $N=1$ can be solved exactly by pen and
paper.
For systems such as helium or lithium atoms ($N=2$ or $3$), the
Hylleraas \textit{ansatz} can be used to provide numerical solutions
that are accurate to dozens of significant figures.
\cite{Hylleraas_1929, Nakashima_2007}
For larger numbers of electrons we may approximate that the wave
function $\Psi$ is an antisymmetrized product (Slater determinant) of
single-particle orbitals, which is the form of wave function that
describes particles that are not coupled by interactions.
Invoking the variational principle of quantum mechanics, we can
generate numerical approximations to the orbitals by minimizing the
energy expectation value $\langle \Psi|\hat{H}|\Psi\rangle/\langle
\Psi|\Psi\rangle$, where $\hat{H}$ is the Hamiltonian.
This is the so-called Hartree-Fock (HF)
approximation. \cite{Slater_1930, Fock_1930, Hartree_1935}
It is important in physics and chemistry because it often provides a
qualitative understanding of electronic behavior and because it is the
starting point for more advanced methods.
For system sizes between $N=4$ and $100$ electrons, quantum chemistry
methods based on expansions in multiple Slater determinants are
computationally tractable and provide highly accurate energies, or at
least energies with consistent and controllable errors.
\cite{Szabo_1989, Shavitt_2009}
An alternative starting point for solving the many-electron
Schr\"odinger equation is provided by a theorem stating that the
ground-state total energy is a unique functional of the electronic
charge density. \cite{Hohenberg_1964}
The electronic charge density can be parameterized using a set of
single-particle orbitals for a fictitious noninteracting system, and
the largest contributions to the variational total energy can be
evaluated using these orbitals. \cite{Kohn_1965}
The remaining contribution to the total energy of the interacting
system can be evaluated as a parameterized functional of the
electronic charge density.
This approach, known as density functional theory (DFT), allows
approximate solutions to the electronic Schr\"odinger equation to be
obtained with up to at least $N=10,000$ electrons.
Despite the many successes of DFT, however, there are certain
important situations in which it performs poorly: in the description
of van der Waals interactions; in the description of electronic
excitations; and in situations in which a cancellation of errors in
energy differences cannot be relied upon. \cite{Burke_2012}
Some of these deficiencies can be remedied to some extent by using DFT
as the starting point for many-body perturbation theory, especially in
the so-called $GW$ approximation. \cite{Hedin_1965}
Nevertheless, it is fortunate that we have alternative, highly
accurate techniques available for solving the electronic Schr\"odinger
equation with up to $N=2,000$ electrons, namely, continuum quantum
Monte Carlo (QMC) methods.

The variational and diffusion quantum Monte Carlo (VMC and DMC)
methods that we discuss in this article use random sampling to solve
the Schr\"odinger equation in the continuous position basis.
\cite{McMillan_1965, Ceperley_1980, Foulkes_2001, Needs_2010}
In the VMC method \cite{McMillan_1965} we evaluate the expectation
value of the Hamiltonian with respect to a trial wave function that
may be of arbitrary complexity, rather than being restricted to an
antisymmetrized product of single-particle orbitals as in
HF theory.
This means that the $3N$-dimensional integral for the energy
expectation value cannot be broken into a series of 3- and
6-dimensional integrals, and so we must use the only general
high-dimensional numerical integration method available, which is
Monte Carlo integration.
In the DMC method \cite{Ceperley_1980} we simulate drifting, random
diffusion, and branching/dying processes governed by a Wick-rotated
time-dependent Schr\"odinger equation in order to project out the
ground-state component of a trial wave function.
Fermionic antisymmetry is enforced for real wave functions by
constraining the nodal surface to remain pinned at that of a trial
wave function.  \cite{Anderson_1976}
QMC results come with a quantifiable random error, which falls off as
$N_{\rm s}^{-1/2}$ with the number of random samples $N_{\rm s}$,
irrespective of the size of problem, together with an unquantifiable,
positive systematic bias due to the error in the trial wave function
(in VMC) or the error in the nodal surface of the trial wave function
(in DMC)\@, the latter of which is referred to as the fixed-node
error.
The random errors are benign, as they are quantifiable, controllable,
and easily propagated to derived results using standard statistical
methods.
The systematic biases are variational in nature (i.e., positive) and
in general are demonstrably much smaller than those in other
variational methods such as HF theory, so that VMC and DMC qualify as
``high-accuracy methods.''
The fact that system sizes of up to $N=2,000$ electrons are accessible
gives QMC methods one of their key unique selling points: they are the
only highly accurate approaches that can be applied to periodic
simulation cells that model meaningful quantities of condensed matter.
For comparison, recent advances in quantum chemistry methods for
solids enable the application of the coupled cluster method to systems
of up to about $N=100$ electrons.  \cite{Gruber_cc_solids_2018,
Gruber_BN_2018, Tsatsoulis_h_si_2018, Liao_hydrogen_2019}

Historically, QMC methods have been applied to model systems in
condensed matter such as the homogeneous electron gas (HEG),
generating the data used to parameterize DFT exchange-correlation
(XC) functionals. \cite{Ceperley_1980}
It remains the case that many of the most interesting, influential,
and experimentally relevant QMC calculations are for simplified models
of interacting particles, e.g., charge carriers within an
effective-mass approximation or positrons immersed in electron gases.
However, \textit{ab initio} QMC calculations can also be performed,
both for benchmarking simpler methods such as DFT, and for directly
generating theoretical predictions against which experiments can be
compared.
A wide variety of QMC methods have been developed over the years,
including path-integral Monte Carlo, \cite{Ceperley_1995} reptation
Monte Carlo, \cite{Baroni_1999} auxiliary-field quantum Monte Carlo,
\cite{Sorella_1989} full-configuration-interaction quantum Monte Carlo
(FCIQMC)\@, \cite{Booth_FCIQMC_2009, Cleland_initiator_2010,
Ghanem_unbias_FCIQMC_2019} orbital-space VMC,
\cite{Sabzevari_orbvmc_2018} and orbital-space VMC-based
geminal-Jastrow optimizers. \cite{Neuscamman_JAGPopt_2016}
In this article we will discuss applications of VMC and DMC to both
model and ``real'' systems.
We will also describe the technical details of the methodology and
discuss the strengths and weaknesses of the approach.
There exist several excellent, actively developed implementations of
continuum QMC methods, including \textsc{champ}, \textsc{qmcpack},
\cite{Kim_2018} \textsc{qwalk}, \cite{Wagner_2009} and
\textsc{turboRVB}.
We will focus on our implementation of the QMC methods in the
\textsc{casino} program, \cite{Needs_2010} and on topics relevant to
users of this software.

In Sec.\ \ref{sec:QMC} we review the basic theory of continuum QMC
methods and their implementation in the \textsc{casino} code.
In Sec.\ \ref{sec:recent} we discuss some recent developments in QMC
methods.
We present a selection of applications of \textsc{casino} in
Sec.\ \ref{sec:applications} to illustrate the capabilities of the QMC
methods.
We discuss future directions for QMC methods and the \textsc{casino}
software in Sec.\ \ref{sec:future}.
Finally we draw our conclusions in Sec.\ \ref{sec:conclusions}.
Except where otherwise stated, we use Hartree atomic units throughout.
We describe the density of a HEG by the density parameter $r_{\rm s}$,
which is the radius of the sphere (or circle in 2D) that contains one
electron on average divided by the Bohr radius.

\section{Continuum quantum Monte Carlo methods}
\label{sec:QMC}

\subsection{Variational quantum Monte Carlo}

\subsubsection{Sampling a trial wave function \label{sec:sampling}}

A familiar example of Monte Carlo methods in undergraduate physics is
the use of the Metropolis algorithm \cite{Metropolis_1953,
  Hastings_1970} to sample the Boltzmann distribution, e.g., in
numerical studies of the Ising model of spins on a lattice.
In this case one repeatedly proposes changes to the spin
configuration.
If a proposed configuration increases the target probability (i.e.,
lowers the energy in the case of the Boltzmann distribution) then it
is automatically accepted; otherwise it is accepted with a probability
given by the ratio of the target probability of the proposed
configuration to the target probability of the current configuration
(assuming the proposal probabilities for forwards and backwards moves
are the same).
If a proposed configuration is accepted then it is added to the list
of configurations sampled and becomes the new ``current
configuration;'' if the proposed configuration is rejected then
another copy of the current configuration is appended to the list of
configurations sampled.
This is an example of a Markov process: each step of the process only
depends on the current configuration.

It is easy to show that once a set of configurations is distributed
according to the target distribution it will continue be distributed
according to that distribution upon application of the Metropolis
algorithm, i.e., the target distribution is a stationary point of the
Metropolis algorithm.
Furthermore, provided the proposal distribution is \textit{ergodic},
allowing any configuration to be sampled after a finite number of
moves from any starting point, the stationary point of the Markov
process is unique and the process will converge to that stationary
point.
Conveniently we do not even need to know the normalization factor of
the target distribution to use this method.

In the VMC method \cite{McMillan_1965, Foulkes_2001} we use the
Metropolis method to sample the continuous distribution of electron
position coordinates distributed according to $|\Psi|^2$, where $\Psi$
is a trial spatial wave function.
The expectation value of an operator $\hat{A}$ can then be evaluated
as
\begin{equation}
  \label{eq:vmc_exp}
  \langle \hat A \rangle
    = \frac{\int \Psi^* \hat{A} \Psi \, {\rm d}{\bf R}}
           {\int |\Psi|^2 \, {\rm d}{\bf R}}
    = \frac{\int |\Psi|^2 \frac{\hat A \Psi}{\Psi} \, {\rm d}{\bf R}}
           {\int |\Psi|^2 \, {\rm d}{\bf R}}
    = \left< A_{\rm L} \right>_{|\Psi|^2},
\end{equation}
where the localized operator $A_{\rm L} \equiv (\hat A\Psi)/\Psi$ can
be evaluated at each configuration ${\bf R}=({\bf r}_1,\ldots,{\bf
  r}_N)$ sampled.
In particular, it is clear from Eq.\ (\ref{eq:vmc_exp}) that the
energy expectation value is estimated by the average of the local
energy $E_{\rm L}=(\hat{H} \Psi)/\Psi$ over the set of
configurations sampled.
The only requirements on the trial wave function are that the integral
for the expectation value should converge and that it should be
possible to evaluate a confidence interval on the average (in
practice, this means that $A_{\rm L}^2$ should also have a
well-defined mean, so that we can compute the standard error in the
mean of $A_{\rm L}$).
In the limit of perfect sampling the expectation value of a Hermitian
operator is real, and hence we usually only evaluate and average the
real part of the local operator $A_{\rm L}^{\rm r} \equiv {\rm
  Re}(A_{\rm L})$, i.e., in practice the VMC estimate is $\langle
A_{\rm L}^{\rm r}\rangle_{|\Psi|^2}$.

In Eq.\ (\ref{eq:vmc_exp}) we have not explicitly mentioned spin.
Consider a many-electron wave function $\Psi_{\rm sp}({\bf X})$, where
${\bf X} = ( {\bf r}_1, \sigma_1, \ldots , {\bf r}_N ,
\sigma_N )$ and ${\bf r}_i$ and $\sigma_i \in \{\uparrow ,
\downarrow \}$ are the position and spin of electron $i$.
Assume that $\Psi_{\rm sp}$ is an eigenfunction of the total spin
operator $\hat{S}_z=\sum_{i=1}^N \hat{s}_{zi}$.
Let the eigenvalue of $\hat{S}_z$ be $(N_\uparrow-N_\downarrow)/2$,
where $N_\uparrow+N_\downarrow=N$.
The electron spins are said to be collinear; we have $N_\uparrow$
spin-up electrons and $N_\downarrow$ spin-down electrons.
This restricts the form of $\Psi_{\rm sp}$, but being an eigenfunction
of $\hat{S}_z$ is a property of exact eigenfunctions of the
Hamiltonian if $[\hat{H},\hat{S}_z]=0$, as is the case for most
problems in condensed matter.
Electrons are fermions, so $\Psi_{\rm sp}$ is antisymmetric under
electron exchange.
The expectation value of a spin-independent operator $\hat{A}$ with
respect to $\Psi_{\rm sp}$ is
\begin{eqnarray} \langle \hat{A} \rangle_{\Psi_{\rm sp}} & = & \frac{\langle
\Psi_{\rm sp} | \hat{A} |\Psi_{\rm sp} \rangle}{\langle \Psi_{\rm sp}
    | \Psi_{\rm sp} \rangle} \nonumber \\ & = & \frac{ \sum_{\bm \sigma} \int
    \Psi_{\rm sp}^\ast({\bf X}) \hat{A}({\bf R}) \Psi_{\rm sp}({\bf
      X}) \, d{\bf R} }{\sum_{\bm \sigma} \int |\Psi_{\rm sp} ({\bf
      X})|^2 \, d{\bf R}}, \label{eq:spin-dep_exp_val}
\end{eqnarray}
where ${\bf R}=({\bf r}_1, \ldots,{\bf r}_N)$ and the sums run over
all spin configurations such that the numbers of spin-up and spin-down
electrons are $N_\uparrow$ and $N_\downarrow$, respectively.
${\bf X}$ may be replaced by ${\bf X}^\prime=\left( {\bf r}_{i_1},
\uparrow , \ldots , {\bf r}_{i_{N_\uparrow}} , \uparrow ,
         {\bf r}_{i_{N_\uparrow}+1}, \downarrow, \ldots , {\bf
           r}_{i_N}, \downarrow \right)$ in
         Eq.\ (\ref{eq:spin-dep_exp_val}) without altering $\langle
         \hat{A} \rangle_{\Psi_{\rm sp}}$, due to the antisymmetry of
         $\Psi_{\rm sp}$.
We may now relabel the dummy integration variables: ${\bf r}_{i_1}
\rightarrow {\bf r}_1$, etc.
The integrals in Eq.\ (\ref{eq:spin-dep_exp_val}) are clearly the same
for each spin configuration ${\bm \sigma}$ such that the number of
spin-up electrons is $N_\uparrow$; hence the spin sums cancel.
So the expectation of $\hat{A}$ may be evaluated as
\begin{equation} \langle \hat{A} \rangle_{\Psi_{\rm sp}} = \frac{\int
\Psi^\ast({\bf R}) \hat{A}({\bf R}) \Psi({\bf R}) \, d{\bf R}}{\int
    |\Psi({\bf R})|^2 \, d{\bf R}}, \end{equation} where the spatial
wave function
\begin{equation} \Psi({\bf R}) \equiv \Psi_{\rm sp}({\bf
  r}_1,\uparrow, \ldots , {\bf r}_{N_\uparrow},\uparrow, {\bf
    r}_{N_\uparrow+1},\downarrow, \ldots, {\bf
    r}_N,\downarrow) \end{equation} is only antisymmetric with respect
to exchanges of positions of same-spin electrons.
In effect, electrons of different spin are treated as distinguishable
fermions. \cite{Hammond_1994}
If we know $\Psi({\bf R})$ then we can recover $\Psi_{\rm sp}({\bf
  X})$ and \textit{vice versa}.
In practice it is much easier just to work with $\Psi$, which will
henceforth just be called ``the (trial) wave function.''
For example, the variational principle is
\begin{equation} \frac{\langle \Psi_{\rm sp}|\hat{H}|\Psi_{\rm sp}\rangle}{\langle
    \Psi_{\rm sp}|\Psi_{\rm sp}\rangle} = \frac{\langle
    \Psi|\hat{H}|\Psi \rangle}{\langle \Psi |\Psi \rangle} \geq
  E_0, \end{equation} where $\hat{H}$ is the Hamiltonian and $E_0$ is
the ground-state energy.
We may apply the variational principle to optimize $\Psi({\bf R})$.

If QMC methods are used to study systems in which the spin and spatial
coordinates are coupled, e.g., due to spin-orbit coupling terms in the
Hamiltonian or because one is examining a broken-symmetry state such
as a spin-density wave then one must return to using an antisymmetric
wave function $\Psi_{\rm sp}$ in which the spatial coordinates are
continuous and the spin coordinates are discrete.
In the case of VMC this results in a trivial modification to the
algorithms presented below.
We briefly discuss some of the implications for DMC in
Sec.\ \ref{sec:soc}.

Typically we find that it is much more efficient to propose and then
accept or reject single-electron moves than whole-configuration moves.
Moving a single electron a given distance results in a much smaller
change in the wave function than moving all $N$ electrons by the same
distance.
The proposed electron move is therefore more likely to be accepted,
and hence the distance traveled by the electrons over $N$ one-electron
moves is greater on average.
Electron-by-electron sampling therefore explores configuration space
more efficiently. \cite{Lee_2010}
In \textsc{casino} we use a Gaussian probability density for proposing
electron moves, with the variance of the Gaussian being chosen such
that the electron acceptance probability is close to 50\%;
empirically, this leads to near-maximum efficiency in exploring
configuration space. \cite{Metropolis_1953, Lee_2010}
The Gaussian probability density is advantageous because it is
symmetric and extremely efficient to evaluate.
It is possible to use more complicated proposal probability densities,
e.g., taking the gradient of the trial wave function into account,
which allow the electrons to travel further on average at each step;
\cite{Umrigar_1993b} however we find that the improvement in the
efficiency of exploring configuration space is easily outweighed by
the additional computational expense of evaluating wave-function
derivatives.

Although the Metropolis algorithm produces configurations with the
correct distribution, those configurations are serially correlated:
successive configurations are similar to each other due to the short
steps taken and the fact that moves are rejected.
For this reason, by default in \textsc{casino} we only evaluate the
local energy and other quantities of interest on every third VMC step
to save the computational expense of evaluating wave-function
derivatives for configurations that are very similar to each other.
Serial correlation complicates the calculation of the standard error
in the mean energy, as there are in effect fewer independent data
points than the number of points sampled.

\begin{figure}[!htbp]
  \begin{center}
    \includegraphics[clip,width=0.45\textwidth]{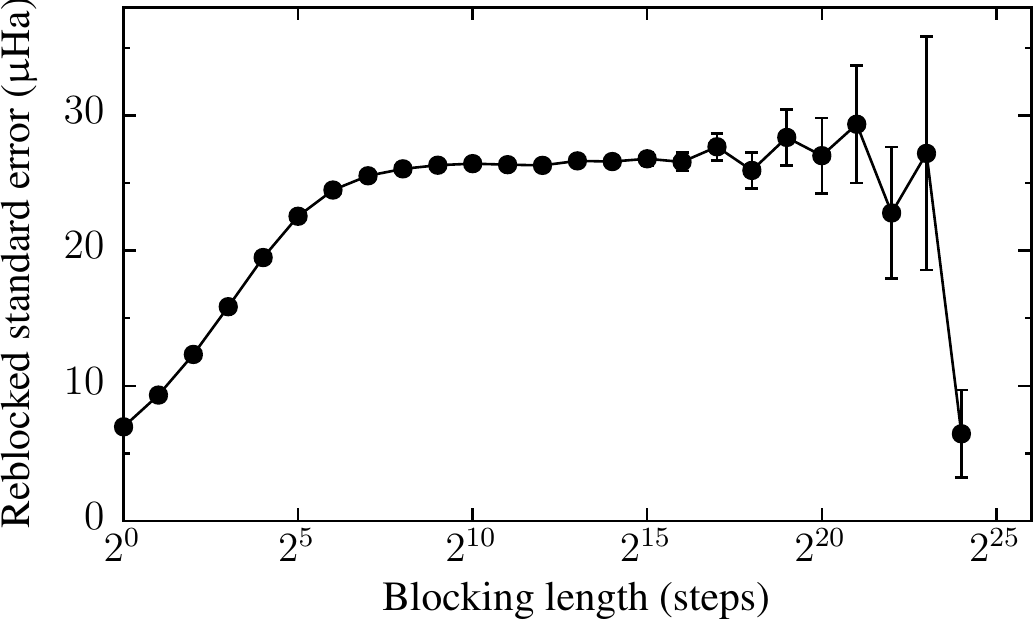}
  \end{center}
  \caption{ Reblocked standard error in the VMC energy of an
    all-electron lithium atom as a function of blocking length.
    The blocking length is the number of successive original data
    points averaged over to obtain each new data point).
    \label{fig:li_reblock}
  }
\end{figure}

A simple, robust method for removing serial correlation is the
reblocking method, \cite{Flyvbjerg_1989, Wolff_2004,
Jonsson_reblock_2018} in which we repeatedly group the data into
successive pairs and average within each pair to produce a new data
set with half as many points, but with a greater degree of
independence between the data points.
This process is illustrated in Fig.\ \ref{fig:li_reblock} using VMC
energy data for a lithium atom.
Evaluating the standard error in the mean using the na\"{\i}ve expression
appropriate for independent data points, corresponding to a blocking
length of $1$, we find that the standard error is too small because
the number of independent data points is overestimated.
As the blocking length increases, the standard error rises and reaches
a plateau when the block averages are essentially independent of each
other.
Specifically, the onset of the plateau occurs when the blocking length
exceeds the decorrelation period of the original data.
Finally, excessively large blocking lengths result in new data sets so
small that the estimate of the standard error is itself subject to
significant noise.
There are various methods to automatically choose the optimal blocking
length for a given data set. \cite{Wolff_2004, Lee_2010,
Jonsson_reblock_2018}
The reblocking method has the advantage that it does not assume the
underlying data to be normally distributed; indeed, local-energy data
generally follow a fat-tailed distribution. \cite{Trail_2008}

\subsubsection{Slater-Jastrow trial wave functions}
\label{sec:slater_jastrow}

The majority of QMC calculations use a Slater-Jastrow trial wave
function,
\begin{equation}
  \Psi({\bf R})
    = e^{J({\bf R})} \Psi_{\rm S}({\bf R})
    = e^{J({\bf R})} D_\uparrow({\bf R}_\uparrow)
                     D_\downarrow({\bf R}_\uparrow) \;,
\end{equation}
in which the Slater part $\Psi_{\rm S}$ is a product of Slater
determinants of single-particle orbitals for spin-up and spin-down
electrons $D_\uparrow$ and $D_\downarrow$, and $e^J$ is a Jastrow
correlation factor.

The single-particle orbitals in the Slater wave function are typically
taken from a HF, DFT, or quantum chemistry calculation
using an appropriate electronic-structure code.
The orbitals may also include free parameters to optimize within QMC
(see Sec.\ \ref{sec:opt}).
Typically the Slater wave function will provide a good description of
the electronic charge density (the one-body electron distribution),
but no description of electronic correlation.
More fundamentally the Slater wave function provides the qualitative
features of the state under consideration; it has the required
fermionic antisymmetry and any other symmetries of the system, and the
topology of the nodal surface of the Slater wave function uniquely
specifies the excited state to be studied.
(For a system without time-reversal symmetry, the phase of the complex
wave function plays the role of the nodal surface of a real wave
function; see Sec.\ \ref{sec:fixed_phase}.)

Any errors in the topology of the nodal surface give rise to so-called
static correlation effects, which are present in nearly all nontrivial
electronic systems and may be energetically important.
\footnote{The use of the terms ``static correlation'' and ``dynamic
  correlation'' in the quantum chemistry literature is less clearcut
  than suggested by our QMC-focused definition in terms of qualitative
  and quantitative errors in the nodal surface of the wave function.
Nevertheless, the definition we have used captures the idea that
static correlations are due to errors in the Hartree--Fock state,
while dynamical correlations are due to the electrons avoiding each
other.}
These errors are unaffected by the Jastrow factor or by the use of a
backflow transformation (see below), and must instead be addressed by
the use of multideterminant or pairing wave functions; see Sec.\
\ref{sec:adv_wfs}.

However, in many cases we are interested in long-range effects and
low-energy effects, e.g., changes to the distribution or occupancy of
the outermost electrons in atoms or molecules.
In such cases, static correlation effects largely cancel out of energy
differences with a single-determinant Slater wave function, and only a
good description of dynamical correlation effects (which do not
involve changing the nodal topology) is required for high accuracy.
Jastrow factors and backflow transformations give accurate and
compact descriptions of dynamical correlations.

A Jastrow factor for a molecular system consisting of $N$ electrons
and $N_{\rm n}$ nuclei is typically a sum of isotropic
electron-electron, electron-nucleus, and electron-electron-nucleus
terms,
\begin{equation}
  \begin{split}
    J({\bf R}) = & \sum_{i<j}^N u_{P_{ij}}(r_{ij}) + \sum_i^N
    \sum_I^{N_{\rm n}}
    \chi_{S_{iI}}(r_{iI}) \\ & + \sum_{i<j}^N \sum_I^{N_{\rm n}}
    f_{T_{ijI}}(r_{iI},r_{jI},r_{ij}) \;,
  \end{split}
\end{equation}
where $u$, $\chi$, and $f$ are parameterized functions of the relevant
interparticle distances, and $P$, $S$, and $T$ are ``channel''
indices such that different parameter sets can be used for, e.g.,
parallel and antiparallel spin electrons or for different atomic
species.

The homogeneous electron-electron dynamical correlations provided by
the $u$ term usually afford the most significant improvement to the
trial wave function, but this alters the accurate one-electron density
encoded in the HF or DFT orbitals.
The $\chi$ function is capable of restoring the one-electron density,
\cite{Boys_1969} and the $f$ term provides inhomogeneous
electron-electron correlations that further refine the wave function.

Jastrow factors are fully symmetric and strictly positive, so they do
not alter the overall symmetry or nodal structure of $\Psi_{\rm S}$.
Therefore the DMC method gives the same result whether a Jastrow
factor is present or not.
However, Jastrow factors do offer a great improvement in the
statistical efficiency of DMC, partly because of their ability to
impose the Kato cusp conditions. \cite{Kato_1957, Pack_1966}

In Coulomb systems the local potential energy diverges whenever two
charged particles coincide, but the exact local energy is a finite
constant everywhere in configuration space, so the local kinetic
energy must diverge so as to cancel the divergence of the local
potential energy.
Let $i$ and $j$ be two particles of charge $q_i$ and $q_j$,
respectively, of reduced mass $\mu_{ij}$, and separated by a distance
$r=|{\bf r}_i-{\bf r}_j|$.
Requiring the local energy to be finite at $r\to 0$ gives
\begin{equation}
  \left( \frac 1 \Psi \frac{\partial \Psi}{\partial r} \right)_{r\to
    0} = \frac{2 q_i q_j \mu_{ij}}{d\pm 1} = \Gamma_{ij} \;, \label{eq:kato_cusp}
\end{equation}
where $d$ is the dimensionality of the system and the plus and minus
signs in the denominator correspond to indistinguishable and
distinguishable particles, respectively.
\footnote{For a strictly one-dimensional system in which the particles
  interact via the Coulomb $1/r$ potential, the wave function must go
  to zero at all coalescence points in order for the energy
  expectation value to be well-defined.
In practice the easiest way of achieving this requirement in a study
of a one-dimensional electron system is to treat all the electrons as
indistinguishable particles. \cite{Lee_2011}
Equation (\ref{eq:kato_cusp}) continues to be valid for a
one-dimensional system provided the plus sign is selected,
irrespective of the spins of the coalescing particles.}

Electron-electron cusps are usually applied on the electron-electron
Jastrow factor by enforcing $\partial u/\partial r_{ij} = \Gamma_{ij}$
at $r_{ij}=0$.
The variance of the local energies encountered in VMC and DMC
calculations is significantly smaller for trial wave functions which
satisfy the electron-electron Kato cusp conditions than for those
which do not.

\begin{figure}[!htbp]
  \begin{center}
    \includegraphics[clip,width=0.45\textwidth]{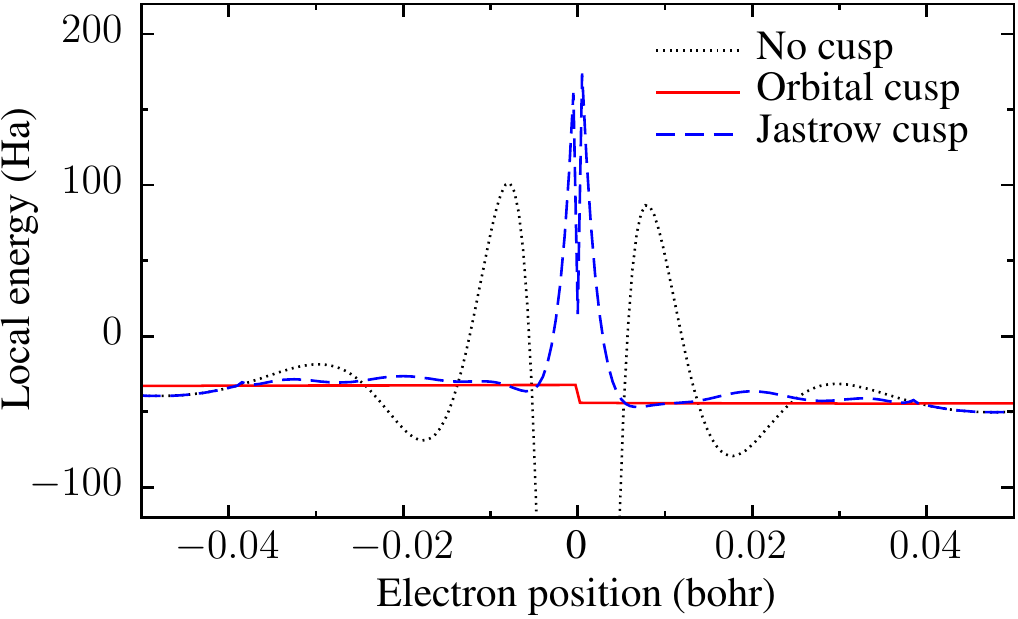}
  \end{center}
  \caption{ Local energy as a function of the $x$ coordinate of an
    electron in a carbon atom using a trial wave function which does
    not satisfy the electron-nucleus cusp condition (``no cusp'') and
    wave functions which impose the electron-nucleus cusp condition by
    modifying the Gaussian orbitals near the nucleus (``orbital
    cusp'') and via the Jastrow factor (''Jastrow cusp'').
    The nucleus is at $x=0$, and the other five electrons are at
    random positions.
    \label{fig:c_atom_cusp}
  }
\end{figure}

When cuspless one-electron orbitals are used in $\Psi_{\rm S}$, such
as orbitals expanded in Gaussians, one can enforce the
electron-nucleus Kato cusp conditions either by modifying the orbital
near the nucleus with an analytically correct form \cite{Ma_cusp_2005}
or by setting $\partial \chi / \partial r_{iI} = \Gamma_{iI}$ at
$r_{iI}=0$.
In practice, cusp-correcting the orbital is found to be the better
approach, because the error in the wave function occurs at a very
short range determined by the basis precision.
We illustrate this by plotting the local energy as a function of the
position of an electron moving through the nucleus of an all-electron
carbon atom in Fig.\ \ref{fig:c_atom_cusp}, where the local energy can
be seen to vary much less with the orbital correction than with a
cusp-enforcing short-ranged $\chi$ function.

At long range the Jastrow factor can be thought of as providing a
slowly varying envelope for the Slater wave function, describing the
behavior of interacting quasielectrons.
Indeed, the very-long-range two-electron contribution to the Jastrow
factor is exactly described by the random-phase approximation (RPA),
i.e., linear response theory within the Hartree approximation.
\cite{Bohm_1953}

In the Drummond-Towler-Needs (DTN) Jastrow factor
\cite{Drummond_jastrow_2004} the $u$, $\chi$, and $f$ functions are
parameterized as natural power expansions,
\begin{equation}
  \begin{split}
    u_P(r) & = t_P(r) \sum_{k} \alpha^P_k r^k \;, \\ \chi_S(r) & =
    t_S(r) \sum_{k} \beta^S_k r^k \;, \\ f_T(r_{ij}, r_{iI}, r_{iJ}) &
    = t_T(r_{iI}) t_T(r_{jI}) \sum_{k,l,m} \gamma^T_{klm} r_{ij}^k
    r_{iI}^l r_{jI}^m \;, \\
  \end{split}
\end{equation}
where the $\{\alpha\}$, $\{\beta\}$, and $\{\gamma\}$ are linear
parameters, and the $\{t(r)\}$ are cutoff functions of the form
\begin{equation}
  t(r) = (r-L)^C \Theta_{\rm H}(r-L) \;,
\end{equation}
where $L$ is an optimizable truncation length, $C$ is an integer
truncation order (typically $C=2$ or $3$), and $\Theta_{\rm H}(x)$ is
the Heaviside step function.

In periodic systems the interparticle distances in the expressions
above are evaluated between nearest periodic images only.
The truncation lengths for the homogeneous two-body terms ($u$) and
one-body terms ($\chi$) must therefore be constrained to be less than
or equal to $L_{\rm WS}$, where $L_{\rm WS}$ is half the nearest-image
distance; otherwise the trial wave function would have a discontinuous
gradient at the surface of the Wigner-Seitz cell of the simulation
cell.
Likewise, the truncation lengths for the inhomogeneous two-body terms
($f$) must be less than or equal to $L_{\rm WS}/2$.
Homogeneous two-body correlations in particular are long-range.
In order to cover the ``corners'' of the simulation cell, the DTN
Jastrow factor provides one- and two-electron cosine expansions.
The form of the two-body cosine expansion is
\begin{equation}
p({\bf r})=\sum_{s=1}{N_p} a_s^{P'} \sum_{{\bf G} \in \bigstar_s^+} \cos({\bf G}\cdot{\bf r}),
\end{equation}
where $\bigstar_s^+$ denotes every second
member of the $s$th star of simulation-cell reciprocal lattice points.
This term is cuspless, but it is not truncated at finite range and
hence it provides a description of long-range correlation effects.
It has the full symmetry of the simulation-cell Bravais lattice.
Since it describes long-range behavior, we generally only need
distinct parameters for particle pairs with different charges; the
same parameters can be used for parallel- and antiparallel-spin pairs.
The one-body cosine expansion has a similar form, but is expanded in
primitive-cell reciprocal lattice points, so that it has the full
symmetry of the primitive-cell Bravais lattice.
It is rarely used, since one-body Jastrow terms are not usually
long-ranged.
A situation in which it might be useful would be in a study of a
single charged defect in a simulation supercell.

Information about \textsc{casino}'s Jastrow factor implementation
beyond the DTN form can be found in Sec.\ \ref{sec:gjastrow}.
The use of a Jastrow factor typically allows VMC to recover about
80--95\% of the DMC correlation energy.

The quality of the DMC energy obtained with a Slater-Jastrow wave
function is limited by the quality of the nodes of $\Psi_{\rm S}$.
However it is possible to alter the nodal surface of the
Slater-Jastrow wave function by the use of a backflow transformation.
\cite{Feynman_backflow_1956, Lee_backflow_1981, Kwon_backflow_1998,
Holzmann_backflow_2003, LopezRios_backflow_2006}
The purpose of the backflow transformation is to provide a
parameterized, smooth transformation of the configuration ${\bf
  X}({\bf R})$ at which $\Psi_{\rm S}$ is evaluated to allow the
optimization of the location of its nodes.

The Slater-Jastrow-backflow wave function is
\begin{equation}
  \Psi({\bf R})
    = e^{\rm J({\bf R})}
      \Psi_{\rm S}\left({\bf X}({\bf R})\right) \;,
\end{equation}
where ${\bf X}=({\bf x}_1,\ldots,{\bf x}_N)$ is a set of quasiparticle
coordinates.
The $i$th quasiparticle coordinate is defined as the position of the
$i$th electron plus a parameterized backflow displacement which depends
on the positions of the other electrons in the system.
In the inhomogeneous backflow \cite{LopezRios_backflow_2006}
implementation in \textsc{casino}, the backflow displacement is,
much like a Jastrow factor, the sum of electron-electron,
electron-nucleus, and electron-electron-nucleus contributions,
\begin{equation}
  \begin{split}
  {\bf x}_i = \; &
      {\bf r}_i
    + \sum_{j\neq i}^N \eta_{P_{ij}}(r_{ij}) {\bf r}_{ij}
    + \sum_{I}^{N_{\rm n}} \mu_{S_{ij}}(r_{iI}) {\bf r}_{iI} \\
  & + \sum_{j\neq i}^N \sum_{I}^{N_{\rm n}} \left[
        \Phi_{T_{ijI}}(r_{ij}, r_{iI}, r_{jI}) {\bf r}_{ij} \right. \\
  &   \quad \quad \quad \quad
      \left. + \Theta_{T_{ijI}}(r_{ij}, r_{iI}, r_{jI}) {\bf r}_{iI}
      \right] \;,
  \end{split}
\end{equation}
where the parameterized functions $\eta$, $\mu$, $\Phi$, and $\Theta$
have the same form as those of the DTN Jastrow factor,
\begin{equation}
  \begin{split}
    \eta_P(r) & = t_P(r) \sum_{k} c^P_k r^k \;, \\
    \mu_S(r) & = t_S(r) \sum_{k} d^S_k r^k \;, \\
    \Phi_T(r_{ij}, r_{iI}, r_{iJ}) &
    = t_T(r_{iI}) t_T(r_{jI}) \sum_{k,l,m} \phi^T_{klm} r_{ij}^k
    r_{iI}^l r_{jI}^m \;, \\
    \Theta_T(r_{ij}, r_{iI}, r_{iJ}) &
    = t_T(r_{iI}) t_T(r_{jI}) \sum_{k,l,m} \theta^T_{klm} r_{ij}^k
    r_{iI}^l r_{jI}^m \;, \\
  \end{split}
\end{equation}
where the $\{c\}$, $\{d\}$, $\{\phi\}$, and $\{\theta\}$ are
optimizable parameters.
Note that since the backflow displacement is a vector function, the
electron-electron-nucleus term requires two separate functions
$\Phi_{\rm T}$ and $\Theta_{\rm T}$ to span the plane defined by the
two electrons and the nucleus.
There exist alternative forms of backflow in the literature, including
analytical, \cite{Holzmann_backflow_2003} recursive,
\cite{Taddei_iter_bf_2015} and orbital-dependent
\cite{Holzmann_orbdep_backflow_2019} formulations.
A deep-neural-network-based wave function has been proposed recently
consisting of a short multideterminant expansion populated with
flexible orbitals and backflow-like features.  \cite{Pfau_2019}

For a demonstration of the effect of backflow on the total energy for
the (very favorable) case of the HEG, see Fig.\ \ref{fig:HEG_mdet} in
Sec.\ \ref{sec:app_heg}, in which backflow recovers over half of the
fixed-node error in DMC\@.
In inhomogeneous systems, backflow can be expected to recover a
smaller but significant fraction of the fixed-node error.
\cite{LopezRios_backflow_2006}

\subsubsection{Optimization of wave-function parameters}
\label{sec:opt}

Two stable and efficient optimization methods are widely used in
modern QMC calculations.
The first approach is based on the
observation that the local energy $E_{\rm L}=(\hat{H}\Psi)/\Psi$ is
constant in configuration space if and only if $\Psi$ is an
eigenfunction of the Hamiltonian $\hat{H}$.
Hence, for any given sampling of points $\{{\bf R}\}$ in configuration
space, we may optimize the wave-function parameters by varying them to
minimize the spread of local energies. \cite{Umrigar_1988,
  Hammond_1994, Kent_1999b, Drummond_2005}

The spread of a set of local energies is most often quantified by the
variance.
In a typical variance-minimization calculation, VMC is used to sample
a set of configurations $\{{\bf R}\}$ distributed as $|\Psi|^2$, then
the free parameters in $\Psi$ are varied to minimize the variance of
the local energies evaluated at the configuration set sampled, then
VMC is used to generate a new set of configurations, and so on.
In practice the algorithm normally converges to the necessary level of
accuracy in 1--3 cycles.
Minimizing the variance of the local energies with respect to
wave-function parameters closely resembles the procedure for
least-squares fitting.

The parameters that recover the largest part of the correlation energy
appear linearly in the Jastrow exponent $J$.
The local energy is a quadratic function of such parameters, and hence
the variance of the set of local energies is a quartic function of
those parameters. \cite{Drummond_2005}
Thus it is possible to minimize the variance rapidly with respect to
those parameters, and also to search for global minima along lines in
parameter space.

Variance minimization is robust and efficient for optimizing
parameters in the Jastrow factor, reflecting the simple form of the
variance as a function of those parameters.
However, variance minimization is much less effective for optimizing
parameters that affect the nodal surface of the wave function.
The local energy of a configuration diverges if that configuration
lies on a node of the trial wave function; hence, as one adjusts a
parameter that affects the nodal surface, the variance of the local
energies diverges each time the nodal surface moves through one of the
configurations sampled.
As a result nonglobal minima are a significant problem, and variance
minimization will often converge very slowly with the number of
cycles.
We find that minimizing the mean absolute deviation from the median
local energy instead of the variance of the local energies can be
advantageous in these situations. \cite{Bressanini_2002}

In Table \ref{table:spread_min_neon} we compare the performance of
optimization methods based on minimizing different measures of the
spread of local energies for an all-electron neon atom.
The results confirm that minimizing the mean absolute deviation from
the median local energy performs relatively well in terms of the
resulting VMC energy, especially when optimizing parameters that
affect the nodal surface.

\begin{table*}[!htbp]
\caption{Average VMC energy and variance for an all-electron neon atom
  over eight cycles of VMC configuration generation/wave-function
  optimization.
  10,000 VMC-sampled configurations were used to perform each
  optimization.
  Four different measures of the spread of the local energies of the
  configurations were minimized to optimize the wave function: the
  unweighted variance of the local energies; the ``filtered''
  variance, in which outlying local energies are excluded from the
  variance estimate; the difference of the upper and lower quartiles
  of the set of local energies; and the mean absolute deviation from
  the median local energy. \label{table:spread_min_neon}}
\begin{center}
\begin{tabular}{lccr@{.}lcr@{.}lc}
\hline \hline

& \multicolumn{2}{c}{Optimizing Jastrow only} &
\multicolumn{3}{c}{Optimizing Jastrow + orbitals} &
\multicolumn{3}{c}{Optimizing Jastrow + backflow} \\

\raisebox{1.5ex}[0pt]{Objective function} & VMC energy (Ha) &
Var.\ (Ha$^2$) & \multicolumn{2}{c}{VMC energy (Ha)} & Var.\ (Ha$^2$)
& \multicolumn{2}{c}{VMC energy (Ha)} & Var.\ (Ha$^2$) \\

\hline

Variance & $-128.8824(6)$ & $1.180$ & ~~$-128$&$8931(5)$ & $0.959$ &
~~~$-128$&$9133(6)$ & $0.405$ \\

``Filtered'' variance & $-128.8826(3)$ & $1.180$ & $-128$&$8936(7)$ &
$0.958$ & $-128$&$9147(5)$ & $0.394$ \\

Quartile difference  & $-128.8837(1)$ & $1.199$ & $-128$&$8953(8)$ & $1.016$ &
$-128$&$91737(9)$ & $0.422$ \\

Mean abs.\ deviation  & $-128.8851(2)$ & $1.195$ & $-128$&$89957(8)$ &
$0.989$ & $-128$&$9178(2)$ & $0.420$ \\

\hline \hline
\end{tabular}
\end{center}
\end{table*}

An alternative approach is to minimize the energy expectation value in
accordance with the variational principle of quantum mechanics.
This works well for both Jastrow parameters and parameters in the
Slater wave function.
It also has the advantage that the importance-sampled DMC algorithm
can be shown to be of maximal efficiency when the VMC energy is
minimized. \cite{Ceperley_1986}

To minimize the energy expectation value, we use a method which
transforms the stochastic optimization problem into the
diagonalization of the Hamiltonian matrix in an approximate finite
basis.  \cite{Nightingale_2001, Toulouse_2007, Umrigar_2007}
Suppose our initial wave function is $\Psi_{{\bm \alpha}_0}$, where
${\bm \alpha}_0$ is a vector of $P$ parameters.
Let us choose our (nonorthogonal) basis set to be the initial wave
function and its derivatives with respect to the parameters,
$\{\Psi_{{\bm \alpha}_0},(\partial\Psi/\partial \alpha_1)_{{\bm
    \alpha}_0},(\partial\Psi/\partial \alpha_2)_{{\bm
    \alpha}_0},\ldots \}$.
We may evaluate the $(P+1) \times (P+1)$ matrix $H$ of the Hamiltonian
with respect to these basis functions, together with the overlap
matrix $S$ between the basis functions by VMC sampling of $|\Psi_{{\bm
    \alpha}_0}|^2$, then solve the resulting generalized eigenproblem
$H{\bf c}={\cal E}S{\bf c}$ to find the ground-state eigenvector ${\bf
  c}$ and corresponding eigenvalue ${\cal E}$.
Suppose we choose the normalization of the $(P+1)$-dimensional
eigenvector ${\bf c}$ such that the coefficient of the first basis
function is 1, i.e., we write ${\bf c}=(1,\Delta \alpha_1,\Delta
\alpha_2,\dots)$, where the second and subsequent elements of ${\bf
  c}$ define a $P$-dimensional vector $\Delta {\bm \alpha}$.
The resulting approximation to the ground-state wave function is
\begin{eqnarray} \Psi_{{\bm \alpha}} & = & \Psi_{{\bm \alpha}_0}+\Delta
\alpha_1(\partial\Psi/\partial \alpha_1)_{{\bm \alpha}_0}+\Delta
\alpha_2(\partial\Psi/\partial \alpha_2)_{{\bm \alpha}_0}+\ldots
\nonumber \\ & = & \Psi_{{\bm \alpha}_0+\Delta {\bm \alpha}}+{\cal
  O}(\Delta \alpha^2). \label{eq:emin_update} \end{eqnarray}
We thus have a new parameter set ${\bm \alpha}={\bm \alpha}_0+\Delta
{\bm \alpha}$, and can repeat the whole process.
Once the process converges, i.e.\ the parameters cease to change over
subsequent cycles, the quadratic ${\cal O}(\Delta \alpha^2)$ term in
Eq.\ (\ref{eq:emin_update}) vanishes, so that we locate a minimum of the energy
(provided our VMC estimates of $H$ and $S$ are sufficiently precise).

A few technical tricks are required to make this work in practice.
Firstly, it can be shown that it is much better not to impose
Hermiticity on the VMC estimates of $H$ and $S$ obtained with a finite
sampling of configurations; \cite{Umrigar_2007} the resulting
algorithm then works exactly both in the limit of an infinite number
of VMC samples and in the limit that the basis functions span an
invariant subspace of the Hamiltonian operator.
One can choose a parameter-dependent normalization for the basis
functions such that they are orthogonal to an appropriate linear
combination of $\Psi_{{\bm \alpha}_0}$ and $\Psi_{\bm \alpha}$.
Finally one can artificially increase the diagonal elements of $H$ to
prevent large steps in parameter space. \cite{Umrigar_2007}
Putting these tricks together results in a stable and effective
energy-minimization algorithm.

\subsection{Diffusion quantum Monte Carlo}

\subsubsection{Fixed-phase approximation}
\label{sec:fixed_phase}

Consider a Hamiltonian $\hat{H}=-\frac 1 2 \nabla^2 + V$, where $V$ is
the total potential energy.
Let $\Phi=|\Phi|e^{i\chi}$ be a complex many-electron spatial wave
function.
By fermionic antisymmetry, the phase $\chi$ must change by $\pi$
whenever two same-spin electrons are exchanged.
Then
\begin{eqnarray}
  e^{-i\chi} \hat{H} \Phi & = & \left[ \frac 1 2 \left(
    -\nabla^2+\left|\nabla\chi\right|^2 \right) + V \right] |\Phi|
  \nonumber \\ &   & {} +i \left[ \frac 1 2 \left( -2\nabla \chi \cdot
    \nabla - \nabla^2 \chi \right) \right] |\Phi| \nonumber \\ &
  \equiv & \left[ \hat{\cal H}_{\chi} + i\hat{\cal K}_{\chi} \right]
  |\Phi|.
\end{eqnarray}
We then find that
\begin{equation}
  \left< \Phi \left| \hat{H} \right| \Phi \right> = \left< |\Phi|
  \left| \hat{\cal H}_\chi \right| |\Phi| \right>.
\end{equation}
So the ground-state eigenvalue of the fixed-phase Schr\"odinger
equation $\hat{\cal H}_\chi \phi_0 = E_0\phi_0$ is equal to the
expectation value of the Hamiltonian $\hat{H}$ with respect to
$|\phi_0|e^{i\chi}$, which is greater than or equal to the fermionic
ground-state energy of $\hat{H}$ by the variational principle,
becoming equal in the limit that the fixed phase $\chi$ is exactly
equal to that of the fermionic ground state. \cite{Ortiz_1993}

In the following we will adopt the fixed-phase approximation,
\cite{Ortiz_1993} in which it is assumed that the phase of wave
function $\Phi$ is fixed to be the same as the phase $\chi$ of the
trial wave function $\Psi$, but its modulus $|\Phi|$ is allowed to
vary.
The fixed-phase approximation is the complex generalization of the
fixed-node approximation. \cite{Anderson_1976}
The latter corresponds to the case in which the trial wave function is
real, so that at each point in configuration space the phase is either
$\chi=0$ or $\pi$.

The fixed-phase approximation with an antisymmetric trial wave
function gives us a variational principle for the lowest-energy
antisymmetric eigenstate. \cite{Hammond_1994}
Likewise, the fixed-phase approximation with a trial wave function
that transforms as a one-dimensional (1D) irreducible representation
of the symmetry group of the Hamiltonian gives a variational principle
with respect to the lowest-energy exact wave function that transforms
as that 1D irreducible representation. \cite{Foulkes_1999}
Thus the DMC energy only satisfies the variational principle for the
ground state and certain excited states.
Nevertheless, DMC always gives the energy of any excited state exactly
if the phase is exact for that state.
Hence we can use fixed-phase DMC to calculate approximate
excited-state energies by using an appropriate trial wave function,
even for excited states for which we do not have the variational
principle.
So we can calculate excitation energies (points on the band structure
for periodic systems) via differences in total energy.
%

\subsubsection{Imaginary-time Schr\"odinger equation}

\begin{figure}[!htbp]
  \begin{center}
    \includegraphics[clip,width=0.45\textwidth]{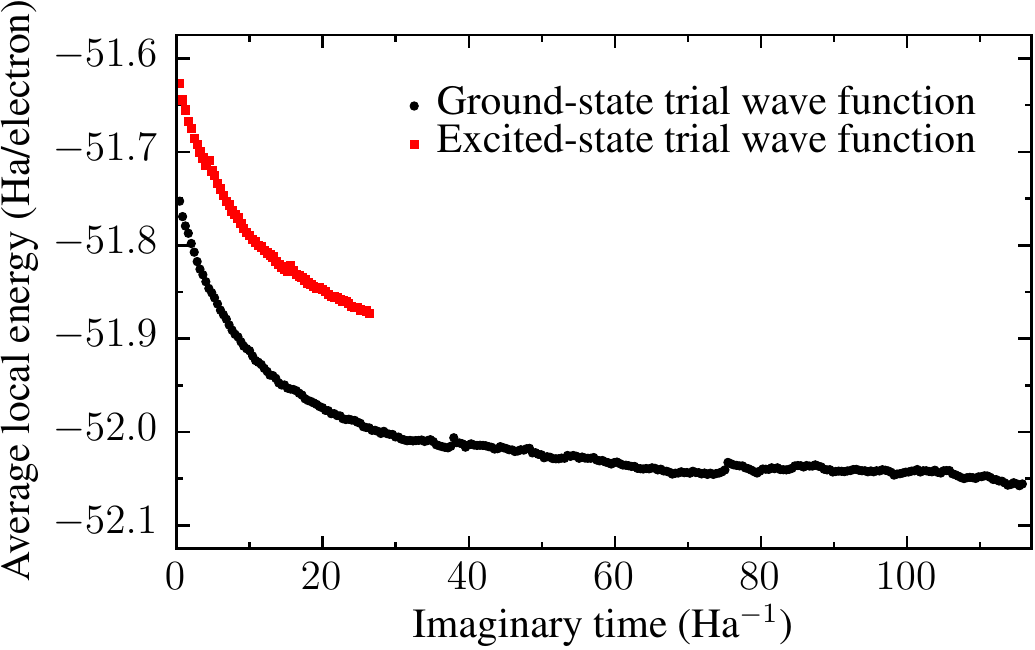}
  \end{center}
  \caption{ Decay of the average local energy of a set of walkers
    whose dynamics is governed by the fixed-phase imaginary-time
    Schr\"odinger equation for the ground state and an excited state
    of a 610-electron 2D HEG at density parameter $r_{\rm s}=10$.
    Results are shown for (i) a real, closed-shell, ground-state,
    Slater--Jastrow trial wave function with plane-wave orbitals and
    (ii) a complex, excited-state wave function in which a single
    electron has been added to a plane-wave orbital outside the Fermi
    surface without reoptimization of the ground-state Jastrow factor.
    The DMC time step is $0.4$ Ha$^{-1}$ and the target population is
    12,000 walkers, so that the statistical error on the energy at
    each time step is small.
    \label{fig:decay}
  }
\end{figure}

Now consider the fixed-phase imaginary-time Schr\"odinger equation,
\begin{eqnarray}
  \label{eq:itse}
  \left[ \hat{\cal H}_\chi - E_{\rm T} \right] |\Phi| & = & \left[
    -\frac 1 2 \nabla^2 + \frac 1 2 |\nabla \chi|^2 + V -E_{\rm T}
    \right] |\Phi| \nonumber \\ & = & -\frac{\partial |\Phi|}{\partial
    t},
\end{eqnarray}
where $\Phi({\bf R},t)=|\Phi|e^{i\chi}$ is an imaginary-time-dependent
wave function whose complex phase $\chi$ is everywhere equal to that
of the trial wave function $\Psi=|\Psi|e^{i\chi}$, and the reference
energy $E_{\rm T}$ is an offset in the energy origin.
The time dependence is separable, so we may write the solution as
\begin{equation}
  \label{eq:im_time_sep}
  |\Phi| = \sum_{n=0}^\infty c_n \phi_n e^{-(E_n-E_{\rm T})t},
\end{equation}
where $E_n$ and $\phi_n$ are the $n$th eigenvalue and eigenfunction of
the fixed-phase Hamiltonian $\hat{\cal H}_\chi$.
Excited states die away exponentially with increasing $t$ compared
with the ground state, as shown in Fig.\ \ref{fig:decay}.
If $E_{\rm T}=E_0$ and the initial conditions have $c_0 \neq 0$ then,
in the limit $t \rightarrow \infty$, $|\Phi|$ is proportional to
$\phi_0$.
The ground-state component of $|\Phi|$ is thus ``projected out.''

Let $f=|\Phi| |\Psi|$ be the \textit{mixed} distribution, where
$\Psi$ is the trial wave function.
Then $f$ is real and positive and Eq.\ (\ref{eq:itse}) leads to
\begin{equation}
  \label{eq:isitse}
  - \frac 1 2 \nabla^2 f + \nabla \cdot \left[ {\bf V}^{\rm r} f
    \right] + \left[ E_{\rm L}^{\rm r} - E_{\rm T} \right] f = -
  \frac{\partial f}{\partial t},
\end{equation}
where ${\bf V}^{\rm r}={\rm Re}[(\nabla \Psi)/\Psi]$ is the real part
of the drift velocity and $E_{\rm L}^{\rm r}={\rm Re}(E_{\rm L})$ is
the real part of the local energy.

Let $\{|{\bf R}\rangle\}$ be many-body position basis vectors,
normalized such that $\langle {\bf R}|{\bf R}'\rangle=\delta({\bf
  R}-{\bf R}')$, and let $|{\bf P}\rangle=(2\pi)^{-3N/2} \int e^{i{\bf
    R} \cdot {\bf P}} |{\bf R}\rangle \, {\rm d}{\bf R}$ be the
corresponding momentum basis vectors.
Let $\hat{\bf R}$ and $\hat{\bf P}$ be the position and momentum
operators.
In Dirac notation the mixed distribution can be written as $f({\bf
  R})=\langle {\bf R}|f\rangle$, and Eq.\ (\ref{eq:isitse}) can be
written as
\begin{equation}
  [ \hat{F} + E_{\rm L}^{\rm r}(\hat{\bf R}) - E_{\rm T} ] |f\rangle =
  - \frac{\rm d}{{\rm d}t}|f\rangle,
\end{equation}
where $\hat{F}=(1/2)\hat{\bf P}^2+i\hat{\bf P} \cdot {\bf V}^{\rm
  r}(\hat{\bf R})$ is the Fokker-Planck operator.

\subsubsection{Propagation in imaginary time \label{sec:propagation}}

The mixed distribution at imaginary time $t+\tau$ can be written as
\begin{equation}
  f({\bf R}, t+\tau) = \int G({\bf R} \leftarrow {\bf R}^{\prime},
  \tau) f({\bf R}^{\prime}, t) \, {\rm d}{\bf R}^{\prime},
\end{equation}
where the Green's function
\begin{eqnarray}
  G({\bf R} \leftarrow {\bf R}^{\prime},\tau) & = & \langle {\bf R} |
  e^{-\tau(\hat{F}+E_{\rm L}^{\rm r}(\hat{\bf R})-E_{\rm T})} | {\bf
    R}^\prime \rangle \nonumber \\ & \approx & \langle {\bf R} |
  e^{-\tau \hat{F}} | {\bf R}^\prime \rangle e^{- \frac \tau 2 \left[
      E_{\rm L}^{\rm r}({\bf R}) + E_{\rm L}^{\rm r}({\bf R}^\prime) -
      2E_{\rm T} \right]} \nonumber \\ & & + {\cal O}(\tau^3).
\end{eqnarray}
is the solution of Eq.\ (\ref{eq:isitse}) satisfying the initial
condition $G({\bf R} \leftarrow {\bf R}^{\prime}, 0) = \delta({\bf R}
- {\bf R}^{\prime})$.

Let $\normord{\hat A}$ be the normal-ordered version of an operator
$\hat A$, in which all $\hat{\bf P}$ operators are moved to the left
of all $\hat{\bf R}$ operators (as if they commuted) within each term.
Note that $\normord{\hat F}=\hat F$, so $\normord{e^{-\tau \hat F}} =
1-\tau \hat{F} + {\cal O}(\tau^2) = e^{-\tau \hat{F}} + {\cal
  O}(\tau^2)$.
Hence
\begin{eqnarray}
  & & \langle {\bf R} | \normord{e^{-\tau \hat F}} | {\bf R}^\prime
  \rangle \nonumber \\ & & \hspace{4em} {} = \int \langle {\bf R} |
          {\bf P} \rangle \langle {\bf P} | \normord{e^{-\tau \left[
                \frac{\hat{\bf P}^2} 2 + i\hat{\bf P} \cdot {\bf
                  V}^{\rm r}(\hat{\bf R}) \right]} } | {\bf R}^\prime
          \rangle \, {\rm d}{\bf P} \nonumber \\ & & \hspace{4em} {}
          = \int
          \langle {\bf R} | {\bf P} \rangle e^{-\tau \left[ \frac{{\bf
                  P}^2} 2 + i{\bf P} \cdot {\bf V}^{\rm r}({\bf
                R}^\prime) \right]} \langle {\bf P} | {\bf R}^\prime
          \rangle \, {\rm d}{\bf P} \nonumber \\ & & \hspace{4em} {}
          = \int
          \frac{e^{i{\bf P} \cdot ({\bf R}-{\bf R}^\prime)}}
               {(2\pi)^{3N}} e^{- \tau \left[ \frac{{\bf P}^2} 2 +
                   i{\bf P} \cdot {\bf V}^{\rm r}({\bf R}^\prime)
                   \right]} \, {\rm d}{\bf P} \nonumber \\ & &
                   \hspace{4em} {}
          = \frac 1 {(2\pi \tau)^{3N/2}} e^{ - \frac 1 {2\tau}
                 \left| {\bf R}-{\bf R}^\prime - \tau {\bf V}^{\rm
                   r}({\bf R}^\prime) \right|^2}.
\end{eqnarray}
The last line is the Langevin or drift-diffusion Green's function,
describing diffusion of particles (``walkers'') in a $3N$-dimensional
fluid of time-independent velocity field ${\bf V}^{\rm r}({\bf
  R}^\prime)$ at small time step $\tau$.
Physically, the approximation of using the normal-ordered Green's
function is equivalent to assuming the drift velocity to be constant
between ${\bf R}$ and ${\bf R}^\prime$.
So the Green's function for Eq.\ (\ref{eq:isitse}) is
\begin{eqnarray}
  G({\bf R} \leftarrow {\bf R}^{\prime},\tau) & \approx & \langle {\bf
    R} | \normord{e^{-\tau \hat{F}}} |{\bf R}^\prime \rangle \, e^{ -
    \frac \tau 2 [ E_{\rm L}^{\rm r}({\bf R}) + E_{\rm L}^{\rm r}({\bf
        R}^{\prime}) - 2 E_{\rm T} ] } \nonumber \\ & & {}      +
  {\cal O}(\tau^2) \nonumber \\ & \equiv & G_{\rm DMC}({\bf R}
  \leftarrow {\bf R}^{\prime}, \tau) + {\cal O}(\tau^2).
\end{eqnarray}
The branching factor $e^{-\frac \tau 2 [ E_{\rm L}^{\rm r}({\bf R}) +
    E_{\rm L}^{\rm r}({\bf R}^{\prime}) - 2 E_{\rm T} ]}$ is the
solution of Eq.\ (\ref{eq:isitse}) without the first two terms on the
left-hand side; it represents exponential growth/decay in the density
of walkers at each point in configuration space.
The DMC Green's function therefore describes the evolution of the
density of a set of walkers drifting, diffusing, and breeding or dying
in a $3N$-dimensional space.

The Trotter-Suzuki expression for the Green's function for a
macroscopic length of imaginary time $M\tau$ is
\begin{widetext}
\begin{eqnarray}
  G({\bf R}\leftarrow{\bf R}^\prime,M\tau) & = & \langle {\bf R} |
  e^{-M\tau(\hat{F}+\hat{E_{\rm L}^{\rm r}}-E_{\rm T})} | {\bf
    R}^\prime \rangle \nonumber \\ & \approx & \int \cdots \int
  \langle {\bf R} | e^{-\tau(\hat{F}+\hat{E_{\rm L}^{\rm r}}-E_{\rm
      T})} | {\bf R}^{\prime \prime} \rangle \cdots \langle {\bf
    R}^{\prime \prime \prime} | e^{-\tau(\hat{F}+\hat{E_{\rm L}^{\rm
        r}}-E_{\rm T})} | {\bf R}^\prime \rangle \,
    {\rm d}{\bf R}^{\prime \prime} \ldots
    {\rm d}{\bf R}^{\prime \prime \prime} + {\cal O}(M\tau^3)
  \nonumber \\ & \approx & \int \cdots \int G_{\rm DMC}({\bf R}
  \leftarrow {\bf R}^{\prime \prime}, \tau) \cdots G_{\rm DMC}({\bf
    R}^{\prime \prime \prime} \leftarrow {\bf R}^\prime,\tau) \,
    {\rm d}{\bf R}^{\prime \prime} \ldots
    {\rm d}{\bf R}^{\prime \prime \prime} + {\cal O}(M \tau^2).
\end{eqnarray}
\end{widetext}
The approximation to the Green's function over a finite interval can
be made arbitrarily accurate by dividing the interval into
sufficiently small slices of imaginary time.
We can therefore use the DMC Green's function to propagate $f$ to
large imaginary time (where $f=|\phi_0| |\Psi|$) using a finite, small
time step $\tau$:
\begin{equation}
  f({\bf R},t) = \int G({\bf R}\leftarrow {\bf R}^\prime,t) f({\bf
    R}^\prime,0) \, {\rm d}{\bf R}^\prime,
\end{equation}
where the propagation to time $t=M\tau$ is carried out in $M$ short
steps of length $\tau$ using $G_{\rm DMC}$.

The use of a nonzero time step continually introduces errors, even as
the evolution in imaginary time projects out the ground-state
component.
At large imaginary time we may write $f=|\phi_0| |\Psi|+\Delta$, where
$\Delta$ is the time-step error in the mixed distribution.
The error in $G_{\rm DMC}$ per time step is ${\cal O}(\tau^2)$, so
error in $f$ is introduced at rate ${\cal O}(\tau)$.
Error is removed at a rate $\sim -\Delta / \tau_{\rm corr}$, where
$\tau_{\rm corr}$ is the decorrelation period in imaginary time (see
Sec.\ \ref{sec:time_steps}).
In steady state, these rates balance.
Hence $\Delta \sim \tau_{\rm corr} \tau$, i.e., the time-step error in
the mixed distribution is ${\cal O}(\tau)$.

At any given moment in a DMC simulation, $f$ is represented by a
discrete population of ``walkers'' in configuration space:
\footnote{The approximation $f \approx \sum_\alpha w_\alpha
  \delta({\bf R}-{\bf R}_\alpha)$ only makes sense under an integral
  sign.
At any instant the error in this approximation is proportional to
$1/\sqrt{W}$, where $W$ is the number of walkers.
The error term is, by construction, zero on average.}
\begin{equation}
  f({\bf R},t) = \left< \sum_\alpha w_\alpha \delta({\bf R} - {\bf
    R}_\alpha) \right>,
\end{equation}
where ${\bf R}_\alpha$ is the position of walker $\alpha$ and
$w_\alpha$ is its weight and the angled brackets denote an ensemble
average.
Ensemble averaging commutes with linear operations such as
differentiation.
The mixed distribution after one time step $\tau$ is
\begin{equation}
  f({\bf R},t+\tau) = \left< \sum_\alpha w_\alpha G_{\rm DMC}({\bf R}
  \leftarrow {\bf R}_\alpha,\tau) \right>.
\end{equation}
It is now clear that the Green's functions can be treated as
transition-probability densities; the ensemble average then has the
correct behavior.
In summary, we must simulate a large population of walkers that, over
the course of one time step, drift by $\tau{\bf V}^{\rm r}({\bf
  R}_\alpha)$ and diffuse (are displaced by a random vector,
Gaussian-distributed with variance $\tau$); finally the branching
factor is absorbed into a new weight for each walker.
This is usually done by a random branching or dying process, such that
the expectation value of the number of unit-weighted walkers that
continue from the walker's current position is equal to the weight of
the walker after the move.
The branching/dying algorithm avoids the situation in which one walker
gathers exponentially more weight than the others.

The Green's function for Eq.\ (\ref{eq:itse}) is $\langle {\bf R} |
e^{-\tau (\hat{\cal H}_\chi-E_{\rm T})} | {\bf R}^\prime \rangle$.
Using the importance-sampling transformation, we can therefore write
the Green's function for Eq.\ (\ref{eq:isitse}) as
\begin{equation}
  G({\bf R} \leftarrow {\bf R}^\prime,\tau) = \Psi({\bf R}) \langle
  {\bf R} | e^{-\tau (\hat{\cal H}_\chi-E_{\rm T})} | {\bf R}^\prime
  \rangle \Psi^{-1}({\bf R}^\prime).
\end{equation}
Now $e^{-\tau (\hat{\cal H}_\chi-E_{\rm T})}$ is Hermitian, so
\begin{equation}
  |\Psi({\bf R}^\prime)|^2 G({\bf R} \leftarrow {\bf R}^\prime,\tau) =
  |\Psi({\bf R})|^2 G({\bf R}^\prime \leftarrow {\bf R},\tau).
\end{equation}
The approximation that ${\bf V}^{\rm r}({\bf R})$ is constant between
${\bf R}^\prime$ and ${\bf R}$ violates this detailed-balance
condition at finite time steps.
We may reimpose this important condition using a Metropolis-style
accept/reject step.
As with VMC, it is more efficient to propose individual-electron moves
than whole-configuration moves.
Although the accept/reject step does not formally change the scaling
of the time-step error in the DMC Green's function, in practice it
enormously reduces its magnitude.

The local energy and the drift velocity diverge as the wave function
goes to zero, which is problematic at finite time steps because the
probability of encountering arbitrarily large drift and branching
terms is nonzero.
Umrigar \textit{et al.}\cite{Umrigar_1993}\ addressed this problem by
replacing the drift velocity of each electron ${\bf v}_i^{\rm r}$ with
\begin{equation}
  {\bar {\bf v}}_i^{\rm r} =
    \frac{-1 + \sqrt{1+2a(v_i^{\rm r})^2 \tau}}
         {a(v_i^{\rm r})^2\tau}
    {\bf v}_i^{\rm r} \;,
\end{equation}
where $0<a<1$ is a parameter of the algorithm, and the local energy in
the branching factor with
\begin{equation}
  \label{eq:unr_branch}
  {\bar E}_{\rm L}^{\rm r} = E_{\rm best} -
                     \frac {\sqrt{\sum_i ({\bar v}_i^{\rm r})^2}}
                           {\sqrt{\sum_i (v_i^{\rm r})^2}}
                     \left( E_{\rm L}^{\rm r} - E_{\rm best} \right) \;,
\end{equation}
where $E_{\rm best}$ is the current estimate of the DMC energy.
The resulting change in the DMC Green's function is ${\cal
  O}(\tau^2)$, and hence does not affect the DMC results in the limit
of zero time step.
This limiting scheme successfully eliminates the stability issues, but
Eq.\ (\ref{eq:unr_branch}) results in size-inconsistent DMC energies
at fixed $\tau$ and substantial time-step errors.
\cite{Zen_timestep_2016}
Instead, Zen \textit{et al.}\ proposed using a hard-limiting scheme,
\begin{equation}
  {\bar E}_{\rm L}^{\rm r} =
  \left\{
  \begin{array}{lcl}
    E_{\rm best} + E_{\rm cut} &,~&
      E_{\rm L}^{\rm r} > E_{\rm best}+E_{\rm cut} \\
    E_{\rm L}^{\rm r}                  &,~&
      \left|E_{\rm L}^{\rm r}-E_{\rm best}\right| < E_{\rm cut} \\
    E_{\rm best} - E_{\rm cut} &,~&
      E_{\rm L}^{\rm r} < E_{\rm best}-E_{\rm cut} \\
  \end{array}
  \right. \;,
\end{equation}
where $E_{\rm cut} = \alpha\sqrt{N/\tau}$ and $\alpha=0.2$, which
restores size consistency and greatly reduces time-step errors.
\cite{Zen_timestep_2016}

Ensuring that the total number of walkers remains near a finite, fixed
target population $W$ can be achieved by adjusting the energy offset
$E_{\rm T}$ according to the instantaneous population.
However population-control mechanisms inevitably introduce bias.
Suppose the local energies are mostly less than $E_0$.
Then the population will try to increase, but the population-control
mechanism counteracts this.
Now suppose the local energies are mostly greater than $E_0$.
Then the population will try to decrease, but the population-control
mechanism counteracts this.
In either case, the average local energy increases as a result, so
that population control introduces a positive bias into the DMC
energy.
Since fluctuations in the average local energy and population are
correlated and proportional to $1/\sqrt{C}$, population-control bias
goes as $1/C$. \cite{Umrigar_1993}

\begin{figure}[!htbp]
  \begin{center}
    \includegraphics[clip,width=0.45\textwidth]{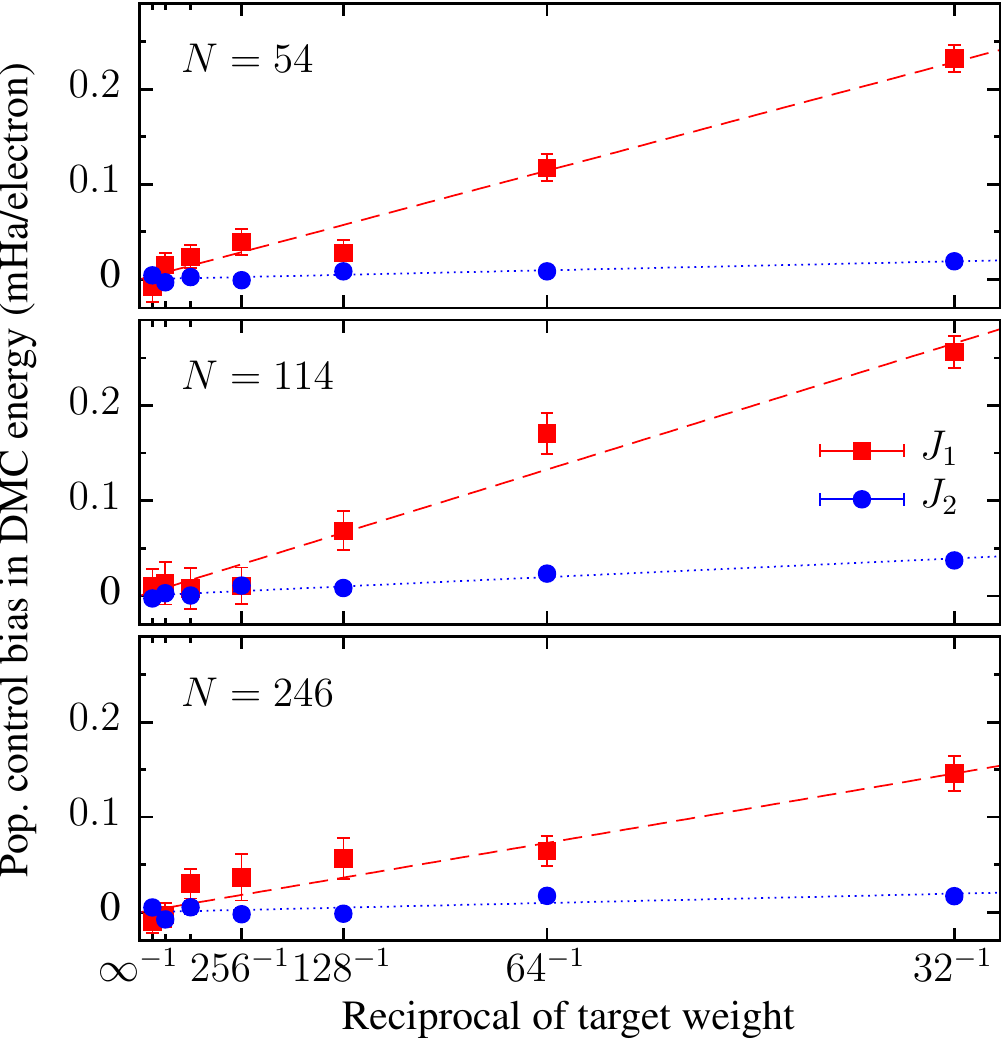}
  \end{center}
  \caption{Population-control bias in the $\Gamma$-point DMC energy
    per electron of a 3D paramagnetic electron gas of density
    parameter $r_{\rm s}=4$ at three different system sizes using a
    fixed time step of $0.1$ Ha$^{-1}$.
    Two different DTN Jastrow factors were used:\ $J_1$, consisting of
    a $u$ term with cutoff length $L=L_{\rm WS}/2$, and $J_2$,
    consisting of a better-quality $u$ term with $L=L_{\rm WS}$.
    The lines are linear fits to the data, and the bias is measured
    with respect to the value of the corresponding fit at infinite
    target population.
    \label{fig:pop_bias_size}
  }
\end{figure}

Population-control bias in the DMC energy per electron is roughly
independent of system size and decreases as the quality of the trial
wave function is improved, because the branching factors remain close
to 1 if the local energy is nearly constant in configuration space.
Figure \ref{fig:pop_bias_size} demonstrates both of these effects, and
also shows that population-control bias is typically small and
can be removed by linear extrapolation for systems of moderate size.
However, at large system size, the statistical correlation of walkers
following branching events causes the target population required to
eliminate population-control bias to scale exponentially with system
size.  \cite{Nemec_2010}
Issues with the convergence of the DMC energy as a function of target
population have been reported for large systems with simple trial
wave functions. \cite{Boninsegni_popbias_2012}
It is therefore important to use good trial wave functions and
restrict the application of DMC to systems with fewer than
$1000$--$2000$ electrons to keep population-control bias at bay.

No explicit steps are required to enforce the fixed-phase
approximation: it is enforced implicitly by our representation of $f$
as a probability density, ensuring it is always real and nonnegative.

A typical DMC calculation has two phases: in the equilibration phase,
we allow excited-state components of $|\Phi|$ to die away; in the
statistics-accumulation phase, we continue to propagate walkers, but
gather energy data.
The equilibration imaginary-time scale is often the same as the
decorrelation time scale $\tau_\text{corr}$ over which the walker
distribution is reset to the ground-state distribution.
In practice we usually equilibrate over a period that is several times
larger than the decorrelation time, to ensure that the fixed-phase
excited-state components of the DMC wave function are exponentially
small.
This is discussed further in Sec.\ \ref{sec:time_steps}.

Noting that $E_{\rm L}^{\rm r}={\rm Re}(\Psi^{-1}\hat{H}\Psi) =
|\Psi|^{-1}\hat{\cal H}_\chi |\Psi|$ and that $\hat{\cal H}_\chi
\phi_0=E_0\phi_0$, the mixed estimate of the energy is equal to the
fixed-phase ground-state energy $E_0$, which is an upper bound on the
fermion ground-state energy $E_0^{\rm F}$:
\begin{eqnarray}
  \left< E_{\rm L}^{\rm r} \right>_{\phi_0|\Psi|} = \frac{\int \phi_0
    |\Psi| E_{\rm L}^{\rm r} \, {\rm d}{\bf R}} {\int \phi_0 |\Psi| \,
    {\rm d}{\bf R}} & = & \frac{\left< \phi_0 \left| \hat{\cal H}_\chi
    \right| |\Psi| \right>} {\left< \phi_0 \left| \right. |\Psi|
    \right>}
  \nonumber \\ & = & E_0 \geq E_0^{\rm F}.
\end{eqnarray}

For operators $\hat{A}$ that do not commute with the Hamiltonian the
mixed estimate $\left< A_{\rm L}^{\rm r} \right>_{\phi_0|\Psi|}$ is
not equal to the pure estimate $\langle \phi_0 | \hat{A} | \phi_0
\rangle / \langle \phi_0 | \phi_0 \rangle$.
In general the difference is linear in the error in the trial wave
function $\Psi$.
However, for local operators $\hat{A}$, the DMC mixed estimate may be
combined with the VMC estimate to give a so-called extrapolated
estimate, $2\left< A_{\rm L}^{\rm r} \right>_{\phi_0|\Psi|}-\left<
A_{\rm L}^{\rm r} \right>_{|\Psi|^2}$, which has an error that is
second order in the error in the trial wave function.
\cite{Ceperley_1979}

\subsubsection{Time step and decorrelation period}
\label{sec:time_steps}

\begin{figure}[!htbp]
  \begin{center}
    \includegraphics[clip,width=0.45\textwidth]{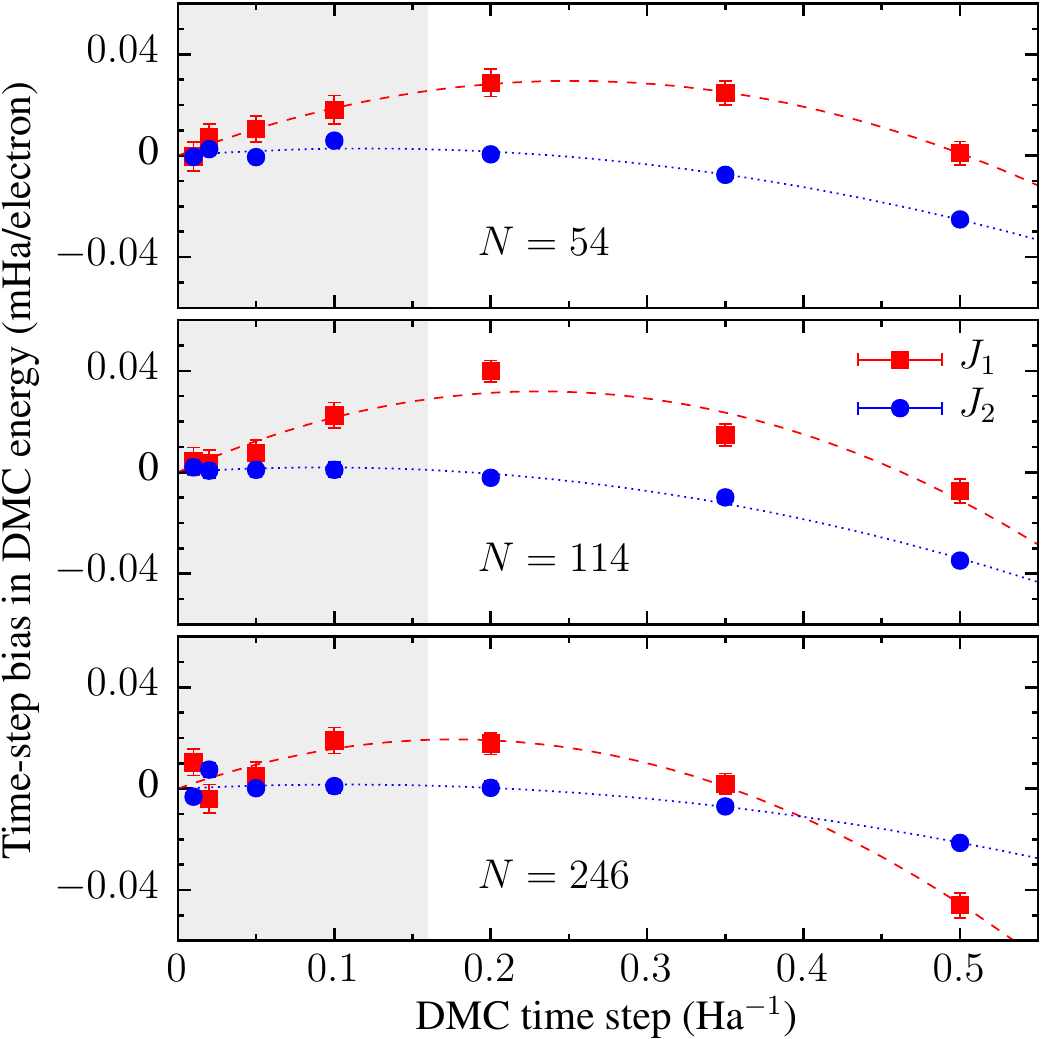}
  \end{center}
  \caption{ Time-step bias in the $\Gamma$-point DMC energy per
    electron of a 3D paramagnetic electron gas of density parameter
    $r_{\rm s}=4$ at three different system sizes using a fixed target
    population of $512$ walkers.
    Two different DTN Jastrow factors were used:\ $J_1$, consisting of
    a $u$ term with a fixed cutoff length $L=7.82$ bohr, which
    equals $L_{\rm WS}/2$ at $N=114$, and $J_2$, consisting of a
    better-quality $u$ term with $L=L_{\rm WS}$.
    The lines are quadratic fits to the data, and the bias is measured
    with respect to the value of the fit at zero time step.
    The shaded areas indicate the range of time steps for which the
    bias is expected to be linear from length-scale considerations
    ($\tau \lesssim 0.01r_{\rm s}^2$).  \cite{Lee_2010}
    \label{fig:dt_bias_size_dep}
  }
\end{figure}

The ${\cal O}(\tau)$ error in the mixed distribution gives an ${\cal
  O}(\tau)$ time-step bias in the mixed estimator.
Time-step bias vanishes in the limit of zero time step and is linear
in the time step for sufficiently small $\tau$.
Figure \ref{fig:dt_bias_size_dep} demonstrates that the time-step bias
per particle is relatively small in this linear regime, does not get
more severe in larger systems, is greatly reduced if the trial wave
function is good, and largely cancels out of energy differences
\cite{Zen_timestep_2016, Hunt_2018} if trial wave functions of similar
quality are used.
In order to obtain accurate total energies one must either (i) use a
sufficiently small time step that the bias is negligible or (ii)
perform simulations at different time steps and extrapolate to zero
time step.

To be sure that we are in the small-time-step, linear-bias regime, the
root-mean-square (RMS) distance diffused by each electron each time
step ($\sqrt{3 \tau}$) should be less than or equal to the
\textit{shortest} length scale in the system.
On the other hand, the RMS distance diffused by each electron over the
equilibration period $\sqrt{N_{\rm eq} \tau d}$ should be greater than
or equal to the \textit{longest} relevant length scale $L$ in the
problem, where $N_{\rm eq}$ is number of equilibration iterations and
$d$ is the dimensionality of the diffusion process ($d=3$ for a
cluster or a bulk material, $d=2$ for a 2D material, and $d=1$ for a
polymer).
DMC is therefore especially challenging for systems with widely
separated length scales, since the shorter length scale determines the
required time step, while the longer length scale determines the
decorrelation period.

From a more rigorous point of view, Eq.\ (\ref{eq:im_time_sep})
implies that the imaginary-time scale $\tau_\text{corr}$ over which
fixed-phase DMC equilibrates is the reciprocal of the difference
between the fixed-phase ground-state energy $E_0$ and the fixed-phase
first-excited-state energy $E_1$.
The fixed-phase energy gap $E_1-E_0$ is an approximation to the gap
between the ground-state energy and the energy of the first excited state
whose wave function transforms as the same 1D irreducible
representation of the symmetry group of the Hamiltonian as the trial
wave function.  $E_1-E_0$ is greater than or equal to the true energy
gap.
As a result, DMC usually equilibrates faster than one would expect on
the basis of the actual energy gap of the material being studied.
\footnote{If the trial wave function does not describe the
  lowest-energy state that transforms as a 1D irreducible
  representation of the symmetry group (e.g., if one were trying to
  calculate the energy of the $2s$ state of a hydrogen atom with an
  approximate trial wave function whose nodal surface is at the wrong
  radius) then the required equilibration imaginary-time period can be
  much larger than the subsequent decorrelation period in the
  statistics-accumulation phase.
In this case the equilibration imaginary-time scale is the time taken
for the walker populations in high-energy nodal pockets to die out,
which is given by the reciprocal of the difference of the pocket
ground-state energy eigenvalues.}
After a DMC simulation has completed, the decorrelation time
$\tau_\text{corr}$ can be estimated as the time step multiplied by the
blocking length at the onset of the plateau in the reblocked error bar
in a reblocking analysis of the DMC total-energy data (see
Sec.\ \ref{sec:sampling}).

It is useful to convert energy gaps to length scales, since the latter
are often more readily accessible for back-of-the-envelope
calculations.
Arguing from the infinite-square-well model, the fixed-phase energy
gap satisfies $E_1-E_0 \sim 3\pi^2/(2L^2)$ if the electrons are
localized on a length scale $L$, and so the decorrelation period must
go as $\tau_\text{corr} \sim 2L^2/(3\pi^2)$.
The imaginary time taken for an electron to diffuse a length $L$ in 3D
is $L^2/3$, which is about five times larger than the estimate of the
decorrelation time based on the infinite-square-well model; hence if we
choose the equilibration period $N_{\rm eq} \tau$ to be such that each
electron diffuses over the longest relevant length scale $L$ then we
will have made the excited-state components of the initial wave
function exponentially small.

For an insulator with little dispersion of the valence and conduction
bands, an infinite-square-well model suggests that this longest length
scale is $L \sim \sqrt{3\pi^2/(2\Delta {\cal E})}$, where $\Delta
{\cal E}$ is the (actual) energy gap; this is the size of the infinite
square well for which the difference between the ground-state and
first-excited-state energy for electrons is equal to the gap (assuming
the effective mass to be the bare mass).
In this case we are simply choosing $N_{\rm eq} \tau \sim 1/\Delta
{\cal E}$, which errs on the side of caution, since $\Delta {\cal E}$
is less than or equal to the fixed-phase energy gap.
\footnote{For a narrow-gap semiconductor, on the other hand, the sharpest
features in the band structure provide the longest length scale.
In the nearly-free-electron model the size of the region in which the
bands ${\cal E}$ deviate from free-electron behavior is $\Delta {\cal
  E}/(\partial {\cal E}/\partial k)_\text{BZ~edge} \approx 2\pi
a\Delta {\cal E}$, where $a$ is the lattice constant of the primitive
cell.
Hence the longest length scale for narrow-gap semiconductors is $L
\sim 1/(a\Delta {\cal E})$.}

In metals the longest length scale diverges in principle.
However, in a finite, periodic simulation cell, the linear size $L
\propto N^{1/d}$ of the cell provides an (unphysical) upper bound on the
longest length scale.
Thus in a metallic system or a narrow-gap semiconductor, the
equilibration time $N_{\rm eq} \tau$ should be sufficiently long that
the electrons can diffuse across the entire simulation cell.
The required number of equilibration steps $N_{\rm eq}$ and the
decorrelation time $\tau_\text{corr}$ therefore increase as the square
of the linear size of the simulation cell.

These estimates of the required number of equilibration steps
generally err on the side of caution.
Provided the number of statistics-accumulation steps $N_\text{s}$ is
more than a couple of times the decorrelation period
$\tau_\text{corr}/\tau$, the statistical error bars, which go as
$1/\sqrt{N_\text{s}}$, are likely to dominate the bias due to limited
equilibration, which goes as $1/N_\text{s}$.
There may be a cancellation of biases in energy differences due to
similar initial walker distributions.
Furthermore, the excited-state components of the initial DMC wave
function are small in magnitude if the trial wave function is
accurate.
For these reasons, DMC calculations with equilibration periods less
than the theoretical decorrelation period may ``work,'' even though
there is in principle insufficient time for all the excited-state
components to die away.
However, this shortcut cannot be taken for granted and should be
tested on a case-by-case basis.

Typically we perform just two DMC calculations with different time
steps.
To minimize the error bar on the DMC energy extrapolated linearly to
zero time step for a given computational effort (neglecting the cost
of equilibration), one should choose the time steps in the ratio 1:4,
and one should gather eight times as many data with the smaller time
step. \cite{Vrbik_1986}
Perhaps the best way of achieving this is by choosing the
corresponding target populations to be in the ratio 4:1 (e.g., 2048
and 512 walkers), so that linear extrapolation to zero time step
simultaneously extrapolates the DMC results to infinite population,
and choosing the number of statistics accumulation steps to be in the
ratio 2:1.
The number of equilibration steps should be chosen to be in the ratio
4:1 to ensure that the equilibration period in imaginary time is the
same for both time steps.
Time steps of 0.01 and 0.04 Ha$^{-1}$ are typical for pseudopotential
calculations.

\subsection{Overview of the CASINO software}

The \textsc{casino} software is written in Fortran 2003 and
parallelized using MPI and OpenMP\@.
To aid portability the program adheres rigidly to the Fortran 2003
standard; the only exception is the optional use of code featuring the
Cray pointers extension to allow wave-function coefficients to be
shared between MPI processes on the same node.
All libraries used by \textsc{casino} are distributed with the code,
although the BLAS and LAPACK libraries can be replaced with optimized
versions where available.
\textsc{Casino} also includes a dummy version of the MPI library that
allows the code to be compiled in serial, which is useful for debugging
purposes.
Porting \textsc{casino} to a new machine is usually extremely
straightforward.

\textsc{Casino} is accompanied by a range of utility programs written
in Fortran 2003, Bash, C, C++, and Python.
These are used for creating and submitting jobs on machines with
queuing systems, managing large numbers of related jobs, and
analyzing and plotting the data generated.
The distribution also includes a detailed manual and a set of
examples.
The examples include a suite of test cases that can be run
automatically and compared with library output files, to check for
regressions.
\textsc{Casino} is also supported by a website, \cite{casino_website}
which includes a discussion forum for users and libraries of
pseudopotentials.
Pseudopotentials are discussed further in Sec.\ \ref{sec:pseudopots}.

\section{Some recent developments in QMC and CASINO}
\label{sec:recent}

\subsection{Advanced wave-function forms}
\label{sec:adv_wfs}

\subsubsection{Generalized Jastrow factors}
\label{sec:gjastrow}

As mentioned in Sec.\ \ref{sec:slater_jastrow}, Jastrow factors
$e^{J({\bf R})}$ are applied to determinantal trial wave functions
$\Psi_{\rm S}$ in order to describe dynamical electronic correlations.
The optimization of wave-function parameters that affect the nodal
surface, such as backflow parameters or determinant coefficients, is
greatly aided by the use of a good Jastrow factor.
Also, there are systems for which it is undesirable to use the DMC
method, e.g., due to computational expense or pseudopotential locality
biases, and cases in which the Jastrow factor itself is the object of
interest; see Sec.\ \ref{sec:fciqmc_tc}.
While the DTN Jastrow factor is adequate in most cases, there are
systems for which it is helpful to use higher-order terms, such as
three- or four-electron terms, or expansions in a basis other than
natural powers or cosines.

In order to cater for these needs, \textsc{casino} implements a
Jastrow-factor construction framework, described in detail in
Ref.\ \onlinecite{LopezRios_Jastrow_2012}, which can express a rich
variety of forms of correlation involving any number of electrons and
nuclei.
Within this framework, a Jastrow exponent term is expressed as an
expansion in electron-electron basis functions
$\Phi_{\nu_{ij}}^{P_{ij}}({\rm r}_{ij})$ and electron-nucleus basis
functions $\Theta_{\mu_{iI}}^{S_{iI}}({\rm r}_{iI})$, where $\nu_{ij}
\in \{1,\ldots,p\}$ and $\mu_{iI} \in \{1,\ldots,q\}$ are expansion
indices, and $p$ and $q$ are expansion orders.
For simplicity, any cutoff functions are factorized into $\Phi$ and
$\Theta$.
An $n$-electron $m$-nucleus Jastrow factor term for a system with $N$
electrons and $N_{\rm n}$ nuclei is thus expressed as
\begin{equation}
  \label{eq:gjastrow}
  \begin{split}
  J_{n,m}({\bf R}) = & \sum_{\bf i}^N \sum_{\bf I}^{N_{\rm n}}
  \sum_{\bm \nu}^p
  \sum_{\bm \mu}^q \lambda_{{\bm \nu}{\bm \mu}}^{{\bf P}{\bf S}} \\ &
  \prod_{\alpha<\beta}^{n} \Phi_{\nu_{i_\alpha i_\beta}}^{P_{i_\alpha
      i_\beta}} ({\bf r}_{i_\alpha i_\beta})
  \prod_{\alpha,\gamma}^{n,m} \Theta_{\mu_{i_\alpha
      I_\gamma}}^{S_{i_\alpha I_\gamma}} ({\bf r}_{i_\alpha I_\gamma})
  \;,
  \end{split}
\end{equation}
where ${\bf i}$ is a vector of $n$ distinct electron indices
$i_1<\ldots<i_n$, ${\bf I}$ is a vector of $m$ distinct nucleus
indices $I_1<\ldots<I_m$, $\bm \nu$ and $\bm \mu$ are arrays of
electron-electron and electron-nucleus expansion indices, ${\bf P}$
and ${\bf S}$ are arrays of electron-electron and electron-nucleus
dependency indices, and $\lambda$ are the linear parameters of the
expansion.
Symmetry and cusp constraints on the linear parameters can be imposed
automatically for any choice of basis functions, expansion orders,
etc., making the implementation of new basis functions a relatively
trivial task.  \cite{Whitehead_jastrow_2016}

\begin{figure}[!htbp]
  \begin{center}
    \includegraphics[clip,width=0.45\textwidth]{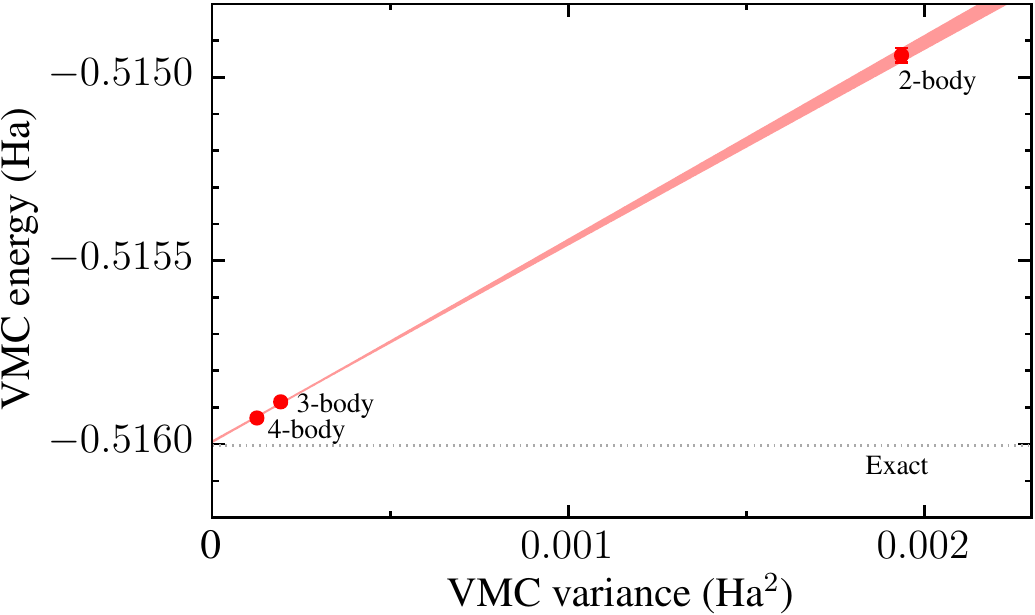}
  \end{center}
  \caption{ VMC energy as a function of VMC variance for Ps$_2$ using
    Jastrow factors including up to 2-, 3-, and 4-body isotropic
    correlations.
    A linear fit is shown as a line of variable width
    indicating the statistical uncertainty in the fit value.
    The exact energy of $-0.5160038$ Ha \cite{Varga_ps2_1998,
    Puchalski_ps2_2008} is shown as a dotted line for reference.
    \label{fig:ps2_evar}
  }
\end{figure}

The DTN Jastrow factor terms can be obtained with
Eq.\ (\ref{eq:gjastrow}) by setting $\Phi$ and/or $\Theta$ to natural
powers of $r$ multiplied by $t(r)$, and higher-order Jastrow factor
terms can be trivially constructed by increasing $n$ and/or $m$.
To illustrate the usefulness of these higher-order terms, in
Fig.\ \ref{fig:ps2_evar} we show the VMC energies and variances
obtained for the positronium dimer, \cite{Hylleraas_ps2_1947} Ps$_2$,
a metastable system consisting of two electrons and two positrons.
We use an electron-positron pairing wave function $\Psi_{\rm S} =
\phi_1(r_{e\uparrow p\uparrow}) \phi_1(r_{e\downarrow p\downarrow})
\phi_2(r_{e\uparrow p\downarrow}) \phi_2(r_{e\downarrow p\uparrow}) +
\phi_2(r_{e\uparrow p\uparrow}) \phi_2(r_{e\downarrow p\downarrow})
\phi_1(r_{e\uparrow p\downarrow}) \phi_1(r_{e\downarrow p\uparrow})$,
where $\phi_1(r)$ and $\phi_2(r)$ are pairing orbitals with
independent parameter values, multiplied by Jastrow factors involving
up to four-body correlations.
Our lowest VMC energy corresponds to $99.53(2)\%$ of the binding
energy of Ps$_2$.
Since the exact wave function is nodeless, an extrapolation to zero
VMC variance \cite{Taddei_iter_bf_2015} can be expected to yield a
reasonable estimate of the ground-state energy.
We find that this extrapolated energy accounts for $99.94(2)\%$ of the
binding energy of Ps$_2$.
Note that the DMC method is exact for this system.

\begin{figure}[!htbp]
  \begin{center}
    \includegraphics[clip,width=0.45\textwidth]{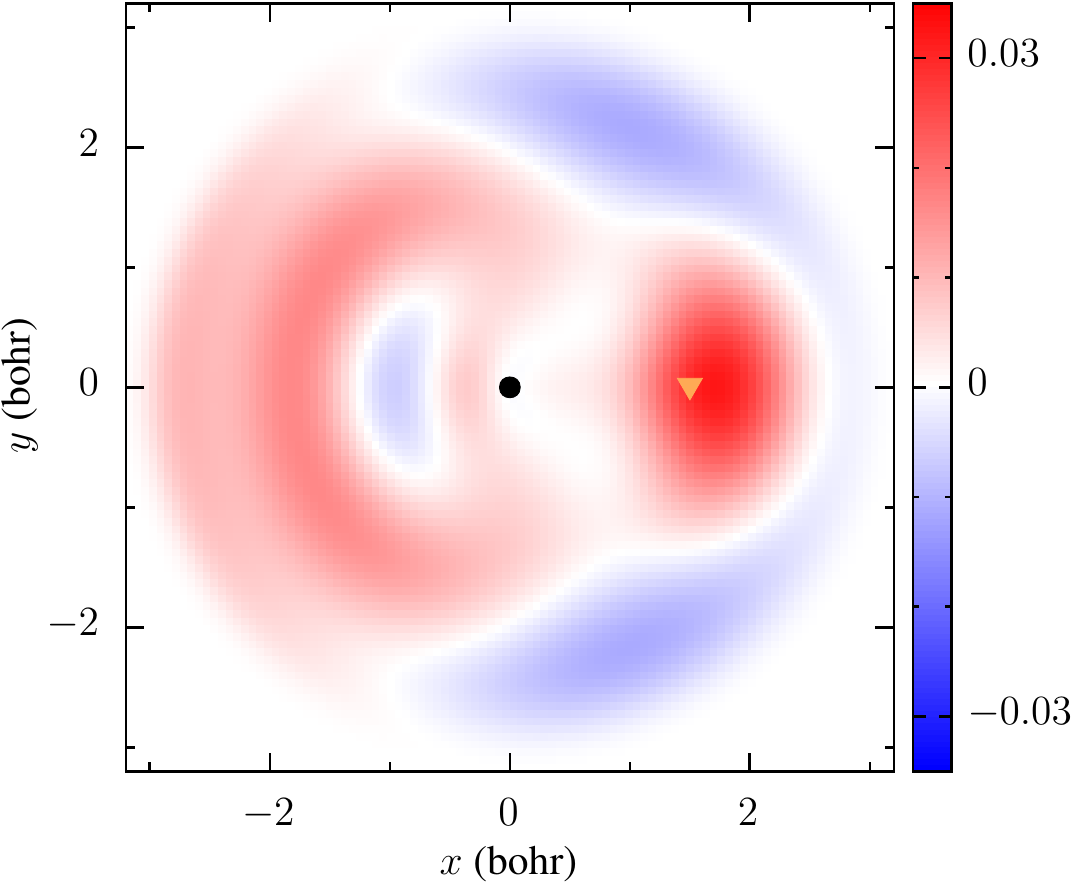}
  \end{center}
  \caption{ Difference between the value of a DTN Jastrow factor and
    one containing an additional electron-electron-nucleus term
    [Eq.\ (\ref{eq:gjastrow_aniso_f})] as an up-spin electron of a
    silicon atom scans the $z=0$ plane.
    The nucleus is at the origin (circle), a down-spin electron is
    fixed at $(1.5,0,0)$ bohr (triangle), and the twelve remaining
    electrons have been placed away from the nucleus so that they do
    not contribute to the value of the Jastrow factor.
    \label{fig:si_atom_iso_aniso}
  }
\end{figure}

Rather more ``exotic'' forms of Jastrow factors can be constructed
with Eq.\ (\ref{eq:gjastrow}), of which
Ref.\ \onlinecite{LopezRios_Jastrow_2012} gives several examples
including electron-nucleus-nucleus terms for polyatomic systems and
electron-electron-nucleus-nucleus terms to describe van der Waals
interactions.
Here we present an additional example of a useful Jastrow factor term,
which is a DTN electron-electron-nucleus term in which $f$ is
augmented by
\begin{equation}
  \label{eq:gjastrow_aniso_f}
  g(r_{ij}, {\bf r}_{iI}, {\bf r}_{jI}) = t(r_{iI}) t(r_{jI})
  \sum_{k,l,m} c_{klm} r_{ij}^k r_{iI}^l r_{jI}^m {\bf r}_{iI} \cdot
      {\bf r}_{jI} \;,
\end{equation}
where $c$ are the linear parameters.
The dot product in Eq.\ (\ref{eq:gjastrow_aniso_f}) can be recovered
in Eq.\ (\ref{eq:gjastrow}) by combining specially crafted
basis functions with constraints on the linear parameters, and
provides the ability to distinguish configurations in which electrons
$i$ and $j$ are on the same side or opposite sides of nucleus $I$,
which is difficult to do with isotropic functions.
We have tested this Jastrow factor on the all-electron silicon atom
using HF orbitals expanded in the cc-pVQZ Gaussian basis set
\cite{Dunning_cc_basis_1989} obtained using the \textsc{molpro} code.
\cite{molpro}
First we optimized the $u$, $\chi$, and $f$ terms in the DTN Jastrow
factor, and then we fixed the parameters in $u$ and $\chi$ and optimized
the $g$ term alongside the $f$ term.
The difference between the DTN and DTN+$g$ Jastrow factors is plotted
in Fig.\ \ref{fig:si_atom_iso_aniso} as a function of the position of
an electron as another is held at a fixed position.
The $g$ term provides a nontrivial correction to the DTN Jastrow
factor, resulting in a total energy reduction of $6.4(9)$ mHa
in this example.

\subsubsection{Multideterminant wave functions}

Multideterminant expansions provide systematic convergence of the
trial wave function, and hence its nodal surface, towards the
ground state.
Let $\{\phi_l({\rm r})\}$ be a set of orbitals that form a complete
basis for 3D functions.
Then the set of all possible $N\times N$ Slater determinants
constructed with these orbitals form a complete basis for
antisymmetric $3N$-dimensional functions.
The configuration interaction (CI) wave function
\begin{equation}
  \label{eq:ci_wfn}
  \Psi_{\rm CI}({\bf R}) = \sum_k^{N_{\rm det}} c_k
     \det\left[\phi_{a_{ik}^\uparrow}({\bf r}_j^\uparrow)\right]
     \det\left[\phi_{a_{ik}^\downarrow}({\bf r}_j^\downarrow)\right]
     \;,
\end{equation}
where $\{a_{ik}^\sigma\}$ are indices selecting which orbitals appear
in the $k$th determinant of spin $\sigma$ and $\{c_k\}$ are unknown
coefficients, is thus guaranteed to converge to the ground-state wave
function in the full-CI limit $N_{\rm det}\to\infty$.

Multideterminantal wave functions have been used in QMC calculations
for many years, \cite{Filippi_mdet_1996} but interest in
multideterminants has surged recently as a potential avenue towards
systematic alleviation of the fixed-node error.
\cite{Giner_mdet_2016, Per_mdet_2017, Scemama_mdet_2018}
Trial wave functions obtained from selected CI approaches appear to
deliver better results than those obtained from complete-active-space
calculations.  \cite{Dash_PSCI_QMC_2018}
Our take on this is discussed in Sec.\ \ref{sec:future_mdet}.
Some of these approaches require very large expansions including
millions of Slater determinants and greatly benefit from the use of
specialized techniques for the rapid evaluation of the trial wave
function and its derivatives.

There are a number of dedicated multideterminant
wave-function-evaluation algorithms.  \cite{Nukala_mdet_2009,
Clark_mdet_2011, Filippi_mdet_2016, Scemama_mdet_2016,
Assaraf_mdet_2017}
\textsc{Casino} implements the compression method of Weerasinghe
\textit{et al.} \cite{Weerasinghe_mdet_2014}
Although this method underperforms some of the other approaches, it
can potentially be combined with them for additional efficiency since
it does not require modification to the evaluation of the wave
function.
The compression algorithm is based on combining determinants differing
by a single column, e.g.,
\begin{equation}
  \label{eq:compress_example}
  \begin{split}
    \left|
      \begin{array}{ccc}
        \phi_1({\bf r}_1) & \phi_3({\bf r}_1) & \ldots \\
        \phi_1({\bf r}_2) & \phi_3({\bf r}_2) & \ldots \\
        \vdots            & \vdots            & \ddots \\
      \end{array}
    \right|
    +
    \left|
      \begin{array}{ccc}
        \phi_2({\bf r}_1) & \phi_3({\bf r}_1) & \ldots \\
        \phi_2({\bf r}_2) & \phi_3({\bf r}_2) & \ldots \\
        \vdots            & \vdots            & \ddots \\
      \end{array}
    \right| & \\
    =
    \left|
      \begin{array}{ccc}
        \phi_1({\bf r}_1) + \phi_2({\bf r}_1)
          & \phi_3({\bf r}_1) & \ldots \\
        \phi_1({\bf r}_2) + \phi_2({\bf r}_2)
          & \phi_3({\bf r}_2) & \ldots \\
        \vdots & \vdots & \ddots \\
      \end{array}
    \right| \;,
    &
  \end{split}
\end{equation}
throughout the expansion.
There are multiple ways in which the $N_{\rm det}$ determinants in the
original expansion can be grouped together, and finding the one that
minimizes the number of determinants $N_{\rm comp}$ in the resulting
wave function is a set-covering problem.
Although the cost of solving a set-covering problem exactly, which can
be done with linear programming, scales nonpolynomially with $N_{\rm
det}$, compressing wave functions containing tens of thousands of
determinants typically takes seconds.
Alternatively, a heuristic ``greedy'' algorithm can be used to
approximately solve the set-covering problem in polynomial time.

The application of compression yields significantly shorter expansions
at the cost of requiring the evaluation of linear combinations of the
orbitals [e.g., $\phi_1+\phi_2$ in Eq.\ (\ref{eq:compress_example})]
when computing the trial wave function and its derivatives.
In practice, this amounts to a tiny fraction of the computer time, and
the overall speedup attained by using compressed expansions is very
close to $N_{\rm det}/N_{\rm comp}$.
Moreover, the size of the compressed expansion is found to scale
sublinearly with the size of the original expansion (i.e., $N_{\rm
  comp} \propto N_{\rm det}^\alpha$ with $\alpha<1$), implying that
the computational cost of the QMC calculation per determinant
decreases with $N_{\rm det}$.
This is also a property of some of the other acceleration methods
reported in the literature.  \cite{Clark_mdet_2011, Scemama_mdet_2016}

\begin{figure}[!htbp]
  \begin{center}
    \includegraphics[clip,width=0.45\textwidth]{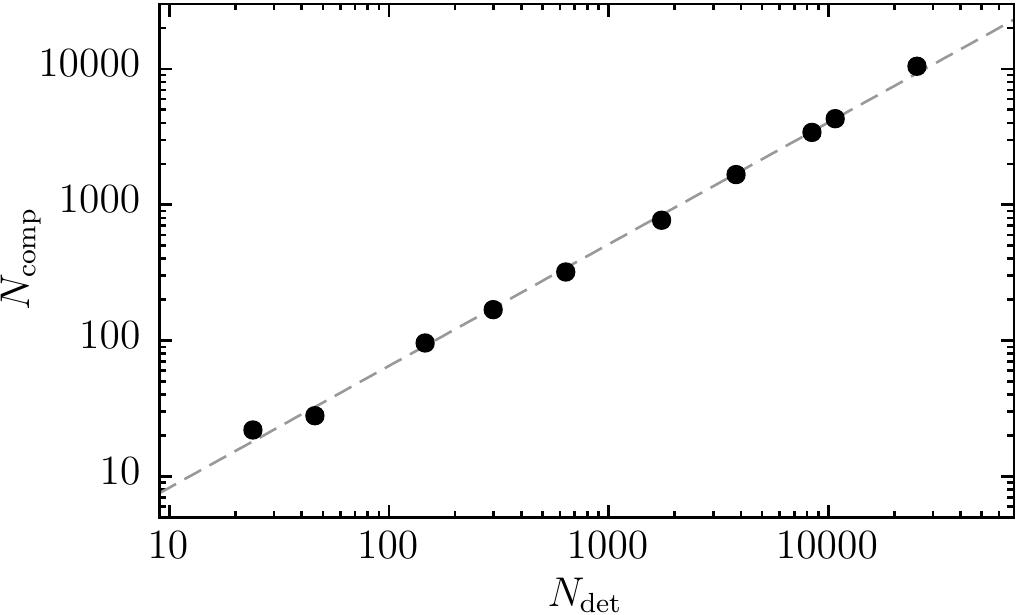}
  \end{center}
  \caption{
    Number of determinants in the compressed expansion $N_{\rm
    comp}$ as a function of the number of determinants in the
    original expansion $N_{\rm det}$ for the C$_2$ molecule at its
    equilibrium geometry.
    A linear fit to the log-log data is shown as a dashed line.
    \label{fig:c2_compress}
  }
\end{figure}

We show an example of the scaling of the compressed expansion size as
a function of $N_{\rm det}$ in Fig.\ \ref{fig:c2_compress} for the
C$_2$ molecule at its equilibrium geometry using HF orbitals expanded
in the cc-pCVTZ basis set \cite{Dunning_cc_basis_1989} obtained using
\textsc{molpro}. \cite{molpro}
The configuration-state-function-based FCIQMC method
\cite{Dobrautz_GUGA_2019} as implemented in the \textsc{neci} package
\cite{Booth_FCIQMC_2009, Cleland_initiator_2010} was used to generate
the original expansion.
Fitting the data yields an exponent $\alpha=0.895<1$, consistent with
the sublinear relation between $N_{\rm comp}$ and $N_{\rm det}$.

\subsubsection{Geminal and pairing wave functions}
\label{sec:pairing}

An alternative way of describing static electronic correlation in the
trial wave function is to replace the Slater determinants with
geminals.
The antisymmetrized geminal power (AGP) wave function
\cite{Hurley_geminal_1956, Nakamura_geminal_1959,
Coleman_geminal_1963} is
\begin{equation}
  \Psi_{\rm AGP}({\bf R}) =
    \det\left( \sum_{l,m} g_{lm} \phi_l({\bf r}_i^\uparrow)
                 \phi_m({\bf r}_j^\downarrow) \right) \;,
\end{equation}
assuming $N_{{\rm e}\uparrow}=N_{{\rm e}\downarrow}$, where
$\{\phi_l({\bf r})\}$ is a set of $n_{\rm orb}$ orbitals and $g_{lm}$
are optimizable coefficients.
If $n_{\rm orb}=N_{{\rm e}\uparrow}$, the AGP wave function trivially
factorizes into up- and down-spin Slater determinants regardless of
the (nonzero) values of the coefficients.
For $n_{\rm orb}>N_{{\rm e}\uparrow}$ this is no longer the case, and
some degree of static correlation is built into the wave function
through pairing of up- and down-spin electrons.
Much in the same spirit, Pfaffians represent an alternative
wave-function form in which the description of correlation is based on
same-spin-electron pairing. \cite{Bajdich_pfaffians_2008}
In QMC calculations, the AGP wave function has been shown to yield a
small advantage over Slater determinants.  \cite{Casula_geminal_2003,
Casula_geminal_2005}

An alternative to the AGP wave function is the use of an
antisymmetrized product of geminals, \cite{Hurley_geminal_1956,
Rassolov_geminal_2002} but the cost of evaluating this wave function
in QMC scales exponentially with $N_{{\rm e}\uparrow}$.
Instead, \textsc{casino} implements a multi-AGP (MAGP) wave function,
\cite{Bugnion_thesis_2014} which is simply a linear combination of
AGPs with independent parameters.
In Ref.\ \onlinecite{Bugnion_thesis_2014}, an MAGP wave function was
constructed to reproduce the coefficients of a CI wave function
including up to double excitations for the HEG, and DMC energies
obtained combining the MAGP wave function with a backflow
transformation were found to be competitive with those from FCIQMC,
\cite{Shepherd_HEG_2012} albeit at significant computational expense.
For a comparison of this result with that obtained using
multideterminantal wave functions, see Fig.\ \ref{fig:HEG_mdet} below.

In the presence of attractive interactions, for instance in a system
consisting of $N_{\rm e}$ negatively charged electrons and $N_{\rm
  h}=N_{\rm e}$ positively charged holes, it is natural to employ a
simpler form of pairing wave function,
\begin{equation}
  \label{eq:pairing_wfn}
  \Psi_{\rm S}({\bf R}) =
    \det\left[\phi({\bf e}_j^\uparrow-{\bf h}_i^\downarrow)\right]
    \det\left[\phi({\bf e}_j^\downarrow-{\bf h}_i^\uparrow)\right] \;,
\end{equation}
where ${\bf e}_j^\sigma$ is the position vector of the $j$th electron
of spin $\sigma$, ${\bf h}_i^\sigma$ is the position vector of the
$i$th hole of spin $\sigma$, and the pairing orbital $\phi$ is
a parameterized function of the difference between the electron and
hole position vectors.
Usually the pairing orbital is assumed to depend solely on the
electron-hole distance, $\phi({\bf r})=\phi(r)$, but \textsc{casino}
also implements a pairing orbital of the form
\cite{Maezono_biexcitons_2013,
LopezRios_thesis_2006}
\begin{eqnarray}
  \label{eq:pairing_orb}
  \phi({\bf r}) = \sum_l p_l \exp(i {\bf G}_l \cdot {\bf r}) + u(r)
    \;,
\end{eqnarray}
where ${\bf G}_l$ is the $l$th shortest reciprocal lattice vector,
$\{p_l\}$ are optimizable coefficients, and $u(r)$ is a parameterized
polynomial of the same form as the $u$ function in the DTN Jastrow
factor.
This orbital confers on the pairing wave function a property of the
AGP wave function: if $u=0$ and the number of plane waves equals the
number of particles per spin, the wave function can be factorized as
the product of electron and hole determinants of plane-wave orbitals.
This pairing wave function is therefore capable of describing both a
two-component fluid and an excitonic fluid, and is found to provide a
particularly good description of the system near the transition
between these two phases.  \cite{LopezRios_thesis_2006}

The pairing wave function of Eq.\ (\ref{eq:pairing_wfn}) can be
adapted to systems with an imbalance in the number of electrons and
holes by completing the pairing matrix with plane waves.
\cite{LopezRios_bilayer_2018}
The extreme case of a single hole immersed in an electron gas is of
particular interest.
For this system \textsc{casino} implements a specialized pairing wave
function, \cite{Spink_trion_2016}
\begin{equation}
  \Psi_{\rm S}({\bf R}) =
    \det\left[\phi_i({\bf e}_j^\uparrow-{\bf h})\right]
    \det\left[\phi_i({\bf e}_j^\downarrow-{\bf h})\right] \;,
\end{equation}
where ${\bf h}$ is the position of the hole and the pairing orbital is
\begin{equation}
  \phi_l({\bf r}) =
    \exp{\left[u_{G_l}(r)\right]}
    \exp\left\{i {\bf G}_l \cdot {\bf r}
               \left[1-\eta_{G_l}(r) / r\right] \right\} \;,
\end{equation}
where $u_{G_l}$ and $\eta_{G_l}$ are parameterized polynomials of the
same functional form as the $u$ function in the DTN Jastrow factor.
This orbital is essentially a plane wave of the electron-hole distance
with an orbital-specific Jastrow-like prefactor and an electron-hole
backflow transformation.

\subsection{QMC calculations for condensed matter}

\subsubsection{Single-particle finite-size effects: momentum
quantization}

Condensed matter physics and materials science are largely concerned
with the bulk properties of extended systems of $10^4$--$10^{26}$
constituent particles.
On an atomic scale such systems are of effectively infinite extent and
most of their properties are well described by equilibrium
thermodynamics.
In studies of condensed matter, the challenge for QMC methods is the
need to obtain results that are valid in the thermodynamic limit of
infinite system size despite only being able to perform calculations
for finite systems with hundreds or thousands of constituent
particles.
In this section we will discuss finite-size effects in the context of
crystalline solids, although everything we state applies equally well
to the case of homogeneous liquids, in which the size of the primitive
unit cell is vanishingly small.
Unlike DFT, QMC is an explicitly correlated wave-function-based
method, so that the electronic Schr\"odinger equation cannot be
reduced to a single periodic unit cell.
Instead one must construct a supercell of multiple primitive cells.
In the following we use $N$ to denote the number of electrons in a
simulation supercell.

One way of approaching the thermodynamic limit, often used in quantum
chemistry, is to study finite clusters of bulk crystalline solids
using hydrogen atoms to passivate dangling bonds on the surface.
In this case the translational symmetry of the bulk crystal is lost
and unwanted surface effects are introduced.
An alternative approach is to use a finite simulation supercell of the
crystal subject to periodic boundary conditions.
This avoids unwanted surfaces and restores translational symmetry, but
introduces spurious interactions between periodic images of particles.
Periodic boundary conditions usually provide a much more reliable
means for extrapolating results to infinite system size.

Simulation supercells are traditionally constructed using the
primitive cell as a repeating unit, with each supercell lattice vector
being an integer multiple of the corresponding primitive-cell lattice
vector.
However one is free to choose supercell lattice vectors to be integer
linear combinations of the primitive-cell lattice vectors.
The use of nondiagonal supercells that maximize the nearest-image
distance of the simulation supercell yields reduced finite-size
effects in energy differences, and grants access to a broader choice
of system sizes.  \cite{Lloyd-Williams_nondiag_2015, Drummond2015,
  Trail_TiO2_2017, Liao_hydrogen_2019}

Consider a supercell of a crystal subject to periodic boundary
conditions.
Translations of individual electrons through supercell lattice vectors
and simultaneous translations of all electrons through primitive
lattice vectors are symmetry operations and therefore commute with the
many-electron Hamiltonian.
Eigenfunctions of the Hamiltonian are therefore eigenfunctions of the
translation operators.
The requirement that the wave function is an eigenfunction of each of
these translation operators leads to two many-body Bloch conditions on
the wave function:
\begin{eqnarray}
  \label{eq:mb_bloch_1}
  \Psi({\bf R}) & = & e^{i{\bf k}_{\rm s}\cdot\sum_{i=1}^N {\bf r}_i}
  U({\bf R}) \\
  \label{eq:mb_bloch_2}
  & = & e^{i{\bf k}_{\rm p}\cdot\sum_{i=1}^N {\bf r}_i / N} W({\bf
    R}),
\end{eqnarray}
where $U$ is invariant under a translation of any electron through any
supercell lattice vector ${\bf R}_{\rm s}$, while $W$ is invariant
under a simultaneous translation of all electrons through any
primitive-lattice vector ${\bf R}_{\rm p}$, and ${\bf k}_{\rm s}$ and
${\bf k}_{\rm p}$ are wavevectors in the first Brillouin zones of the
simulation cell and the primitive cell, respectively.
\cite{Rajagopal_1994, Rajagopal_1995}

As usual, we impose these symmetry requirements via the Slater part of
the trial wave function, and choose our Jastrow factor to be fully
symmetric.
In particular we can simultaneously satisfy
Eqs.\ (\ref{eq:mb_bloch_1}) and (\ref{eq:mb_bloch_2}) by choosing the
orbitals in each Slater determinant to be of Bloch form, with the
Bloch wavevectors lying on a grid of simulation-cell reciprocal
lattice points offset from $\Gamma$ by ${\bf k}_{\rm s}$, for any
${\bf k}_{\rm s}$ in the first Brillouin zone of the simulation cell.

In the infinite system, the Bloch wavevectors of the orbitals form a
continuum and in a single-particle theory the total energy per
particle can be expressed in terms of integrals over wavevector ${\bf
  k}$.
In the finite supercell subject to periodic boundary conditions the
integrals are replaced by sums over a discrete set of ${\bf k}$
points.
The difference between the discrete sums for the finite supercell and
the integrals for the infinite system varies in a quasirandom manner
with system size.
These quasirandom fluctuations are particularly large in metallic
systems, where the number of occupied shells of ${\bf k}$ points
changes as the simulation cell is increased and hence the grid of
${\bf k}$ points gets finer.
In the context of a single-particle theory, these finite-size effects
are regarded as ${\bf k}$-point sampling errors.
To a first approximation, when looking at energy differences,
including electron-electron interactions and hence correlations in the
wave function simply renormalizes the effective mass of the
electrons. \cite{Landau_1957a, Landau_1957b, Landau_1959}
Hence similar oscillations in energy per particle as a function of
system size are seen in QMC simulations as in uncorrelated
single-particle theories, but usually with a different amplitude.
One approach for reducing momentum-quantization finite-size errors is
to treat the DFT energy or HF kinetic energy obtained using the ${\bf
  k}$-point grid corresponding to a QMC supercell relative to the DFT
energy with a fine ${\bf k}$-point mesh as a covariate in a fit to the
QMC energy as a function of system size; such a fit allows
extrapolation to infinite system size. \cite{Ceperley_1978}

In general, however, it is better to average over supercell Bloch
wavevectors ${\bf k}_{\rm s}$ (i.e., to average over offsets to the
grid of single-particle ${\bf k}$ points) for a given supercell size;
this process is referred to as \textit{twist averaging}.
\cite{Lin_2001}
Averaging over a large number of offsets to the grid of ${\bf k}$
points turns a sum over discrete ${\bf k}$ into an integral.
For a metallic system, the shape of the twist-averaged Fermi surface
is a polyhedron with (in the case of the HEG) the same volume as the
real Fermi surface; the incorrect shape leads to a relatively small,
positive, quasirandom finite-size effect.
It is possible to use DFT energies calculated using exactly the same
${\bf k}$ points as the QMC calculations relative to the DFT energy
with a fine ${\bf k}$-point grid as a covariate in a fit to the QMC
energies as a function of twist; this simultaneously removes both
noise due to the finite number of twists and the residual bias in the
twist-averaged energy.

\subsubsection{Long-range finite-size effects \label{sec:fs_lr}}

There are additional finite-size effects in a correlated QMC
calculation that are absent from a DFT calculation. (Strictly speaking
this is because the XC functional in a DFT calculation is, by
construction, appropriate for an infinite system; thus DFT provides an
incorrect description of a finite cell subject to periodic boundary
conditions.)
The physical origins of these finite-size effects lie in (i) the
treatment of the Coulomb interaction in a finite supercell and (ii)
the fact that long-range correlation effects cannot be described
correctly in a finite supercell.

Let us first consider Coulomb finite-size effects.
The usual treatment of the Coulomb interaction in a periodic cell is
to write the interaction between each pair of charges using the Ewald
interaction $v_{\rm E}({\bf r})$, which is the periodic solution of
Poisson's equation around a point charge, with Fourier components
$4\pi/k^2$. \cite{Ewald_1921}
In addition, the electrostatic energy of the lattice of periodic
images of each charge in the supercell, known as the Madelung
constant, is included in the energy.
In practice the Ewald interaction is evaluated as the sum of two
rapidly convergent series in supercell lattice vectors and in
supercell reciprocal lattice vectors.
The use of the periodic solution to Poisson's equation amounts to the
assumption that there is no macroscopic electric field within the
crystal; this could be guaranteed by embedding a macroscopic sample of
the crystal in a perfect metal so that there are no surface
polarization charges.

The expectation value of the potential-energy operator can be
decomposed into an interaction between the fixed nuclei and the
electronic charge density, plus a Hartree energy given by the
classical electrostatic potential energy of the electronic charge
density, plus the Coulomb XC energy, which is everything else:
\begin{widetext}
\begin{eqnarray}
  \langle \hat{V} \rangle & = & \sum_i \sum_I -Z_I\langle v_{\rm
    E}(\hat{\bf r}_{iI}) \rangle + \sum_{i>j} \langle v_{\rm
    E}(\hat{\bf r}_{ij}) \rangle + \frac{v_{\rm M}} 2 \left(N+\sum_I
  Z_I^2\right) \nonumber \\ & = & - \sum_I Z_I \int v_{\rm E}({\bf
    r}-{\bf r}_I) \rho({\bf r}) \, {\rm d}{\bf r} + \frac{v_{\rm M}} 2
  \sum_I Z_I^2 + \frac 1 2 \iint v_{\rm E}({\bf r}-{\bf r}') \rho({\bf
    r}) \rho({\bf r}') \, {\rm d}{\bf r} \, {\rm d}{\bf r}' \nonumber
    \\ &   & {} + \frac N 2 \int \left[v_{\rm E}({\bf r}) -
    v_{\rm M}\right] \frac 1 N \int \left[ \rho_2({\bf r}'+{\bf r},
    {\bf r}') - \rho({\bf r}'+{\bf r}) \rho({\bf r}') \right] \,
    {\rm d}{\bf r}' \, {\rm d}{\bf r},
\end{eqnarray}
\end{widetext}
where the charge density is $\rho({\bf r})=\left< \sum_i \delta({\bf
  r}-\hat{\bf r}_i) \right>$ and the pair density is $\rho_2({\bf
  r},{\bf r}')=\left< \sum_{i\neq j} \delta({\bf r}-\hat{\bf
  r}_i)\delta({\bf r}'-\hat{\bf r}_j) \right>$.
The Coulomb XC energy per electron is therefore the Coulomb
interaction between a point charge at the origin and the
system-averaged XC hole $\rho_{\rm xc}({\bf r})\equiv (1/N)\int \left[
  \rho_2({\bf r}'+{\bf r},{\bf r}') -\rho({\bf r}'+{\bf r})\rho({\bf
    r}') \right] \, {\rm d}{\bf r}'$, which is the average effective
charge density around each electron due to correlations with all the
others.
The long-range behavior of the XC hole is well-described by the random
phase approximation, i.e., linear-response theory of the
noninteracting electron system in the rest frame of one of the
electrons.

The electronic charge density $\rho$ has the periodicity of the
primitive unit cell and generally converges extremely quickly with
simulation-cell size; hence there is very little finite-size error in
the electron-nucleus potential energy and in the Hartree energy.
However, the finite-size errors in the Coulomb XC energy can be
substantial.

One source of error is that the XC hole $\rho_{\rm xc}({\bf r})$ is
long-ranged, with the systematic part falling off as $r^{-8}$ in an
infinite system; \cite{Gori_2002} this tail is truncated at finite
range in a periodic cell.
On top of this there are long-range Friedel oscillations in the XC
hole, which result in very small quasirandom finite-size errors in the
energy when the oscillations are forced to be commensurate with the
periodic simulation supercell.
Evaluating the Coulomb interaction between the electron at the origin
and the missing $r^{-8}$ tail of the surrounding XC hole in a finite
cell, we find that the resulting finite-size error per particle falls
off as $N^{-2}$.

However a much more significant source of finite-size error is the
form of interaction between the electron and the XC hole.
In the infinite-system limit this should simply be $1/r$; in the
finite cell it is $v_{\rm E}({\bf r})-v_{\rm M}$.
A power series expansion of the Ewald interaction \cite{Makov_1995,
  Fraser_1996} shows that $v_{\rm E}({\bf r})-v_{\rm M}=1/r+{\cal
  O}(r^2/N)$.
The correction term in this expansion results in an error that goes as
$N^{-1}$ in the Coulomb XC energy per particle.

Remarkably, it is possible to eliminate the leading-order finite-size
errors in the Coulomb XC energy by replacing the Ewald interaction
with a model periodic Coulomb interaction that causes the Hartree
energy to be evaluated using the Ewald potential while the Coulomb XC
energy is evaluated using the $1/r$ interaction within the simulation
cell. \cite{Fraser_1996, Williamson_1997, Kent_1999}
The residual finite-size errors in the XC energy per particle are
therefore ${\cal O}(N^{-2})$ due to the truncation of the XC hole.

An alternative perspective on the problem of finite-size errors is
found by transforming to reciprocal space. \cite{Chiesa_2006}
The XC energy per particle can be written as
\begin{eqnarray}
  & & \frac 1 2 \int \left[ v_{\rm E}({\bf r}) - v_{\rm M}\right]
  \rho_{\rm xc}({\bf r}) \, {\rm d}{\bf r} \nonumber \\ & & {} =
  \frac 1 2 \left( \frac 1 {\Omega_{\rm s}} \sum_{{\bf G}_{\rm s}
  \neq {\bf 0}} \frac{4\pi} {G_{\rm s}^2} \left[ S({\bf G}_{\rm s})
  - 1\right] + v_{\rm M} \right),
\end{eqnarray}
where the static structure factor is $S({\bf k})=1+\int \rho_{\rm
xc}({\bf r}) e^{-i{\bf k}\cdot{\bf r}} \, {\rm d}{\bf r}$ and
$\Omega_{\rm s}$ is the volume of the simulation supercell.
Since the XC hole is rapidly convergent, so is the structure factor;
hence the leading-order finite-size correction to the potential energy
per particle is due to the difference between an integral and sum:
\begin{eqnarray}
  \frac{\Delta V_{\rm xc}} N & = & \frac 1 2 \left( \frac 1 {(2\pi)^3}
  \int \frac{4\pi} {k^2} \left[ S({\bf k})-1\right] {\rm d}{\bf k}
  \right) \nonumber
  \\ & & {} - \frac 1 2 \left( \sum_{{\bf G}_{\rm s} \neq {\bf 0}}
  \frac{4\pi}{\Omega_{\rm s} G_{\rm s}^2} \left[ S({\bf G}_{\rm
      s})-1\right] + v_{\rm M} \right) \nonumber \\ & = & \frac 1
       {4\pi^2} \int \frac{S({\bf k})} {k^2} \, {\rm d}{\bf k} -
       \frac{2\pi} {\Omega_{\rm s}} \sum_{{\bf G}_{\rm s}\neq{\bf 0}}
       \frac{S({\bf G}_{\rm s})} {G_{\rm s}^2} \nonumber \\ & \approx
       & \frac{2\pi} {\Omega_{\rm s}} \lim_{{\bf k} \to {\bf 0}}
       \frac{\bar{S}(k)} {k^2},
\end{eqnarray}
where $\bar{S}$ is the spherical average of the structure factor and
we have used the fact that
\begin{equation}
  v_{\rm M} = \lim_{\epsilon \to 0} \left( \frac{4\pi} \Omega_{\rm s}
  \sum_{{\bf G}_{\rm s}\neq {\bf 0}} \frac{e^{-\epsilon G_{\rm s}^2}}
      {G_{\rm s}^2} - \frac 1 {2\pi^2} \int \frac{e^{-\epsilon
          k^2}}{k^2} \, {\rm d}{\bf k} \right).
\end{equation}
The spherically averaged structure factor is quadratic at short range,
and $\lim_{{\bf k}\to{\bf 0}} \bar{S}(k)/k^2$ may be evaluated using
the structure factor accumulated in a QMC calculation.
Indeed for simple systems such as the electron gas, $\lim_{k\to
0}S(k)/k^2$ can be calculated analytically using the RPA\@.
\cite{Gaskell_1961}
If a system has cubic symmetry then the error in the approximation to
the finite-size correction is the ${\cal O}(N^{-2})$ error due to the
neglected tail of the XC hole; in noncubic systems the finite-size
correction using the spherical average of the structure factor does
not fully remove the leading-order error.

\begin{figure}[!htbp]
  \begin{center}
    \includegraphics[clip,width=0.45\textwidth]{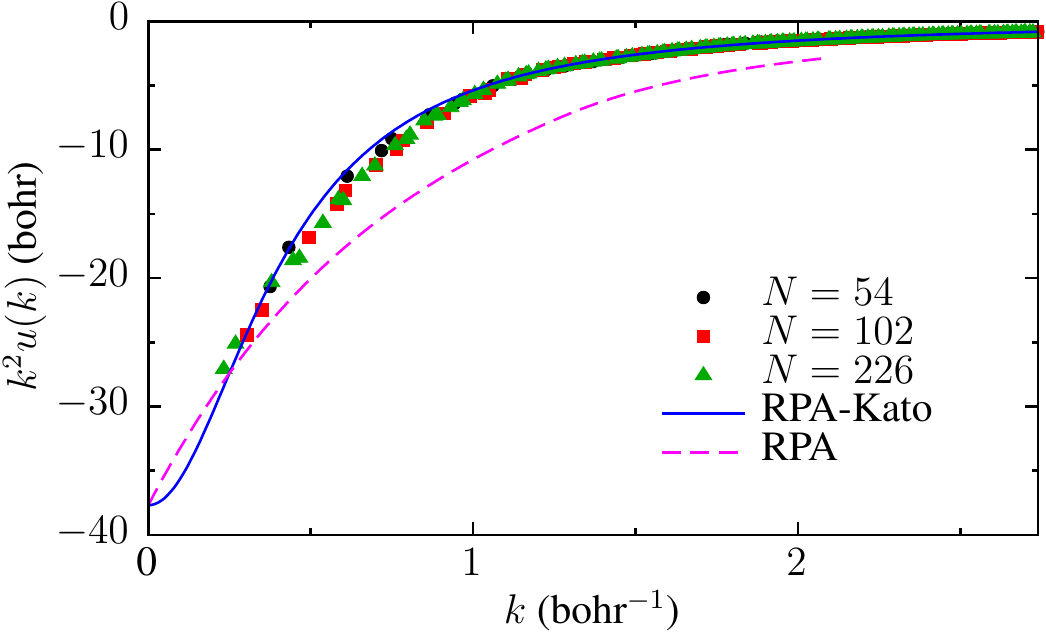}
  \end{center}
  \caption{ Fourier transform of the optimized two-body Jastrow factor
    for a paramagnetic HEG of density parameter $r_{\rm s}=3$ at
    different system sizes $N$, together with the RPA result, and an
    interpolation between the leading-order RPA behavior and the
    short-range Kato behavior.
    \label{fig:u_of_k}
  }
\end{figure}

We now turn our attention to the neglect of long-range correlation
effects.
The two-body correlations described by the Jastrow factor are
long-range.
They are restricted in a finite simulation cell, leading to bias in
kinetic energy.
We may correct for this by interpolating the Fourier transformation of
the two-body Jastrow factor. \cite{Chiesa_2006}
Let us write $\Psi$ as the product of a long-range two-body Jastrow
factor, which has the periodicity of the simulation cell and inversion
symmetry, and a part consisting of everything else, $\Psi_{\rm s}$:
\begin{eqnarray}
  \Psi & = & \Psi_{\rm s} \exp\left( \sum_{i>j} u({\bf r}_i-{\bf r}_j)
  \right), \nonumber \\ & = & \Psi_{\rm s} \exp\left( \sum_{{\bf
      G}_{\rm s} \neq {\bf 0}} \frac{u({\bf G}_{\rm s})
    \hat{\rho}^\ast ({\bf G}_{\rm s}) \hat{\rho} ({\bf G}_{\rm s})}
        {2\Omega_{\rm s}} + K \right).
\end{eqnarray}
The kinetic energy per particle may be evaluated as the average of
\begin{eqnarray}
  \frac{\hat{T}} N & = & \frac{-1}{4N} \nabla^2 \ln(\Psi) \nonumber
  \\ & = & \frac{{\hat T}_{\rm s}} N - \frac 1 {8 N\Omega_{\rm s}}
  \sum_{{\bf G}_{\rm s} \neq {\bf 0}} u({\bf G}_{\rm s}) \nabla^2
  \left[ {\hat \rho}^\ast ({\bf G}_{\rm s}) \hat \rho ({\bf G}_{\rm
      s}) \right],
\end{eqnarray}
where $\hat{T}_{\rm s}/N=-\nabla^2 \ln(\Psi_{\rm s})/(4N)$.
Noting that
\begin{equation}
  \nabla^2 \left[ {\hat \rho}^\ast ({\bf G}_{\rm s}) \hat \rho ({\bf
      G}_{\rm s}) \right] = - 2 G_{\rm s}^2 \left[ \hat{\rho}^\ast
    ({\bf G}_{\rm s}) \hat{\rho} ({\bf G}_{\rm s}) - N \right],
\end{equation}
we find that
\begin{equation}
  \frac{\langle \hat{T} \rangle} N = \frac{\langle \hat{T}_{\rm s}
    \rangle} N + \frac 1 {4N\Omega_{\rm s}} \sum_{{\bf G}_{\rm s} \neq
    {\bf 0}} G_{\rm s}^2 u({\bf G}_{\rm s}) \left[ \left<
    \hat{\rho}^\ast ({\bf G}_{\rm s}) \hat{\rho} ({\bf G}_{\rm s})
    \right> - N \right].
\end{equation}
Now $\rho({\bf k})$ is only nonzero for reciprocal lattice vectors of
the primitive lattice.
Assuming the sum runs only over small ${\bf G}_{\rm s}$,
\begin{eqnarray}
  \label{eq:T_longrange}
  \frac{\langle \hat{T} \rangle} N & = & \frac{\langle \hat{T}_{\rm s}
    \rangle} N + \frac 1 {4 \Omega_{\rm s}} \sum_{{\bf G}_{\rm s} \neq
    {\bf 0}} G_{\rm s}^2 u({\bf G}_{\rm s}) S^\ast ({\bf G}_{\rm s})
  \nonumber \\ & & {} - \frac 1 {4 \Omega_{\rm s}} \sum_{{\bf G}_{\rm
      s}\neq {\bf 0}} G_{\rm s}^2 u({\bf G}_{\rm s}).
\end{eqnarray}
$u({\bf k})$ has same form at different $N$.
It diverges as $k^{-2}$, \cite{Bohm_1953, Gaskell_1961} so
$\lim_{k\rightarrow0} k^2\bar{u}(k)$ exists: see
Fig.\ \ref{fig:u_of_k}.

In the infinite-system limit, the sum over ${\bf G}_{\rm s}$ should be
replaced by an integral in Eq.\ (\ref{eq:T_longrange}).
\cite{Chiesa_2006}
The leading-order finite-size corrections are due to the differences
between these integrals and sums.
The RPA suggests that we can approximate $\bar{u}(k)$ by
$-4\pi(A/k^2+B/k)$ at small $k$ in 3D, where $A$ and $B$ are
constants. \cite{Gaskell_1961}
In \textsc{casino} the constants are determined by fitting this model
of $\bar{u}$ to the optimized two-body Jastrow terms at the first few
nonzero stars of ${\bf G}_{\rm s}$ vectors.
We evaluate the finite-size correction to the kinetic energy per
electron in 3D as
\begin{equation}
  \frac{\Delta T} N = \frac{\pi A} \Omega_{\rm s} + \frac{C_{\rm 3D} B}
       {\Omega_{\rm s}^{4/3}},
\end{equation}
where
\begin{equation}
  C_{\rm 3D} = \frac {\Omega_{\rm s}^{4/3}} 4 \lim_{\alpha \rightarrow 0}
  \left[ \frac 1 {\pi \alpha^2} - \frac {4\pi} \Omega_{\rm s} \sum_{{\bf
        G}_{\rm s} \neq {\bf 0}} G_{\rm s} \exp(-\alpha G_{\rm s}^2)
    \right]
\end{equation}
is a lattice-specific constant, which can be evaluated by brute force.
The leading-order finite-size error in the kinetic energy per electron
falls off as ${\cal O}(N^{-1})$. \cite{Drummond_2008}
Note that it has been shown that the use of backflow alters the
coefficient of the ${\cal O}(N^{-1})$ contribution to the finite-size
error in the energy per electron.  \cite{Holzmann_2016}

The work that has been done in extrapolating HEG results to infinite
system size can be used to correct finite-size errors in inhomogeneous
systems using DFT with a special finite-size local density
approximation (LDA) functional. \cite{Kwee_2008}
In the finite-size LDA calculations, the XC energy per particle at a
point in space with density $\rho$ is given by the XC energy per
particle of a finite electron gas of density $\rho$ in the supercell
obtained by unfolding the ${\bf k}$ points.
The finite-size correction is the difference between a the DFT energy
with the usual (infinite-system) LDA and a fine ${\bf k}$-point mesh
and the finite-size LDA energy with a ${\bf k}$-point sampling
corresponding to the QMC calculation to be corrected.
Single-particle and long-range finite-size errors are corrected
simultaneously in this approach.
It is straightforward to use this method in conjunction with twist
averaging.
In practice, the available parameterization of the finite-system LDA
means that long-range effects due to the shape of the simulation cell
are neglected. \cite{Kwee_2008}

\subsubsection{Finite-size effects in properties other than the total
  energy}

Most expectation values, such as the pair-correlation function and
structure factor, can be twist averaged to reduce single-particle
finite-size effects.
However, for some expectation values such as the momentum density, the
choice of simulation-cell Bloch wavevector merely defines the set of
${\bf k}$ points at which the momentum density can be evaluated.
Performing calculations at multiple twists therefore gives additional
points on the momentum density.
The MD of metallic systems has been shown to exhibit slowly decaying
finite-size errors near the Fermi edge. \cite{Holzmann_2009}
In expectation values such as charge densities and pair-correlation
functions, there are finite-size errors associated with the fact that
Friedel oscillations are forced to be commensurate with the simulation
cell.
Finite-size effects in excitation energies are discussed in
Sec.\ \ref{sec:fs_effects_gaps}.

\subsection{Pseudopotentials for correlated methods
\label{sec:pseudopots}}

Pseudopotentials or effective core potentials play an important role
in almost all \textit{ab initio} electronic structure methods.
In essence they replace the influence of core electrons on valence
electrons with an effective potential, so reducing each atom to a
pseudo-atom composed of valence electrons only.
For all-electron DMC calculations the scaling of the cost with atomic
number $Z$ is $Z^{5}$--$Z^{6.5}$. \cite{Ceperley_1986,
  Ma_all-electron_2005}
The use of pseudopotentials reduces the effective value of $Z$, making
QMC calculations feasible for all atoms.

Replacing the dynamic interaction of valence and core electrons with a
potential is necessarily approximate, but the error can be controlled.
A pseudopotential with a usably small error is nonlocal, and is
defined to reproduce physical properties of the valence electrons.
The physical properties usually chosen are either core scattering (to
first order) or ionization and excitation energies.
Pseudopotentials generated using these two approaches are
referred to as ``shape consistent'' and  ``energy consistent,''
respectively.

The generation of both types of pseudopotentials in the context of DFT
or HF theory is well established, with pseudopotential error often
less than the error inherent in the underlying theory.
However, using such pseudopotentials in many-body methods introduces a
significant, unnecessary, and uncontrolled error due to the
inconsistent application of theory.
For HF pseudopotentials the error is due to the absence of
core-valence correlation.
A DFT pseudopotential suffers from a deeper inconsistency between the
Kohn-Sham orbitals they are designed to reproduce and the many-body
wave functions to which they are applied.
Note that neither the Kleinman-Bylander nor the ultrasoft
pseudopotential forms are useful for many-body methods.

For a pseudopotential to be accurate in QMC it must be constructed to
reproduce scattering and/or excitation properties of the many-body
Hamiltonian.
Pseudopotentials generated using DFT or HF do not have this property
and, perhaps most importantly, the accompanying error is difficult to
estimate.
No error-correction strategies are available.

Trail and Needs have produced several generations of pseudopotentials,
including HF and Dirac-Fock pseudopotentials (TNDF),
\cite{Trail_ppot_2005, Trail_norm_2005}, correlated-electron
pseudopotentials (CEPPs), \cite{Trail_ppot_2013, Trail_ppot_2015} and
energy-consistent correlated-electron pseudopotentials (eCEPPs).
\cite{Trail_ppot_2017}
eCEPPs provide the best accuracy among these, and are the most
sophisticated pseudopotentials currently available due to their
accurate description of correlated electron systems.
CEPPs and eCEPPs are given in the Supplementary Material of Refs.\
\onlinecite{Trail_ppot_2013, Trail_ppot_2015, Trail_ppot_2017} and can
also be obtained from Ref.\ \onlinecite{ppot_website}.
Other groups have also developed pseudopotentials for correlated
electronic calculations. \cite{Annaberdiyev_ppot_2018}

The eCEPP construction process reproduces three different aspects of
the core-valence electron interaction.
Electron correlation is included throughout, and the independent
electron approximation is not used.

First, an all-electron atomic calculation is used to generate a
nonlocal potential that reproduces the first-order scattering
properties of the core.
This is achieved by partitioning the density matrix into a core and a
valence part, disposing of the core part, and redefining the valence
part close to the atomic nuclei.
The nonlocal potential so generated is a correlated-electron
generalization of the well-established norm-conserving DFT
pseudopotential.

Second, the long-range core-polarization interaction is reproduced by
these potentials and is represented using the standard (two-body)
core-polarization potential form. \cite{Shirley_core-valence_1993}
At the DFT level no long-range core polarization occurs.

In the third stage of the eCEPP generation process we use the fact
that the pseudopotentials defined by the first and second stages are
not unique.
This allows us to search the space of pseudopotentials that reproduce
scattering properties to find those that also reproduce atomic
ionization and excitation energies.

All three parts of the eCEPP construction process involve many-body
Hamiltonians and wave functions, and employ explicitly correlated
atomic calculations, namely multiconfiguration self-consistent field
and coupled-cluster calculations with single, double, and perturbative
triple excitations [CCSD(T)] with accurately corrected basis-set
errors.

\begin{figure*}[!htbp]
  \begin{center}
    \includegraphics[clip,width=0.95\textwidth]{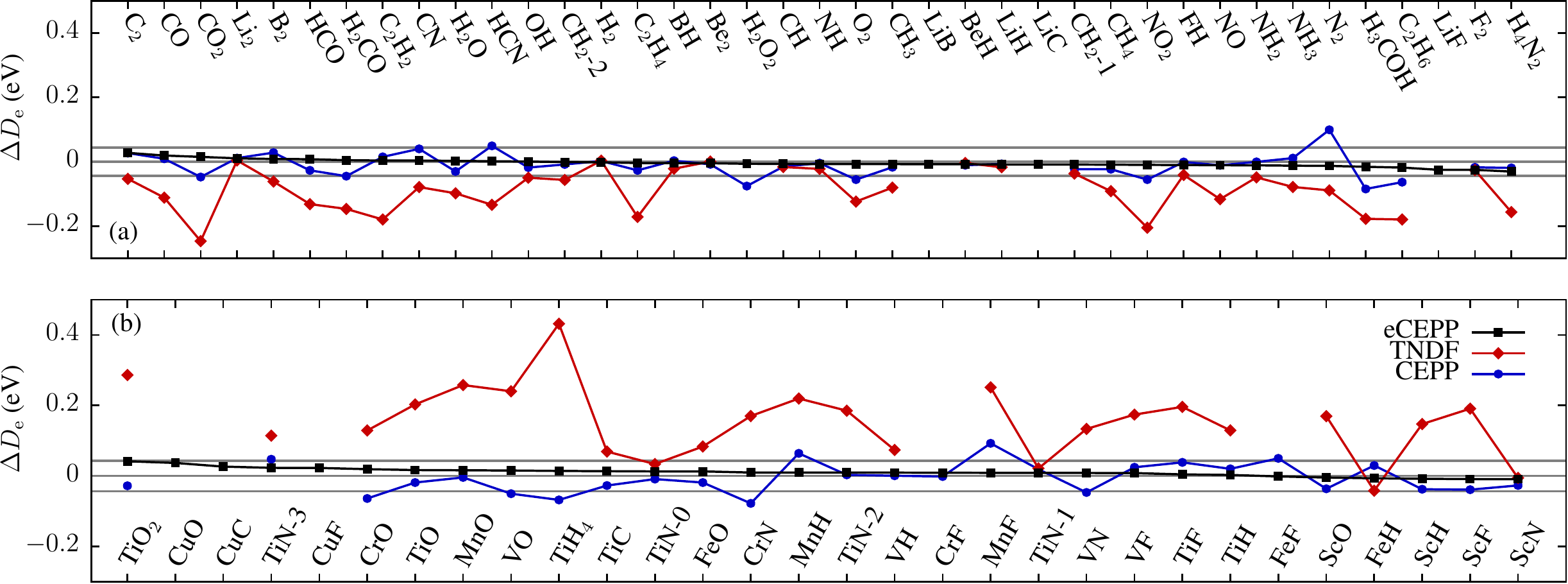}
  \end{center}
  \caption{
  \label{fig:eCEPP}
  The difference between molecular dissociation energies $D_{\rm e}$
  from pseudopotential and all-electron CCSD(T) calculations.
  The top panel shows differences for molecules containing first-row
  atoms only, and the bottom panel shows differences for molecules
  containing $3d$ transition-metal atoms.
  Data for the eCEPP, CEPP, and TNDF pseudopotentials are from Refs.\
  \onlinecite{Trail_ppot_2013, Trail_ppot_2015, Trail_ppot_2017}.
  The region between the gray lines at $\pm 43$ meV contain all data
  points that reproduce all-electron energies to within chemical
  accuracy.
}
\end{figure*}

The \textit{ab initio} eCEPPs so generated reproduce the scattering of
valence electrons by the core (norm conservation), as well as the
ionization and excitation energies for a number of states (energy
conservation), and long-range polarization interactions.
Consequently the eCEPPs are unique in that they are inherently
correlated and are both shape- and energy-consistent.
Better than chemical accuracy was demonstrated for these
pseudopotentials applied to a moderately large set of small molecules
with CCSD(T), as shown in Fig.\ \ref{fig:eCEPP}.
At this level of accuracy some care was required in
controlling for basis-set error; the direct use of standard contracted
all-electron CCSD(T) Gaussian basis sets is insufficient.
Contracted Gaussian basis sets generated to be accurate when used with
the eCEPPs are available. \cite{Trail_ppot_2017}

There is room for further research to improve pseudopotential theory
for QMC\@.
While the above eCEPPs consistently achieve better than chemical
accuracy for all test cases, they are available for first-row and
transition-metal atoms only.
There is nothing to prevent the method from being applied to the rest
of the periodic table in the future.

A bias arises in DMC due to the nonlocality of pseudopotentials.
The many-body electronic wave function $\Phi$ is not explicitly
available (rather, it is implicit in the probability distribution that
arises from the dynamics), so the nonlocal pseudopotential operator
$\hat V$ cannot be applied to it.

A locality approximation can be used to avoid this difficulty by
replacing $\hat{V}\Phi$ with $(\Psi^{-1} \hat{V} \Psi ) \Phi$, where
$\Psi$ is the trial wave function.
This approximation replaces a sum of nonlocal one-body potentials with
a local many-body potential, and is exact if $\Psi=\Phi$.
The error introduced by this approximation is of second order in the
error in $\Psi$, is system-dependent, is nonvariational, and can be
large relative to the fixed-node error.

An alternative partially local ``T-move'' scheme is also available.
\cite{Casula_ppot_2006, Casula_ppot_2010}
This approximation is more robust and obeys a variational principle
that ensures that the bias is positive, though not necessarily smaller
in magnitude than the locality approximation error.
Crucially, the T-move scheme removes the negative divergences in the
local energies that occur at the nodes of the trial wave function in
the localized pseudopotential operator $\Psi^{-1} \hat{V} \Psi$; these
divergences can lead to population explosions in DMC\@.
The T-move scheme leads to stable DMC calculations that satisfy the
variational principle across the entire periodic
table. \cite{Drummond_2016}
The recently proposed ``determinant locality approximation'' in which
$\hat{V}\Phi$ is simply approximated by $(\Psi_{\rm S}^{-1} \hat{V}
\Psi_{\rm S} ) \Phi$, where $\Psi_{\rm S}$ is the Slater part of the
wave function, has also been shown to deliver enhanced stability and
resilience of the results against noise due to optimization of the
Jastrow factor.  \cite{Zen_DLA_2019}
Both of these schemes are available in \textsc{casino}.

Plane-wave DFT calculations normally use a Kleinman-Bylander
representation of pseudopotentials.
However, Kleinman-Bylander representations can introduce ghost states,
which are unphysical, low-energy bound states of the pseudopotential.
Problems associated with ghost states for the TNDF pseudopotentials
can generally be avoided by choosing the $s$ angular-momentum channel
to be local in the pseudopotentials in the plane-wave DFT calculation.
In subsequent QMC calculations, which do not use Kleinman-Bylander
representations, the local channel can be chosen to be the highest
angular-momentum component present in the nonlocal
pseudopotential. \cite{Drummond_2016}
A second challenge in plane-wave DFT calculations is to choose the
plane-wave cutoff energy to be sufficiently high.
The convergence of the DFT energy to chemical accuracy is a sufficient
condition for convergence of the DMC energy.
Suggested plane-wave cutoffs for TNDF pseudopotentials can be found in
Ref.\ \onlinecite{Drummond_2016}.

\subsection{QMC for excited-state properties}

\subsubsection{Quasiparticle and excitonic gaps \label{sec:qp_ex_gaps}}

QMC provides a powerful methodology for calculating excitation
energies, which are obtained as differences in the total energy.
The quasiparticle band at Bloch wavevector ${\bf k}$ is ${\cal
  E}({\bf k})=E_{N+1}({\bf k})-E_N$ for unoccupied states, where $E_N$
is the $N$-electron ground-state total energy and $E_{N+1}({\bf k})$
is the $(N+1)$-electron total energy, in which a single electron has
been added to an orbital at ${\bf k}$.
In a finite cell, the same twisted boundary conditions must be used
for the ground state and excited state.
For occupied states, the quasiparticle band is defined as ${\cal
  E}({\bf k})=E_N-E_{N-1}({\bf k})$, where $E_{N-1}({\bf k})$ is the
total energy when a single electron is removed from a state at ${\bf
  k}$.
We are implicitly assuming that the wave function of the
noninteracting electron system is adiabatically connected to the wave
function of the interacting system, so that we can use the wavevectors
of the noninteracting orbitals to label the energy levels of the
interacting system.
QMC thus provides a direct and physical description of quasiparticle
excitations: the excited state in question is determined by the
occupancy of the single-particle orbitals in the Slater determinant,
with the Jastrow factor, the backflow function, and fixed-node DMC
providing increasingly good descriptions of dynamical correlation
effects while leaving the topology of the nodal surface unchanged
relative to the noninteracting case.
At finite temperatures, the quasiparticle bands should be defined in
terms of differences in the Helmholtz free energy when an electron is
added to or removed from a particular state.
This could be achieved by adding DFT vibrational free energies to the
static-nucleus total energies obtained using QMC calculations.

The Schr\"odinger equation for the $(N+1)$-electron excited-state wave
function is $\hat{H}\Psi_{\bf k} = [E_N+{\cal E}({\bf k})]\Psi_{\bf
  k}$, where $\Psi_{\bf k}=e^{i{\bf k}\cdot \sum_{i=1}^{N+1} {\bf r}_i
  / (N+1)} W_{\bf k}({\bf R})$ is the wave function according to
Eq.\ (\ref{eq:mb_bloch_2}).
$W_{\bf k}$ is invariant under simultaneous translation of all the
particles through a primitive lattice vector.
It is easy to show that $\hat{\cal H}_{\bf k}W_{\bf k} = [E_N+{\cal
    E}({\bf k})]W_{\bf k}$, where
\begin{equation}
  \hat{\cal H}_{\bf k} = -\frac{1}{2} \left( \nabla^2 + \frac{2i}{N+1} {\bf
    k} \cdot \sum_{i=1}^{N+1} \nabla_i - k^2 \right) + V.
\end{equation}
Noting that
\begin{equation}
  \hat{\cal H}_{{\bf k}+\delta {\bf k}} = \hat{\cal H}_{\bf k} + \frac
  1 {N+1} \delta {\bf k} \cdot {\bf k} - \frac i {N+1} {\bf k} \cdot
  \sum_{i=1}^{N+1} \nabla_i + {\cal O}(k^2),
\end{equation}
we may use first-order perturbation theory to show that
\begin{equation}
  {\cal E}({\bf k} + \delta {\bf k}) = {\cal E}({\bf k}) + \delta{\bf
    k} \cdot \left< \frac 1 {N+1} \sum_{i=1}^{N+1} \hat{\bf p}_i
  \right> + {\cal O}(\delta k^2),
\end{equation}
where $\hat{\bf p}=-i\nabla_i$ is the momentum operator for electron
$i$.
Hence the mean momentum per electron is given by the gradient of the
quasiparticle energy band with respect to ${\bf k}$, confirming that
${\bf k}$ continues to be the wavevector of the excitation in the
interacting many-electron system.
In a crystalline solid, the quasiparticle bands should be extrapolated
to the thermodynamic limit $N \to \infty$.

The quasiparticle band gap $\Delta_{\rm qp}$ of a semiconductor or
insulator may be evaluated as the difference between the quasiparticle
bands at the conduction-band-minimum (CBM) and the valence-band
maximum (VBM):
\begin{eqnarray} \Delta_{\rm qp} & = & {\cal E}_{\rm CBM}-{\cal E}_{\rm VBM}
\nonumber \\ & = & E_{N+1}({\bf k}_{\rm CBM})+E_{N-1}({\bf k}_{\rm VBM})-2E_N.
\end{eqnarray}
Again, this should be extrapolated to infinite system size as
discussed in Sec.\ \ref{sec:fs_effects_gaps}.
The quasiparticle band gap is the amount of energy required to produce
a free electron and a free hole.
This gap determines the thermal concentrations of charge carriers, and
hence can be determined by examining the temperature-dependence of the
conductance of a sample.
The occupied and unoccupied quasiparticle bands can be measured
directly using photoemission and inverse photoemission spectroscopies,
or can be probed using X-ray absorption near-edge structure
measurements.

A second type of gap plays a crucial role in the interaction of
semiconductors with light, namely the excitonic gap.
This is defined as the difference between the ground-state and
excited-state energies of the $N$-electron system: $\Delta_{\rm
  ex}=E'_N({\bf k},{\bf k}')-E_N$, where $E'_N({\bf k}',{\bf k}')$ is
the energy of an excited state in which an electron has been promoted
from an occupied state at wavevector ${\bf k}$ to an unoccupied state
at wavevector ${\bf k}'$.
Again, assuming adiabatic connection from the noninteracting system,
we can select the excited state by the occupancy of the Slater
determinant in a QMC wave function; the Jastrow factor, backflow
function, and the use of fixed-node DMC provide dynamical correlation
without changing the qualitative state of the system.
The excitonic gap is the lowest energy at which the electronic system
can absorb or emit a photon.
This is the gap that is observed in photoluminescence measurements.
Physically, the excitonic gap is less than the quasiparticle gap
because photoexcitation produces an electron-hole pair, which bind to
form an exciton; thus the difference between the quasiparticle and
excitonic gaps at zero temperature is given by the exciton binding
energy.
Since photon momenta are vanishingly small compared with electrons,
single-photon absorption or emission only occurs for direct gaps,
where ${\bf k}={\bf k}'$.
Otherwise there must be additional processes involved, such as
emission of phonons, in order to satisfy the conservation of momentum,
and so we do not have a corresponding sharp peak in absorption or
emission.
Excitonic gaps can be calculated for cases where ${\bf k}-{\bf k}'$ is
commensurate with the simulation cell.
When we take into account nuclear motion, a few further issues arise.
Geometry differences between the ground state and excited state lead to
differences between the excitonic gaps for optical absorption and
emission, respectively; this difference is referred to as a Stokes
shift.
Photon absorption can take place anywhere in the crystal, and it
takes place instantaneously on the time scale of nuclear motion.
Therefore, the excitonic gap should be averaged over the thermal and
quantum distributions of nuclear coordinates; this gives the energy of
the excitonic peak in the photoluminescence spectrum.
This is discussed further in Sec.\ \ref{sec:vib_corr}.

In an isolated atom or molecule, the electron affinity is ${\cal
  E}_{\rm ea}=E(N)-E(N+1)$, where $E(N)$ and $E(N+1)$ are the total
ground-state energies of the $N$-electron and $(N+1)$-electron
systems, while the ionization potential is ${\cal E}_{\rm
  ip}=E(N-1)-E(N)$, where $E(N-1)$ is the total ground-state of the
$(N-1)$-electron system.
These total energies should in principle include static correlation
effects.
The great challenge here is to use multideterminant wave functions of
equivalent accuracy for the neutral atom or molecule and the positive
and negative ions.

\subsubsection{Finite-size effects in gaps}
\label{sec:fs_effects_gaps}

QMC studies of condensed matter usually involve performing
calculations for finite simulation cells subject to periodic boundary
conditions.
Nevertheless, the excitation energies obtained in a finite simulation
cell differ from those in the thermodynamic limit of infinite system
size.
Fortunately, the phenomenological quasiparticle picture described in
Sec.\ \ref{sec:excitonics} can be used to understand the dominant
finite-size effects in \textit{ab initio} QMC
calculations. \cite{Hunt_2018}

Consider the quasiparticle bands of an insulator or semiconductor.
As described in Sec.\ \ref{sec:qp_ex_gaps}, an unoccupied band is
defined via the difference in the total energy relative to the ground
state when an electron is added to the system.
Within the band effective mass approximation, the additional electron
moves like a negatively charged quasielectron whose mass is determined
by the curvature of the conduction band.
Similarly, excitations from the valence bands behave as positively
charged quasiholes whose mass is determined by the curvature of the
valence band.
The Coulomb interactions between quasiparticles are screened by the
response of the other electrons in the crystal.

When one adds an electron to a periodic simulation supercell to
calculate the CBM, the effect is to create a lattice of quasielectrons
repeated throughout space.
The leading-order finite-size error is therefore the unwanted Madelung
energy $v_{\rm M}/2$ of this lattice of quasielectrons, which must be
subtracted from the energy $E_{N+1}$ of the $(N+1)$-electron system.
Similarly the Madelung energy of the unwanted lattice of quasiholes
must be subtracted from the energy $E_{N-1}$ of the $(N-1)$-electron
system.
So the leading-order finite-size correction to the quasiparticle gap
$\Delta_{\rm qp}=E_{N+1}+E_{N-1}-2E_N$ in a finite cell is that the
supercell Madelung constant $v_{\rm M}$ must be subtracted from the
gap. \cite{Hunt_2018}
Note that the supercell Madelung constant must be evaluated using the
screened Coulomb interaction.
In the case of a crystal of cubic symmetry the screened Madelung
constant is simply $v_{\rm M}=v_{\rm M}'/\epsilon$, where $v_{\rm M}'$
is the unscreened Madelung constant and $\epsilon$ is the
static-nucleus (high-frequency) permittivity.
In the general 3D case, the screened Madelung constant is found by a
coordinate transformation to the eigenbasis of the permittivity tensor
$\epsilon$: $v_{\rm M}({\bf a}_1,{\bf a}_2,{\bf a}_3)=v_{\rm
  M}'(\epsilon^{-1/2}{\bf a}_1,\epsilon^{-1/2}{\bf
  a}_2,\epsilon^{-1/2}{\bf a}_3)/\sqrt{\det(\epsilon)}$, where ${\bf
  a}_1$, ${\bf a}_2$, and ${\bf a}_3$ are the supercell lattice
vectors.
For a given shape of simulation cell, the finite-size error in the
uncorrected gap falls off slowly as $N^{-1/3}$.
Moreover, since $v_{\rm M}$ is often negative, the uncorrected
quasiparticle gap may be absurdly small, even negative.
In the case of a layered or 2D material, the Madelung constant should
be evaluated using the Rytova-Keldysh interaction. \cite{Rytova_1965,
  Keldysh_1979}
In another example of the symbiotic relationship between QMC and DFT,
the permittivity tensor can usually be evaluated to sufficient
accuracy using density functional perturbation theory calculations.
Alternatively the permittivity tensor can be evaluated using the
small-${\bf k}$ limit of the static structure factor, which can be
evaluated using ground-state QMC calculations. \cite{Yang_2019}

Beyond-leading-order finite-size effects in the quasiparticle bands
and therefore quasiparticle gap arise due to the fact that the
charge-density distributions in the quasiparticles have quadrupole
moments in general; the resulting charge-quadrupole interactions give
an ${\cal O}(N^{-1)}$ finite-size error in 3D materials.
The addition of charges to a finite simulation cell also causes
long-range oscillatory behavior in the electronic pair density.
These oscillations are forced to be commensurate with the finite
periodic simulation cell, resulting in quasirandom fluctuations in the
gap with system size.
Hence a suggested procedure for calculating QMC quasiparticle gaps is
to perform calculations at a range of system sizes, correct the
leading-order errors by subtracting the screened Madelung constant,
then extrapolate the resulting gaps to infinite system size assuming a
$1/N$ error; this simultaneously removes residual systematic effects
and averages out quasirandom finite-size errors. \cite{Hunt_2018}

To calculate an \textit{ab initio} exciton binding energy, the
supercell must be significantly larger than the exciton Bohr radius.
In smaller cells, the quasielectron and quasihole are effectively
unbound, and hence the excitonic gap is similar to the quasiparticle
gap.
If the simulation supercell is large enough to contain well-formed
excitons then interactions between periodic images are relatively
unimportant, since excitons are neutral.
However, the interaction between the quasielectron and quasihole is
different in a finite cell and an infinite system.
From the Taylor expansion of the 3D Ewald interaction $v_{\rm E}$ for
a periodic cell, the leading-order difference from the Coulomb $1/r$
interaction is ${\cal O}(N^{-1})$. \cite{Makov_1995,Fraser_1996}
This leads to an ${\cal O}(N^{-1})$ finite-size error in the exciton
binding energy; hence, provided the simulation cell is large enough,
the excitonic gap can be extrapolated to the thermodynamic limit.
The conclusion is modified in 2D or layered systems, where the
interaction is of Rytova-Keldysh form, leading to ${\cal
  O}(N^{-2/3})$ scaling of the finite-size error that eventually
crosses over to ${\cal O}(N^{-1})$ behavior on near-macroscopic
length scales. \cite{Hunt_2018}

\subsubsection{Intraband excitations}

QMC methods can be used to study intraband excitations in metallic
systems.
An example of this is the calculation of the renormalization of the
electron mass by electron-electron interactions in the 2D HEG, which
is discussed in Sec.\ \ref{sec:qem_2D_HEG}.

\subsection{Computational efficiency of QMC}

\subsubsection{Parallelization of QMC algorithms \label{sec:parallelisation}}

QMC methods are intrinsically well-suited to massively parallel
architectures.
\textsc{Casino} makes use of message-passing interface (MPI) for an
``outer'' distributed-memory level of parallelism and OpenMP for an
``inner'' shared-memory level of parallelism.
In practice the MPI parallelism is more important and much more widely
used.
However, where possible, data such as wave-function coefficients are
shared between MPI processes within each processor node.

Assuming equilibration to be a negligible fraction of the total
central-processing-unit (CPU) time, the VMC method is ``embarrassingly
parallel.''
Multiple independent VMC calculations may be run in parallel, each
with a different random seed, to generate an amount of data that is
proportional to the total computational effort.
The only required communication between MPI processes is the broadcast
of the initial distribution of geometry and wave-function data, the
sending of the initial random seeds, and the summation of the final
set of expectation values to average.
No interprocess communication is required during the VMC calculation.
In wave-function optimization, VMC-sampled configurations are divided
between MPI processes, which independently evaluate local energies and
other required quantities.
A ``master'' process gathers reduced data such as the variance of the
local energies in order to choose a new parameter set, which must then
be broadcast.

The natural parallelization strategy for DMC calculations is to divide
the walker population between MPI processes, so that each process has
its own population of walkers.
Unlike VMC, communication is required during the simulation, as a
master process has to decide on a reference energy in order to control
the overall population.
Furthermore, because the walkers randomly branch and die, it is
necessary constantly to balance the load on different MPI processes by
transferring walkers from processes with larger populations to
processes with smaller populations.
The cost of a DMC time step is determined by the MPI process with the
largest population of walkers.
The amount of branching varies from one system to another and is
reduced when the wave function is more accurate, thereby improving the
parallel performance.
Since fractional fluctuations in the walker population fall off as the
inverse square root of the population, the fractional cost of
transferring walkers can be made arbitrarily small (and hence the
parallel efficiency of the statistics-accumulation phase can be made
arbitrarily large) by choosing a large number of walkers per process.
However, increasing the population leads to additional expense due to
the need to equilibrate the large population, as discussed in
Sec.\ \ref{sec:Nscaling}.

The secondary OpenMP parallelism is used to accelerate loops over the
electrons when evaluating interparticle interactions, Jastrow terms,
and orbitals.

\begin{figure}[!htbp]
\begin{center}
\includegraphics[clip,width=0.45\textwidth]{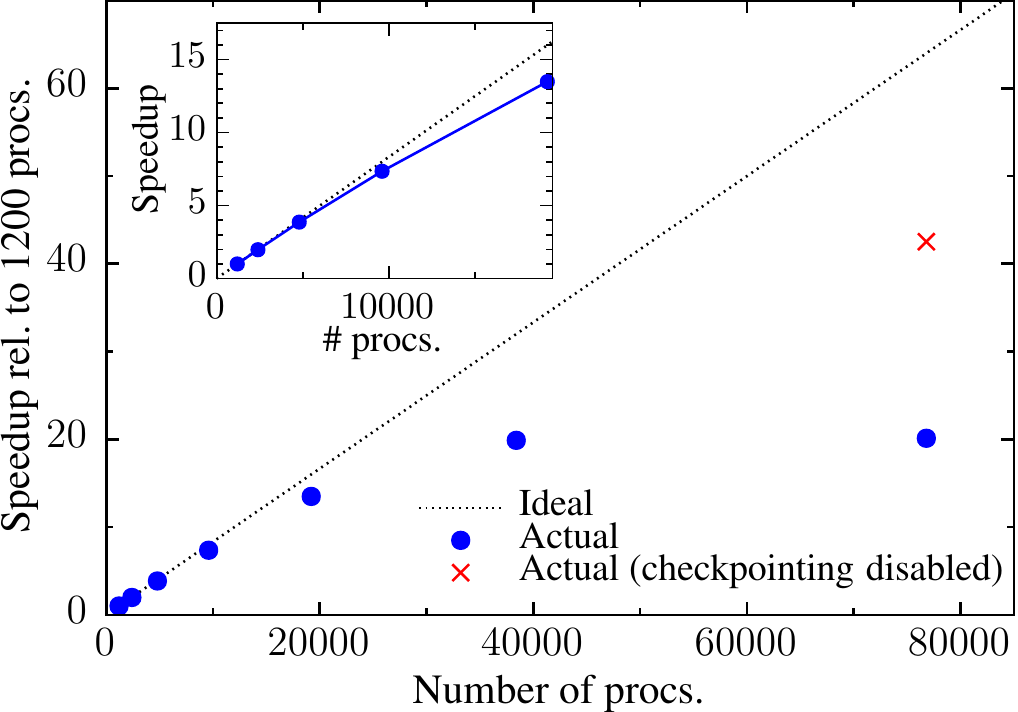}
\caption{Strong-scaling parallel speedup for DMC calculations with
  \textsc{casino} for a 54-atom supercell of silicon ($N=216$
  electrons), performed on Jaguar.
  The inset shows the scaling with small processor numbers in greater
  detail. \label{fig:si_scaling}}
\end{center}
\end{figure}

Figure \ref{fig:si_scaling} shows the strong-scaling parallel speedup
when carrying out a block of 50 time steps in a \textsc{casino} DMC
calculation against the number of processors.
The calculations were performed on Oak Ridge Leadership Computing
Facility's Jaguar supercomputer.
The system studied was a 54-atom supercell of silicon and the number
of electrons in the simulation was $N=216$.
The SJ wave function, time step, etc., were all realistic.
The only unrealistic element of the calculation was that the
calculation was very short, with \textsc{casino} writing out
checkpoint data at the end of the 50 iterations.
The time taken to assemble the checkpoint data on the master processor
and write it to disk is the bottleneck for large numbers of cores, as
evidenced by the fact that if checkpointing is disabled the time taken
falls significantly, and the apparent scaling with processor number is
greatly improved.
The time spent checkpointing with 76,800 cores is about 100 seconds.
In a real calculation, we typically checkpoint once every 20--30
minutes, so this bottleneck is an artifact of the short test runs used
in the scaling test.
It is clear that DMC calculations with \textsc{casino} scale very well
up to and beyond 10,000 cores.
In fact \textsc{casino} exhibited virtually perfect linear scaling on
up 124,416 cores on Jaguar, 131,072 cores on an IBM BGQ, and over half
a million cores on the K computer, due to modifications described in
Ref.\ \onlinecite{Gillan_2011}.

\begin{figure}[!htbp]
\begin{center}
\includegraphics[clip,width=0.45\textwidth]{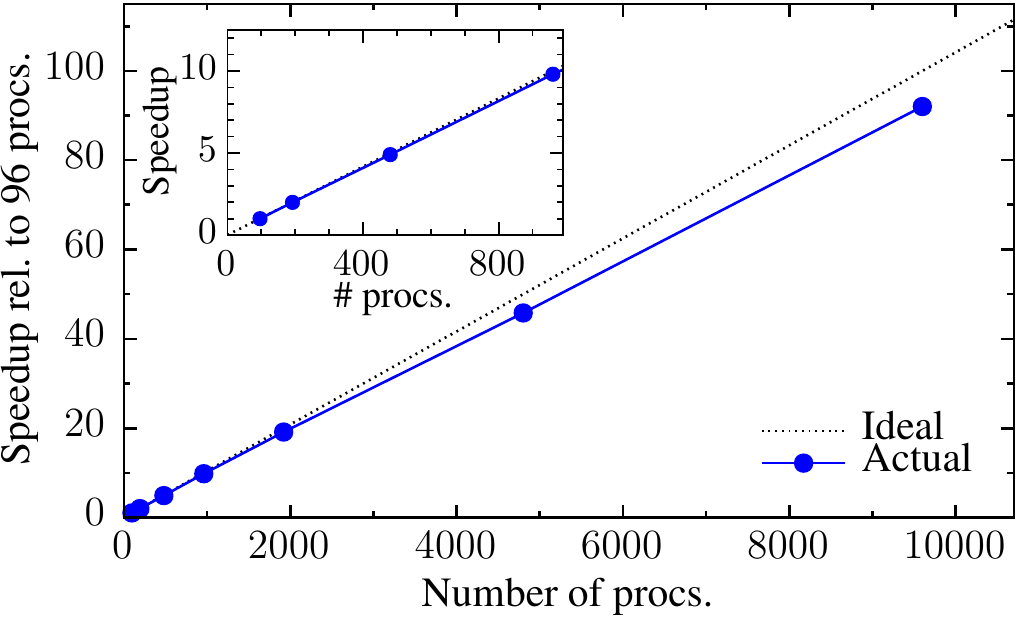}
\caption{Strong-scaling parallel speedup for DMC calculations with
  \textsc{casino} for a 144-atom supercell of bilayer graphene
  ($N=576$ electrons), performed on ARCHER\@.
  The inset shows the scaling with small processor numbers in greater
  detail. \label{fig:blg_scaling}}
\end{center}
\end{figure}

Figure \ref{fig:blg_scaling} shows a similar parallel scaling
analysis, this time for a 144-atom supercell of bilayer graphene,
carried out on the supercomputer ARCHER\@.
Again it is clear that the speedup for a fixed number of DMC time
steps and target population is extremely good, up to at least 10,000
cores.

\subsubsection{Scaling of QMC with system size \label{sec:Nscaling}}

A basic analysis of system-size scaling must consider the cost of
moving each walker, and the number of such moves that are required to
achieve a result of a given precision (target error bar).
We shall consider the cost when a Slater-Jastrow wave function with a
fixed number of determinants is used.

Consider moving an $N$-electron walker; this involves proposing a move
of each electron in turn, then evaluating the local energy at the
resulting configuration at the end of the move.
(In the case of VMC we only need to evaluate the energy once every few
moves; in the case of DMC we must evaluate it after every walker move,
as it is needed to evaluate the branching factor.)
There are several ${\cal O}(N^2)$ contributions to the time taken:
electron-electron interactions must be evaluated; the Jastrow factor
nearly always contains long-range two-body terms of
system-size-independent complexity between each pair of electrons; and
each orbital must be evaluated at each electron position.
We have implicitly assumed the use of a localized basis set for the
orbitals; if an extended basis set is used then there is an extra
factor of $N$ in the scaling of orbital evaluation, since the number
of basis functions for each orbital at each position increases as $N$
in this case.
For this reason, when orbitals are generated using plane-wave DFT
calculations, they are re-represented on a B-spline (blip) grid in
real space before use in QMC calculations;
\cite{Williamson_2001,Alfe_2004} instead of evaluating ${\cal O}(N)$
plane waves at each electron position, we evaluate twelve B-spline
functions.
Likewise, Gaussian basis functions are rigorously truncated to zero
beyond a certain radius, so they can be treated as a localized basis.
In studies of insulators, the cost of evaluating the orbitals can be
reduced to ${\cal O}(N)$ by performing a linear transformation to a
set of (nonorthogonal) localized orbitals that can be truncated to
zero outside predefined localization regions.
\cite{Williamson_2001,Alfe_2004b}
To minimize the bias arising from the truncation, we maximize the
overlap of the localized orbitals with the set of localization
regions.
To avoid bias, the localization regions for valence bands need to be
significantly larger than the length scales estimated in
Sec.\ \ref{sec:time_steps}.
For core states, the localization regions can have radii of order 1
bohr.

Evaluating a Slater determinant given all the orbital values is an
${\cal O}(N^3)$ operation.
In an electron-by-electron algorithm Slater determinants must be
updated after each accepted single-electron move.
Unless backflow is used, a single-electron move only affects a single
column of the Slater matrix.
Such updates can be performed in a time that scales as ${\cal O}(N^2)$
by storing the matrix of cofactors of the Slater matrix and updating
this using the Sherman--Morrison formula, resulting in an ${\cal
  O}(N^3)$ cost of updating the Slater determinant over a walker
move. \cite{Ceperley_1977}
In practice, however, we find that the ${\cal O}(N^3)$ operations have
negligible cost and, over the practical range of system sizes, the
cost of each walker move scales as ${\cal O}(N^2)$.
The costs of evaluating the Jastrow factor and the orbitals are
usually comparable, with the cost of evaluating electron-electron
interactions being significantly less.

For a given wave-function form the variance of the total energy scales
as ${\cal O}(N)$, while the variance in the energy per particle falls
off as ${\cal O}(N^{-1})$.
The standard error in the mean energy is given by the square root of
the variance divided by the square root of the number of independent
samples (see Sec.\ \ref{sec:sampling}).
For insulators at least, the number of independent samples is
proportional to the number of actual samples for a given time step.
Hence the number of configurations sampled must increase as ${\cal
  O}(N)$ to achieve a given error bar on the total energy, while the
number of samples required to achieve a given error bar on the energy
per particle falls off as ${\cal O}(N^{-1})$.
Similar comments apply to the scaling of the QMC estimate of any
extensive or intensive quantity, respectively.
In the commonly encountered situation in which the cost of
equilibration is negligible and the cost of each walker move is ${\cal
  O}(N^2)$, this leads to the well-known ${\cal O}(N^3)$ scaling of
the QMC method.  \cite{Foulkes_2001}

We now consider the scaling of the DMC method in more detail, taking
into account the need to equilibrate the walker population.
Suppose we are interested in using DMC to study a real,
$d$-dimensional metallic system using pseudopotentials to represent
the atoms.
The linear size of an $N$-electron simulation cell is $L=N^{1/d}r_{\rm
  s}$, where $r_{\rm s}$ is the density parameter for the valence
electrons.
The valence electron density parameter in a real system is typically
$r_{\rm s}=2$--$5$.
The RMS distance diffused by each electron in equilibration needs to
be greater than $L$ in order to ensure that equilibration errors are
exponentially small.
So the number of equilibration steps $N_{\rm eq}$ must satisfy
\begin{equation}
  \label{eq:diff_dist}
  \sqrt{N_{\rm eq} \tau d} \gtrsim N^{1/d}r_{\rm s}.
\end{equation}
Note that there is little point in using a huge simulation cell if one
does not equilibrate sufficiently long to describe the long-range
behavior correctly.
If a DMC simulation does not extend over a few decorrelation periods
then we will not be able to observe a plateau in reblocking analysis,
and it will be difficult to obtain accurate error bars.

Let $W$ be the total walker population and $P$ be the number of MPI
processes.
To minimize equilibration time, the number of walkers per MPI process
$W/P$ will be a fixed small number.
It is preferable not to have fewer than two walkers per MPI process,
otherwise parallel efficiency is reduced.
If the number of walkers per MPI process is fixed, the wall-clock time
taken to perform $N_{\rm eq}$ time steps is essentially independent of
the number of processes.

For a Slater-Jastrow wave function with B-spline (blip) orbitals, the
time in seconds taken to perform $N_{\rm eq}$ DMC iterations is
\begin{equation}
  \label{eq:eq_time}
  T_{\rm eq} = c_0 N^2 N_{\rm eq} W/(P\Theta),
\end{equation}
where $\Theta$ is the number of OpenMP threads and the constant $c_0$
depends on the system, the wave-function quality, and the computer
hardware.
In practice OpenMP parallelism is useful provided $N/\Theta \gtrsim 100$.
Empirically, $c_0 \sim 10^{-5}$ s for a range of first-row atoms with
Slater-Jastrow wave functions and various Intel processors from around
the year 2010.

Combining Eqs.\ (\ref{eq:diff_dist}) and (\ref{eq:eq_time}), we have
\begin{equation}
  N \lesssim \left( \frac{\tau T_{\rm eq} P \Theta d} {c_0 r_{\rm s}^2 W}
  \right)^{d/8}.
\end{equation}
If we have $W/P=2$ walkers per MPI process, a single thread per
process, $r_{\rm s}=3$ (a typical valence electron density), the
maximum equilibration time $T_{\rm eq}=4$ days ${}\approx 3 \times
10^5$ s, and the usual time step $\tau=0.01$ Ha$^{-1}$ for
pseudopotential calculations, we obtain
\begin{equation}
  N \lesssim 800.
\end{equation}
So performing DMC for more than $N=800$ electrons with a time step of
0.01 Ha$^{-1}$ and a single thread per MPI process will be difficult,
because one will have to equilibrate for many days, no matter how big
a computer one uses.
There is a bit of leeway here if one is willing to use larger time
steps (e.g., assuming cancellation of biases), or is willing to accept
incomplete equilibration (because the resulting bias is somehow known
to be negligible compared with a target error bar), or is willing to
equilibrate for many days.
We have also assumed that we are looking at a metallic system, which
is a worst-case scenario; in a system in which electrons are highly
localized, the equilibration time could be much shorter.
Alternatively, to evade this limit, one may use multiple threads per
MPI process.
With $\Theta=16$ threads per process, the upper limit on the number of
electrons grows to about 2000.

For the largest system sizes that one might like to consider using
$\Theta=1$ thread ($N \approx 800$ electrons), the equilibration
period lasts $T_{\rm eq}=4$ days.
Let us assume the processor number is sufficiently large that we are
in the regime where we have a fixed number of walkers per MPI process.
The scaling with processor number is good if and only if equilibration
is a small fraction of the total run time.
The point at which the scaling with processor number becomes poor is
when the time spent on statistics accumulation is about the same as
the time spent on equilibration.
After that, increasing the number of MPI processes makes increasingly
little difference to the total run time, because the equilibration
time remains stuck at $T_{\rm eq}$.
So, for the largest number of MPI processes it is worth using, the
number of statistics accumulation steps $N_{\rm s}$ is such that the
time taken for statistics accumulation is equal to the time taken for
equilibration:
\begin{equation}
  \label{eq:stats_eq}
  c_0 N^2 N_{\rm s} W/(P\Theta) = T_{\rm eq}.
\end{equation}

For insulators at least\footnote{For metals the longest length scale
  grows with system size; however, shorter length scales make the
  largest energy contributions.} the error bar on the DMC total energy
per electron in pseudopotential calculations goes as
\begin{equation}
  \label{eq:errbar_per_elec}
  \Delta = \frac {k_0} {\sqrt{WNN_{\rm s}\tau}},
\end{equation}
where $k_0$ depends on the system and wave-function quality.
Empirically, $k_0 \sim 0.1$ Ha$^{1/2}$ for a range of first-row atoms
with Slater-Jastrow wave functions.

Suppose the target accuracy is $\Delta_0$.
By using Eq.\ (\ref{eq:errbar_per_elec}) to eliminate the number of
statistics-accumulation steps from Eq.\ (\ref{eq:stats_eq}), the
maximum useful number of MPI processes is
\begin{equation}
  P_{\rm max} = \frac{k_0^2 c_0 N} {\Delta_0^2 T_{\rm eq} \tau
    \Theta}.
\end{equation}
Putting in $\Theta=1$ thread, $N=800$ (the largest size for which
equilibration is feasible), $\tau=0.01$ Ha$^{-1}$, $T_{\rm eq}=4$ days and
a very stringent target accuracy of $\Delta_0=2 \times 10^{-6}$
Ha per electron (i.e., $\Delta_0=8 \times 10^{-6}$ Ha per atom
when there are 4 electrons per atom) gives the maximum useful number
of MPI processes as $P_{\rm max}=6000$.
When running with the maximum useful number of MPI processes, the
total run time is $2T_{\rm eq}$ (i.e., eight days).

In practice there are usually trivial sources of parallelism in twist
averaging, running at different time steps, running for different
systems and system sizes, etc.

Now suppose we are interested in total energies, e.g., to calculate a
band structure or a defect formation energy.
In that case the error in the total energy is
\begin{equation}
  \label{eq:errbar_tot}
  \Delta_{\rm tot} = k_0 \sqrt{\frac N {WN_{\rm s}\tau}}.
\end{equation}
Suppose the target accuracy is $\Delta_{0{\rm tot}}$.
By Eqs.\ (\ref{eq:stats_eq}) and (\ref{eq:errbar_tot}), the maximum
useful number of MPI processes is
\begin{equation}
  P_{\rm max} = \frac{k_0^2 c_0 N^3} {\Delta_{0{\rm tot}}^2 T_{\rm eq}
    \tau \Theta}.
\end{equation}
A target accuracy for calculating a band structure might be 0.001 Ha.
Putting in $\Theta=1$, $N=800$, and $\tau=0.01$, we find that $P_{\rm
  max} \approx 10^4$.

The same sort of analysis can be applied to VMC, although in VMC the
time step is much larger, so that equilibration remains negligible for
much larger system sizes.

If we return to a situation in which the equilibration period can be
assumed to be negligible, e.g., because extremely small error bars are
required, then a fundamental difference between VMC and DMC emerges.
VMC with a Slater-Jastrow wave function scales as
$\mathcal{O}(N^3+\epsilon N^4)$, while VMC with a
Slater-Jastrow-backflow wave function scales as $\mathcal{O}(N^4)$.
DMC shows the same scaling at small $N$, but asymptotically crosses
over to exponential scaling. \cite{Nemec_2010}
The issue is related to the system-size dependence of
population-control bias, discussed in Sec.\ \ref{sec:propagation}.
When a walker branches, the resulting daughter walkers are highly
correlated, reducing the effective walker population.
The daughter walkers decorrelate at a rate given by the inverse of the
decorrelation period $\tau_{\rm corr}$.
It is clear from the form of the branching factors that the rate at
which branching takes place is proportional to the spread of local
energies that appear in the branching factors.
The spread of local energies may be quantified by the standard
deviation of the local energy, which scales as the square root of
system size.
The spread of local energies is smaller for better wave function
forms.
Nevertheless, for any given wave-function form the branching rate
eventually overtakes the decorrelation rate as the system size is
increased.
This occurs when the standard deviation of the local energies exceeds
the fixed-phase energy gap.
In this case the effective population size decreases to a single
walker.
This provides an exponential hard wall on the system-size scaling of
DMC, which for Slater-Jastrow wave functions is typically reached with
of order $1000$--$2000$ electrons. \cite{Nemec_2010}

In summary, the ${\cal O}(N^3)$ scaling of DMC total-energy
calculations breaks down at system sizes of around $N=1000$--$2000$
electrons, and the excellent scaling of QMC with processor number does
not overcome this problem.
When studying condensed matter, we must use careful analysis of
finite-size effects to obtain accurate results from system sizes of
fewer than $2000$ electrons.

\section{Some recent applications of CASINO}
\label{sec:applications}

\subsection{Model systems}

\subsubsection{Excitonic complexes \label{sec:excitonics}}

QMC methods are powerful techniques for solving the many-electron
Schr\"{o}dinger equation in \textit{ab initio} chemistry and
condensed-matter physics.
However, the same methods can be used to solve the Schr\"{o}dinger
equation in models that capture the key physics of a material or
device.
Indeed it is often much easier to make direct connection with
experiments in semiconductor physics by means of such model
calculations.

An idea of fundamental importance in solid-state physics is the band
effective mass approximation.
Interactions of electrons with external fields that are slowly varying
on the scale of the primitive unit cell are described by quasiparticle
excitations, whose wave functions provide envelopes for the underlying
Bloch orbitals.
These envelope wave functions are solutions of the Schr\"{o}dinger
equation for quasiparticles whose effective mass is determined by the
curvature of the underlying electronic band structure.
In semiconductors, the most important quasiparticles are charge
carriers, i.e., the small number of electrons found near the CBM and
the small number of holes found near the VBM\@.
Under the assumption of locally quadratic bands, the effective masses
of the electrons and holes are the reciprocals of the second
derivatives of the conduction and valence bands with respect to
wavevector at their minimum and minimum, respectively.
Most of the complexity of the many-electron problem is buried in the
effective masses; the remaining challenge is to solve the
Schr\"{o}dinger equation for the quasiparticles interacting with
external fields, with defects in the lattice, and with each other.

Quasiparticles interact via Coulomb interactions that are screened by
the response of the electrons and ions in the crystal.
In a bulk semiconductor of cubic symmetry, the screening is described
by a scalar permittivity $\epsilon$, and the interaction between
charge carriers is of the usual $1/r$ form scaled down by $\epsilon$.
If the exciton energy is small compared with the optical phonon
energies then it is reasonable to assume that the ions have time to
relax as the charge carriers move, so that the permittivity $\epsilon$
should be the static permittivity.
By contrast, in materials with large exciton binding energies compared
to the optical-phonon energies, the ions do not have time to respond
to the electronic motion, and hence the high-frequency
(static-nucleus) permittivity should be used.
In a bulk semiconductor of lower symmetry, $\epsilon$ is a tensor and
hence the interaction $1/|\epsilon {\bf r}|$ between charges is
anisotropic.
In a layered or 2D material, the form of the interaction is altered
more radically due to the fact that the material is only polarizable
in-plane.
This leads to the so-called Rytova-Keldysh form of interaction,
\cite{Rytova_1965, Keldysh_1979} which is of Coulomb $1/r$ form at
long range but is logarithmic at short range.

At very short range the effective-mass approximation breaks down, and
local exchange and correlation effects lead to effective contact
interaction potentials between charge carriers.
For weakly bound Mott-Wannier excitons and excitonic complexes these
contact interactions are often negligible, and they largely cancel out
of the binding energies of larger complexes.

Charge-carrier complexes are bound states of electrons, holes, and
charged defects in a semiconductor.
The simplest bound complexes are excitons (bound states of a single
electron and a single hole), donor atoms (bound states of an electron
and a positively charged defect), and acceptor atoms (bound states of
a hole and a negatively charged defect).
Three-particle complexes include positive and negative trions (two
holes and an electron, and two electrons and a hole, respectively).
Four-particle complexes include biexcitons (two electrons and two
holes).
Valley degeneracy may lead to multiple distinguishable species of
electron or multiple species of hole, and hence the possibility of
even larger complexes.
In many cases the fermionic statistics obeyed by the charge carriers
are irrelevant, because there is only one member of each
distinguishable species present in each stable complex, and the
long-range Hamiltonian does not depend on spin or valley; hence the
DMC method is exact.
Where this has been investigated in 2D materials, it has been found
that biexcitons with a pair of indistinguishable charge carriers are
unstable except when the masses of the indistinguishable charge
carriers are much greater than the masses of the other two charge
carriers. \cite{Mostaani_2017}

Charge-carrier complexes play a crucial role in the interaction
between semiconductors and light at low temperature, as recombination
takes place from electrons and holes bound in complexes rather than
from a gas of free charge carriers.
At low temperature, peaks in photoluminescence spectra due to each
stable species of charge carrier complex may be visible, dependent on
selection rules.
The energy of a carrier complex provides the energy of a peak in the
photoluminescence spectrum relative to the quasiparticle band gap.
The energy to remove an exciton from a complex provides the position
of the peak due to the complex relative to the exciton peak.
The binding energy relative to the most energetically favorable
daughter complexes also provides an estimate of the temperature at
which the peak due to a particular complex will disappear.
Depending on the doping of the semiconductor, free electrons or holes
may be present at the time of photoexcitation, allowing the formation
of charged complexes such as trions.
On the other hand, where charge carriers are produced purely by laser
photoexcitation of an undoped semiconductor, the charge-carrier
complexes are overwhelmingly likely to be neutral (excitons or
biexcitons).

QMC methods have been used to evaluate the binding energies of
isotropic 3D excitons and ideal 2D biexcitons interacting via the
Coulomb interaction, \cite{Bressanini_1998} biexcitons and trions in
ideal 2D bilayers modeling coupled quantum-well heterostructures of
III-V semiconductors, \cite{Tan_2005,Lee_2009,Witham_2018} 3D
biexcitons and trions in type-II superlattices of III-V
semiconductors, \cite{Tsuchiya_1998,Tsuchiya_2000,Tsuchiya_2004} and
3D biexcitons and trions in quantum wells, \cite{Tsuchiya_1998b}
quantum dots, \cite{Tsuchiya_2000b} and type-II quantum rings
\cite{Thomas_2019}.
The issues studied have included not just the prediction of peaks in
photoluminescence spectra at low temperature, but also the inhibition
of Bose-Einstein condensation of excitons in 2D coupled quantum
wells.
Charge-carrier complexes play a particularly important role in the
optoelectronic properties of layered and 2D materials.
In the field of 2D materials, the DMC-calculated binding energies of
trions \cite{Ganchev_2015,Mayers_2015,Szyniszewski_2017} and the
binding energies of larger complexes such as biexcitons and quintons
(charged biexcitons) \cite{Mostaani_2017,Barbone_2018} have been
reported, as have the DMC-calculated binding energies and
VMC-calculated recombination rates of ion-bound charge carrier
complexes in heterobilayers of transition-metal
dichalcogenides. \cite{Danovich_2018,Vialla_2019}

\subsubsection{Ground-state energy of the HEG}
\label{sec:app_heg}

Ruggeri \textit{et al.}\cite{Ruggeri_heg_2018}\ obtained the correlation
energy of the ferromagnetic HEG at high densities to unprecedented
accuracy thanks to two key elements.

\begin{figure}[!htbp]
  \begin{center}
    \includegraphics[clip,width=0.45\textwidth]{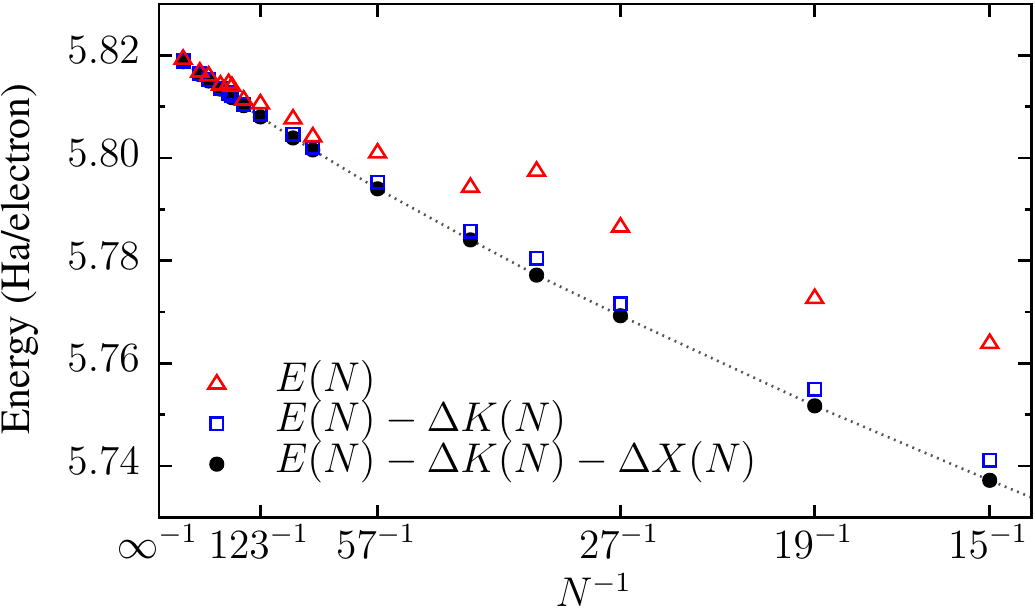}
  \end{center}
  \caption{ Twist-averaged SJ-DMC energy per electron of the $r_{\rm
      s}=0.5$ HEG as a function of the reciprocal of system size with
    and without the $\Delta K(N)$ and $\Delta X(N)$ corrections
    discussed in the text.
    \label{fig:heg_extrap}
  }
\end{figure}

First, advances in the understanding of finite-size effects enabled an
extremely reliable extrapolation of the fixed-node energy to the
thermodynamic limit.
While it is customary to correct the total energy $E(N)$ at system
size $N$ for quasirandom fluctuations by subtracting $\Delta K(N) =
K(N) - K(\infty)$ from it, where $K(N)$ is the HF kinetic energy at
system size $N$, the analogous correction using the HF exchange energy
$X(N)$ is avoided because it introduces a slowly varying dependence on
$N$ into the corrected energy which complicates the extrapolation
process.
However, Drummond \textit{et al.}\cite{Drummond_2008}\ determined
the prefactor of this slowly-varying term, and the correction
\begin{equation}
  \Delta X(N)
    = X(N) - X(\infty)
    + \frac{3 C_{\rm HF}}{8\pi} r_{\rm s}^{-1} N^{-2/3} \;,
\end{equation}
where $C_{\rm HF} = 2.837297479$ for simple-cubic simulation cells,
essentially suppresses quasirandom fluctuations in the fixed-node DMC
energy per electron while keeping its leading-order behavior at
${\cal O}(N^{-1})$.
The effect of these corrections is illustrated in Fig.\
\ref{fig:heg_extrap} for the HEG at density parameter $r_{\rm s}=0.5$.

The prefactor of the $N^{-1}$ term is also known \cite{Chiesa_2006}
provided backflow is not used; \cite{Holzmann_2016} hence the use of
Slater-Jastrow wave functions enables us to fit the
finite-size-corrected DMC energies per electron to a constant plus a
power expansion in $N^{-1/3}$ of leading order $N^{-4/3}$.
This procedure yields extremely reliable thermodynamic limits for the
fixed-node energy.

The second key element in the work of Ruggeri \textit{et al.}\ is the
use of FCIQMC to obtain essentially exact energies for the HEG at
system sizes $N=15$--$33$.
This allows the evaluation of the fixed-node error at these system
sizes, which can be fitted to a low-order power expansion containing
$N^0$, $N^{-1}$, and $N^{-4/3}$ terms, \cite{Holzmann_2016,
Ruggeri_heg_2018} under the assumption that the fixed-node error is a
smoother function of $N^{-1/3}$ than the total energy, and
extrapolated to $N\to\infty$.

\begin{figure}[!htbp]
  \begin{center}
    \includegraphics[width=0.45\textwidth,clip]{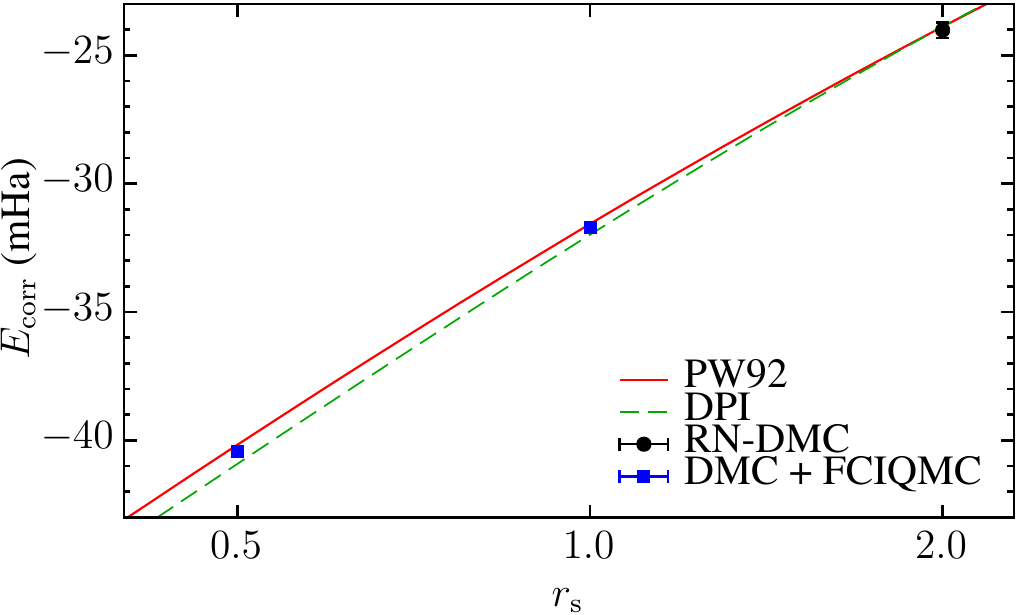}
    \caption{
      Correlation energy of the ferromagnetic electron gas as a
      function of density parameter $r_{\rm s}$ (note the logarithmic
      scale of the $r_{\rm s}$ axis).
      Results obtained by combining DMC and FCIQMC data,
      \cite{Ruggeri_heg_2018} as described in the text, are shown,
      along with the release-node DMC result of Ceperley and Alder
      \cite{Ceperley_1980} at $r_{\rm s}=2$, the parameterization of
      Perdew and Wang (PW92) \cite{Perdew_PW92_1992} based on the data
      of Ceperley and Alder, and the ``density parameter
      interpolation'' (DPI) parameterization of Bhattarai \textit{et
      al}.\@ \cite{Bhattarai_DPI_2018} which uses minimal input from
      QMC calculations.
      The correlation energies from DMC and FCIQMC resolve the
      discrepancy between the different parameterizations at high
      densities.
      \label{fig:heg_ecorr}}
  \end{center}
\end{figure}

The energies of the high-density ferromagnetic HEG obtained by adding
the extrapolated fixed-node energy and the extrapolated fixed-node
error, plotted in Fig.\ \ref{fig:heg_ecorr}, are estimated to be
accurate to $1$ meV/electron.
This is much smaller than the uncertainty in the release-node DMC
energy obtained by Ceperley and Alder \cite{Ceperley_1980} at the
highest density they considered ($r_{\rm s}=2$ for the ferromagnetic
HEG)\@.
These results offer the possibility of refining existing
parameterizations of the correlation energy of the HEG
\cite{Perdew_PW92_1992, Bhattarai_DPI_2018} used in the construction
of exchange-correlation functionals for DFT\@.
Furthermore, knowledge of the magnitude of the fixed-node error in the
energy of HEG as a function of density is potentially very useful; we
discuss this further in Sec.\ \ref{sec:dmc_assist_fciqmc}.

\subsubsection{Phase diagram of electron-hole bilayers}

Model systems with attractive interactions are also tractable with
QMC\@.
Traditionally, different wave functions have been used to study the
two-component fluid and excitonic phases of the electron-hole system,
using the energy associated with each wave function to determine which
phase is stable at each set of system parameters.
\cite{DePalo_bilayer_2002}
Using the wave function of Eqs.\ (\ref{eq:pairing_wfn}) and
(\ref{eq:pairing_orb}) enables a much more consistent description of the
system across phase boundaries, although this requires the evaluation
of expectation values other than the energy to determine the phase of
the system.
Maezono \textit{et al.}\cite{Maezono_biexcitons_2013}\ used this
approach to compute the phase diagram of the equal-mass, equal-density
electron-hole double layer as a function of density and interlayer
distance, using the condensate fraction to discriminate the
two-component fluid from the Bose-Einstein condensate (BEC) phase in
which electrons and holes form localized excitons.
Maezono \textit{et al.}\ found that the trial wave function was also
capable of describing a biexcitonic phase, with exciton-exciton
binding being captured by the DTN Jastrow factor.
The biexcitonic phase was found at small interlayer separations and
low carrier densities; the pair-correlation function was used to
distinguish this phase from the two-component fluid.

\begin{figure}[!htbp]
  \begin{center}
    \includegraphics[width=0.45\textwidth,clip]{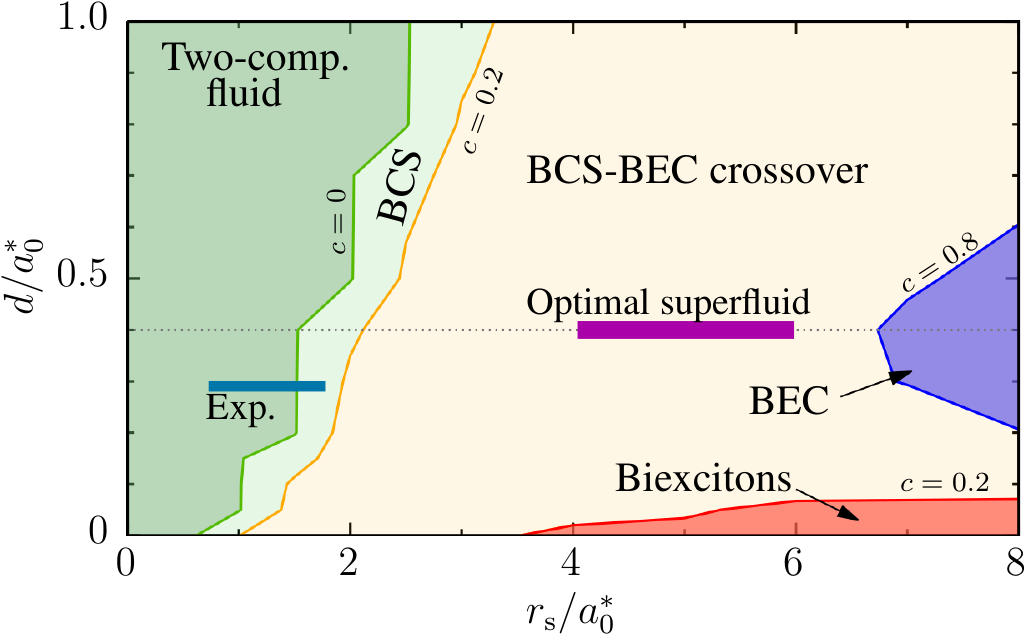}
    \caption{
      Phase diagram of the electron-hole double layer according to the
      value of the condensate fraction $c$.
      \cite{Maezono_biexcitons_2013}
      The length unit is the exciton Bohr radius $a_0^* =1/\mu$, where
      $\mu=m_{\rm h}/(1+m_{\rm h})$ is the reduced mass of the
      electron-hole pair and $m_{\rm h}$ is the hole mass.
      The dotted line marks the interlayer separation below which
      biexciton formation is possible, \cite{Lee_2009} and the region
      labeled ``Optimal superfluid'' corresponds to the maximum
      superfluid gap. \cite{LopezRios_bilayer_2018}
      The discrepancy with the density range in which superfluidity
      has been experimentally observed \cite{Burg_graphene_2018}
      (labeled ``Exp.'')\ is likely due to the omission of 2D
      polarization and multivalley effects in the QMC calculations.
      \label{fig:eh_bilayer}}
  \end{center}
\end{figure}

The identification of the system parameters at which the excitonic
phase of electron-hole double layers is theoretically stable is
important in the experimental search for superfluidity, but so is the
characterization of the system in this regime.
L\'opez R\'ios \textit{et al.}\cite{LopezRios_bilayer_2018}\ obtained
the superfluid parameters of the symmetric electron-hole double layer
as a function of density, fixing the interlayer distance to the
smallest value such that biexciton formation is precluded.
\cite{Lee_2009}
This was done by subtracting the energy of the neutral system plus the
chemical potential from the energy of the system with an additional
electron of associated wavevector ${\bf k}$, and fitting the results to
the Bardeen-Cooper-Schrieffer (BCS) dispersion relation,
\begin{equation}
  \varepsilon(k) = \sqrt{ \left( {k^2}/{2m^*} - \mu \right)^2
                          + \Delta^2} \;,
\end{equation}
where $m^*$, $\mu$, and $\Delta$ are the effective mass, chemical
potential, and superfluid gap of the electron quasiparticle.
Knowledge of these parameters revealed the density range in which the
superfluid can be expected to be most stable.
The phase diagram of the electron-hole double layer is shown in
Fig.\ \ref{fig:eh_bilayer}.

Superfluidity has been demonstrated experimentally in double bilayer
graphene encapsulated in a few layers of WSe$_2$.
\cite{Burg_graphene_2018}
This experimental system exhibits multivalley effects, and describing
the strong dielectric response of the effectively 2D encapsulating
material requires the use of the Rytova-Keldysh interaction instead of
the bare Coulomb interaction.
Given that neither of these aspects was considered in
Ref.\ \onlinecite{LopezRios_bilayer_2018}, the density range over which
the superfluid is predicted to be stable is in reasonable
(order-of-magnitude) agreement with experiment.

The behavior of the superfluid parameters of the symmetric
electron-hole double layer exhibit similarities to that of non-Coulomb
systems across the BCS, BCS-BEC crossover, and BEC regimes,
\cite{Perali_contact_2011, Hosono_iron_2015} suggesting the existence
of universal physical behavior which does not depend on the details of
the microscopic interaction.  \cite{LopezRios_bilayer_2018}
The peculiarity of the electron-hole double layer in this regard is
the near suppression of the regime of stability of the BCS phase.
\cite{LopezRios_bilayer_2018}

\subsubsection{Photoexcitation in doped semiconductors}

\begin{figure}[!htbp]
\begin{center}
\includegraphics[width=0.45\textwidth,clip]{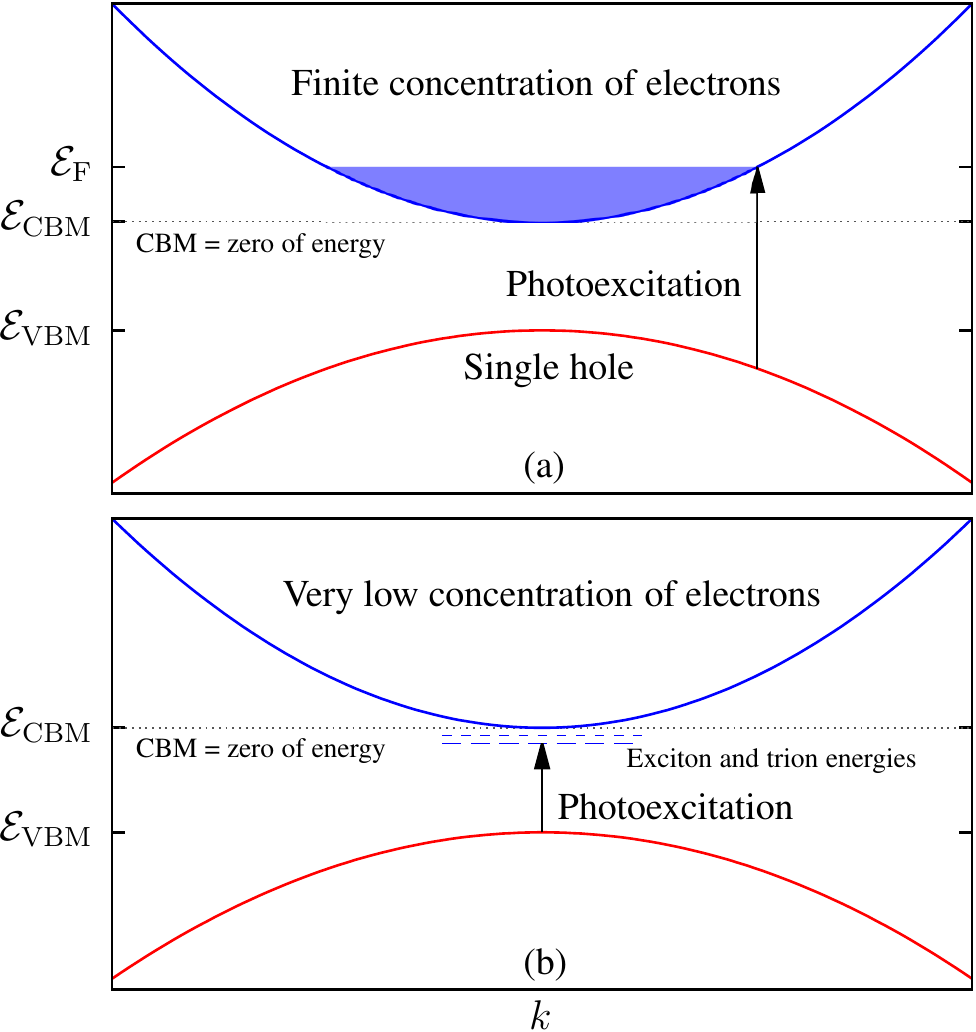}
\caption{Photoexcitation of an n-doped direct semiconductor at (a)
  high electron concentration and (b) low electron
  concentration. \label{fig:photoex_doped_sc}}
\end{center}
\end{figure}

Consider the problem of photoexciting a valence electron in an n-doped
semiconductor with a finite concentration of conduction electrons.
The situation is illustrated in Fig.\ \ref{fig:photoex_doped_sc}.
The minimum photon energy for photoexcitation is the quasiparticle gap
plus the Fermi energy of the electron gas relative to the CBM, plus
the energy of the isolated hole at the Fermi wavevector, plus the
electron-hole correlation energy.

The crossover between high-density Fermi-edge-singularity behavior
dominated by many-body correlations and low-density behavior
characterized by the presence of isolated excitons and trions has been
investigated experimentally in high-mobility 2D HEGs in GaAs/AlGaAs
and InGaAs/InAlAs
heterostructures. \cite{Huard_2000,Yusa_2000,BarJoseph_2005,Yamaguchi_2013}
Rapid changes in line shapes and transition energies in absorption and
photoluminescence spectra of a gated modulation-doped quantum well
allow experimentalists to locate a critical ``crossover'' density.

\begin{figure}[!htbp]
\begin{center}
\includegraphics[clip,width=0.45\textwidth]{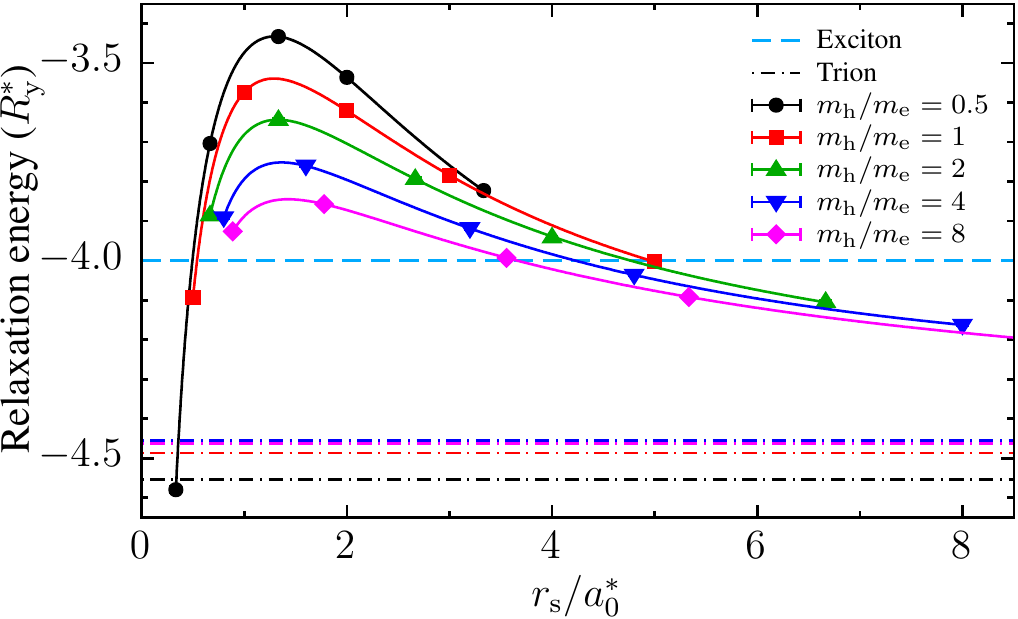}
\caption{Electron-hole correlation energy for a single hole immersed
  in a 2D HEG of density parameter $r_{\rm s}$.
The length unit is the exciton Bohr radius $a_0^* =1/\mu$, where
$\mu=m_{\rm h}/(1+m_{\rm h})$ is the reduced mass of the
electron-hole pair and $m_{\rm h}$ is the hole mass.
The energy unit is the exciton Rydberg $R_{\rm y}^\ast =
\mu/2$.
\label{fig:2D_relax}}
\end{center}
\end{figure}

\begin{figure}[!htbp]
\begin{center}
\includegraphics[clip,width=0.45\textwidth]{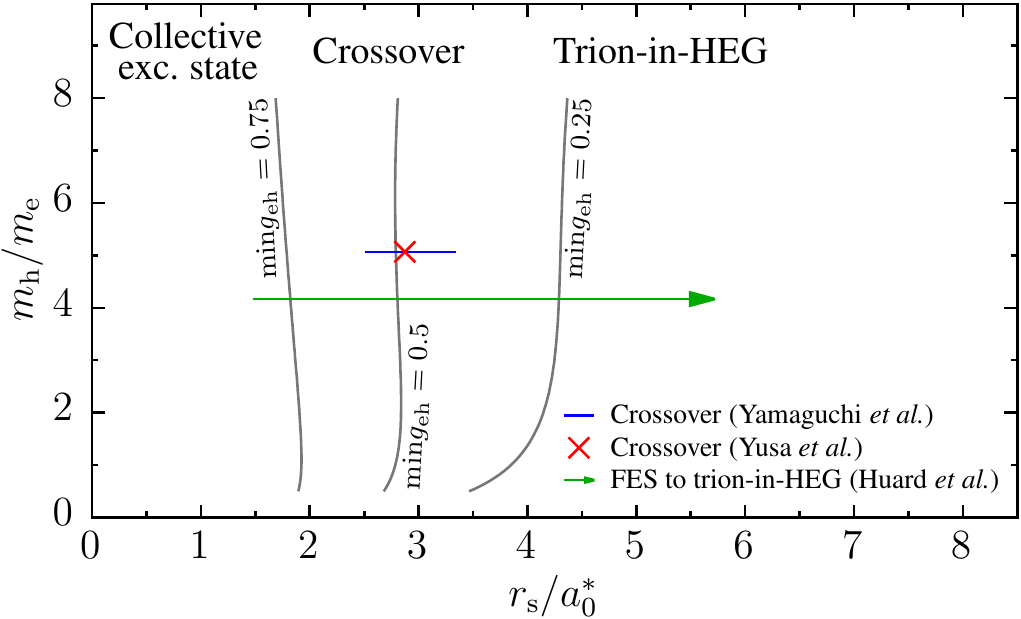}
\caption{Crossover between collective exciton behavior and isolated
  trion behavior.
  The gray curves identify contours of the minimum value of the
  electron-hole pair-correlation function $g_{\rm eh}$, used by
  Spink \textit{et al}.\cite{Spink_trion_2016}\ to identify
  the phases of the system.
  The arrow marks the density range over which an evolution from a
  collective state characterized by a Fermi-edge singularity (FES)
  to a trion-in-HEG phase was experimentally detected,
  \cite{Huard_2000} and the other symbols flag regions of phase space
  which have been experimentally characterized as being in the
  crossover regime.  \cite{Yusa_2000, BarJoseph_2005, Yamaguchi_2013}
  \label{fig:mahan_crossover}}
\end{center}
\end{figure}

An idealized model of the situation is a quantum impurity problem: a
single hole immersed in a zero-temperature 2D
HEG\@. \cite{Spink_trion_2016}
To solve this model Spink \textit{et al.}\ used VMC and DMC as
implemented in \textsc{casino} to simulate $N_{\rm e}=86$ electrons (a
closed-shell configuration) plus a single hole in a periodic hexagonal
cell.
The specialized pairing trial wave function described in
Sec.\ \ref{sec:pairing} was developed for this problem.
A cell area of $(N_{\rm e}-1)\pi r_{\rm s}^2$ was used, where $r_{\rm
  s}$ is the HEG density parameter, so that the electron density far
from the hole is correct.
The electron-hole correlation energy was calculated by subtracting the
energy of a pure HEG of the same number of electrons in the same cell.
The results are shown in Fig.\ \ref{fig:2D_relax}.
The low-density limit of the electron-hole correlation energy is the
energy of an isolated trion.
The gradual crossover between collective exciton and isolated trion
behavior occurs in a parameter range consistent with the absorption
and photoluminescence spectra seen experimentally, as shown in
Fig.\ \ref{fig:mahan_crossover}.

The Fermi energy of the 2D HEG is
\begin{equation}
  {\cal E}_{\rm F}
  = \left(\frac{{\rm d}E}{{\rm d}N}\right)_V
  = \frac{\rm d}{{\rm d}n}(\epsilon n)
  = \epsilon(r_{\rm s}) - \frac{r_{\rm s}}{2}
    \frac{{\rm d}\epsilon(r_{\rm s})}{{\rm d} r_{\rm s}},
\end{equation}
where $n$ is the number density and $\epsilon$ is the energy per
electron as a function of HEG density parameter $r_{\rm s}$.
The HF contribution to the energy per particle is readily calculated
by pen and paper.
DMC was used to calculate the correlation contribution to the energy
per particle $\epsilon_{\rm c}(r_{\rm s})$ of the 2D HEG as a function
of density. \cite{Drummond_2009}
Hence Spink \textit{et al.}\ were able to calculate the correction to
the quasiparticle gap due to the finite concentration of electrons in
the conduction band and the resulting electron-hole correlation.

\subsubsection{Quasiparticle effective mass of the 2D HEG
\label{sec:qem_2D_HEG}}

QMC methods have been used to study intraband excitations in the 2D
HEG\@.
Landau's phenomenological Fermi liquid theory \cite{Landau_1957a,
  Landau_1957b, Landau_1959} predicts the lifetime of quasiparticle
excitations to diverge at the Fermi surface due to Pauli blocking of
plane-wave states, so that quasiparticles of a particular momentum
${\bf k}$ near the Fermi surface correspond to a well-defined
excited-state energy.
Hence the quasiparticle energy band can be calculated as described in
Sec.\ \ref{sec:qp_ex_gaps}, and is valid in the vicinity of the Fermi
surface.
This corresponds to the assumption that there is an adiabatic
connection between the energy eigenstates of the noninteracting and
interacting electron systems, i.e., the nodal topology of the
interacting and noninteracting wave functions is the same, or at least
sufficiently similar for electronic states in the relevant range of
energies.
The energy band ${\cal E}({\bf k})$ therefore provides the dispersion
relationship for long-lived quasielectrons in the all-important energy
range near the Fermi surface.

Further from the Fermi surface the difference of total-energy levels
ceases to correspond to a quasiparticle excitation due to finite
lifetime effects.
Nevertheless we can evaluate the energy band ${\cal E}({\bf k})$ over
a broad range of ${\bf k}$.
We may make a linear approximation to the quasiparticle energy band
near the Fermi surface, ${\cal E}({\bf k}) = {\cal E}_{\rm F}+(k_{\rm
  F}/m^\ast)(k-k_{\rm F})$, where ${\cal E}_{\rm F}$ is the Fermi
energy, $k_{\rm F}$ is the Fermi wavevector, and $m^\ast$ is the
quasiparticle effective mass.
This effective mass describes the renormalization of the electron mass
by electron-electron interactions (on top of the band effective mass
approximation).

The quasiparticle effective mass of a paramagnetic 2D HEG has been the
subject of controversy over the decades.
Some experiments \cite{Smith_1972,Pudalov_2002} found a large
enhancement of $m^\ast$ at low density; other experiments
\cite{Tan_2005b,Padmanabhan_2008} contradicted this.
$GW$ calculations give a range of possible results depending on the
choice of effective interaction. \cite{Giuliani_2005}
Previous QMC studies have predicted (i) much less \cite{Kwon_1994} and
(ii) much more \cite{Holzmann_2009} enhancement of $m^\ast$ than found
in recent experiments.
Experiment \cite{Padmanabhan_2008} and theory \cite{Zhang_2005}
suggest that the effective masses in paramagnetic and ferromagnetic
HEGs behave quite differently as a function of density.

\begin{figure}[!htbp]
\begin{center}
\includegraphics[clip,width=0.45\textwidth]{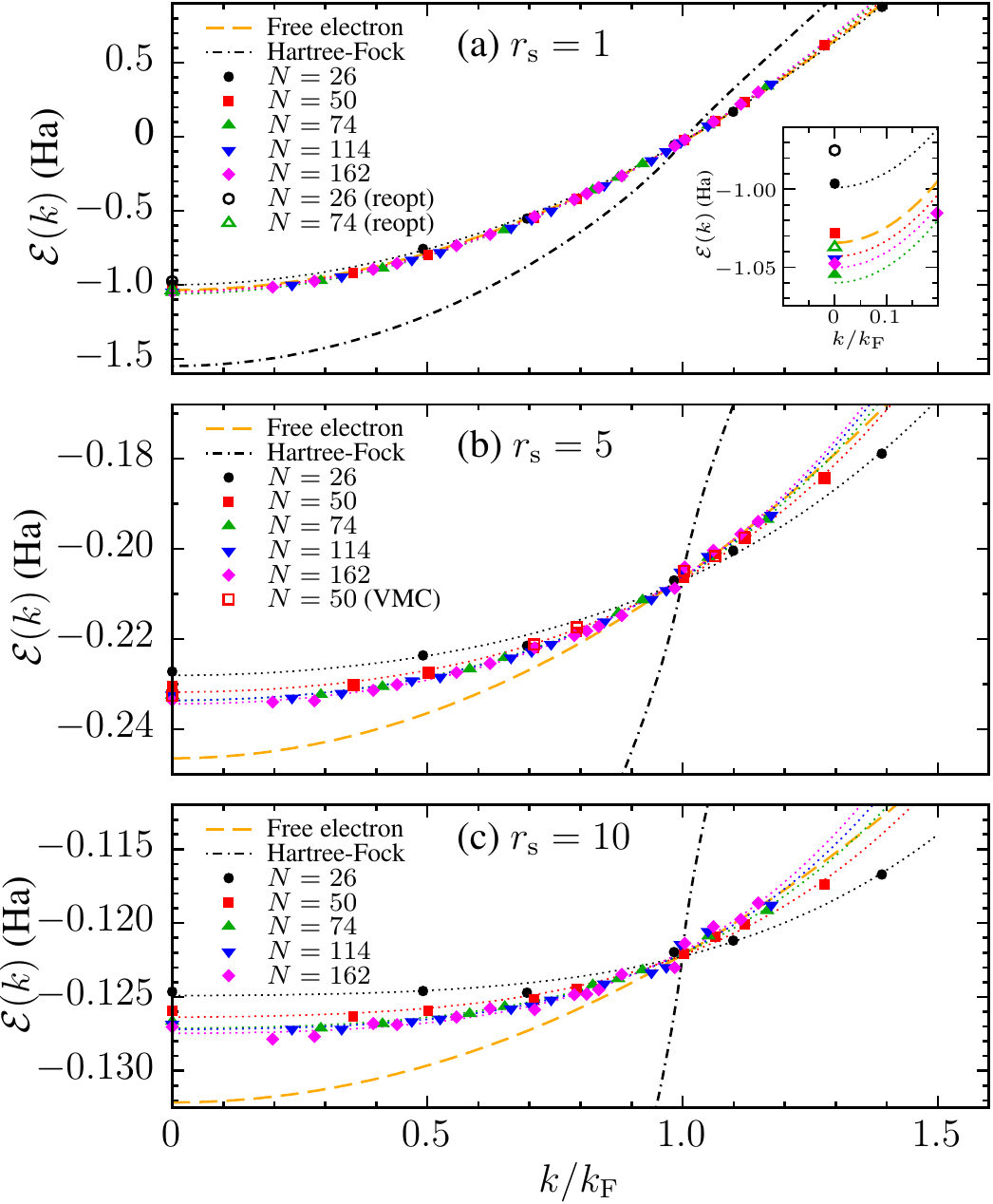}
\caption{QMC-calculated energy bands of paramagnetic 2D HEGs at three
  different densities, for different system sizes $N$. Also shown are
  the free-electron ($k^2/2$) and HF bands, offset to match the DMC
  band at the Fermi wavevector $k_{\rm
    F}$.  \label{fig:heg_eband_para}}
\end{center}
\end{figure}

\begin{figure}[!htbp]
\begin{center}
\includegraphics[clip,width=0.45\textwidth]{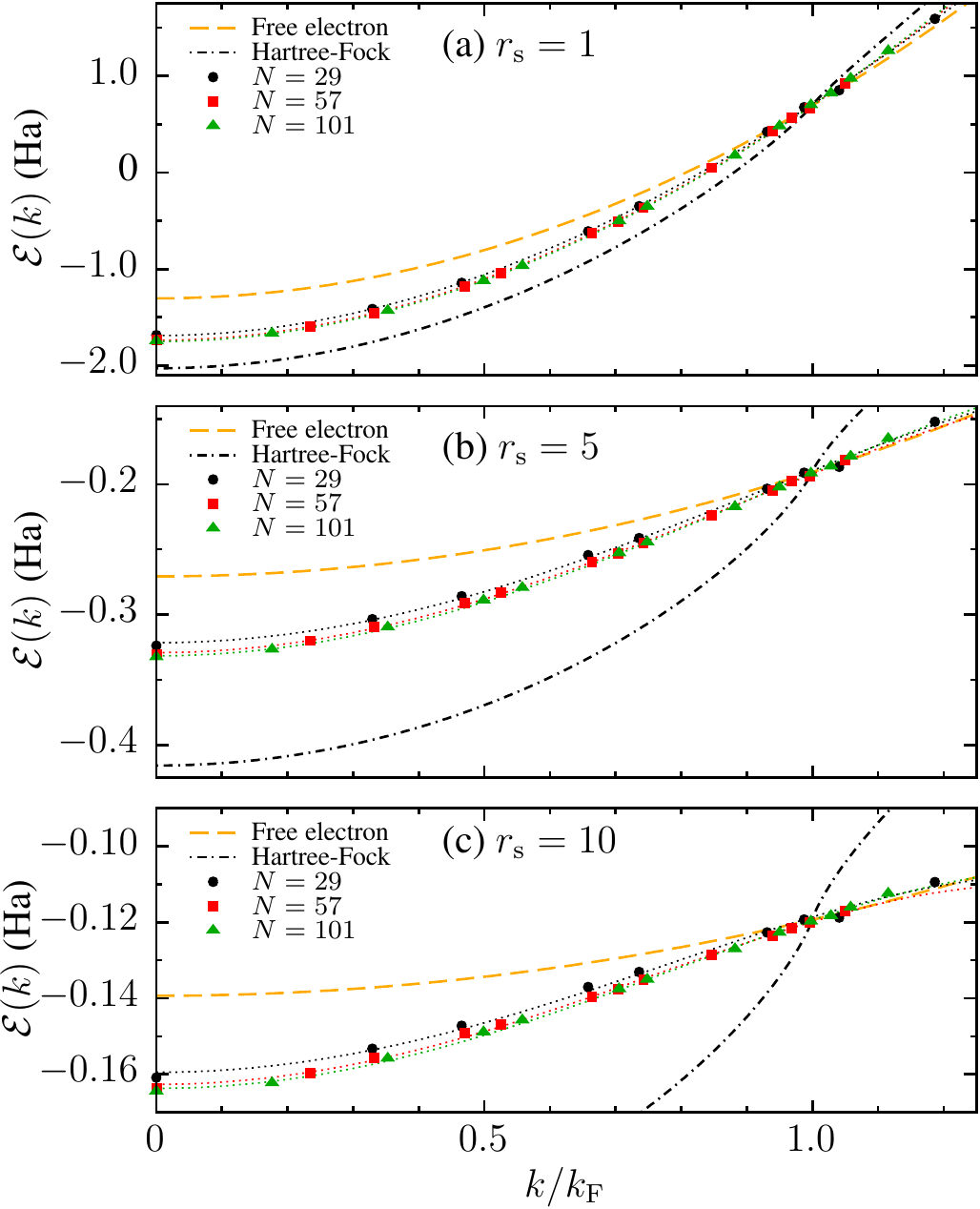}
\caption{As Fig.\ \ref{fig:heg_eband_para}, but for full
  spin-polarized HEGs. \label{fig:heg_eband_ferro}}
\end{center}
\end{figure}

To calculate the quasiparticle effective mass, the DMC energy band
${\cal E}(k)$ was determined at a range of $k$ by taking the energy
difference when an electron is added to or removed from a closed-shell
ground-state. \cite{Drummond_2009b, Drummond_2013}
The resulting energy bands for non-spin-polarized and fully
spin-polarized 2D HEGs are plotted in Figs.\ \ref{fig:heg_eband_para}
and \ref{fig:heg_eband_ferro}, respectively.
A quartic ${\cal E}(k)=\alpha_0+\alpha_2 k^2+\alpha_4 k^4$ was fitted
to the energy band values, then the effective mass was calculated as
$m^\ast=k_{\rm F}/({\rm d}{\cal E}/{\rm d}k)_{k_{\rm F}}$.
The calculations were performed in finite cells subject to periodic
boundary conditions.
Finite-size effects are the major sources of error and bias in the QMC
results.
In our finite simulation cell subject to twisted periodic boundary
conditions, the available momentum states fall on the (offset) grid of
reciprocal lattice points.
This restricts the ${\bf k}$ values that we can consider.
There are also finite-size errors in the excitation energies due to
the neglect of long-range interactions and correlations.
These errors have been shown to fall off slowly, as $N^{-1/4}$, near
the Fermi surface. \cite{Holzmann_2009}

In the infinite-system limit, the exact energy band is
smooth and well-behaved at the Fermi surface.
However, the HF band is pathological.
In the infinite-system limit its derivative has a logarithmic
divergence at the Fermi surface.
In finite systems it behaves very badly.
DMC does not entirely remove the pathological behavior from HF
theory.
Hence we need to consider excitations away from the Fermi surface to
obtain the gradient of the energy band at $k_{\rm F}$.

\begin{figure}[!htbp]
\begin{center}
\includegraphics[clip,width=0.45\textwidth]{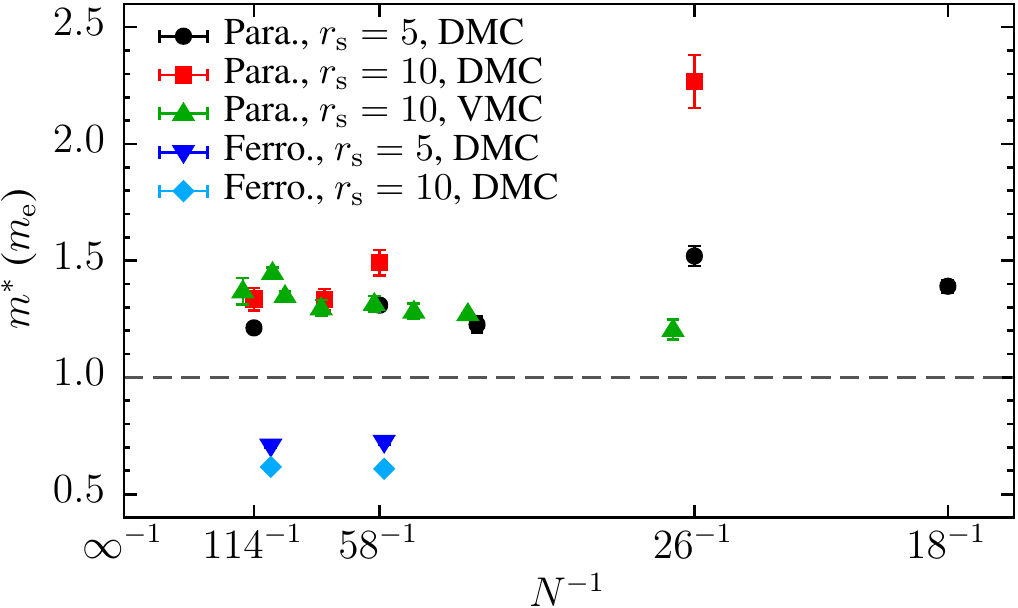}
\caption{Quasiparticle effective mass calculated using VMC
  \cite{Holzmann_2009} and DMC \cite{Drummond_2009b} against system
  size $N$ for paramagnetic and fully ferromagnetic 2D HEGs.
  \label{fig:qem_v_N}}
\end{center}
\end{figure}

The occupied bandwidth is $\Delta {\cal E}={\cal E}({k_{\rm F}})-{\cal
  E}(0)=E_-(0)-E_-(k_{\rm F})$.
The DMC bandwidth is an upper bound: assuming DMC retrieves the same
fraction of the correlation energy in the ground and excited states,
the occupied bandwidth will lie between the HF value $E_-^{\rm
  HF}(0)-E_-^{\rm HF}(k_{\rm F})$, which is too large, and the exact
result $E_-^{\rm exact}(0)-E_-^{\rm exact}(k_{\rm F})$.
Extrapolating the VMC energy with different trial wave functions to
zero variance suggests that our DMC calculations retrieve more than
99\% of the correlation energy, and that the fraction retrieved is
similar in both the ground and excited states.
The free-electron bandwidth is greater than or approximately equal to
the exact bandwidth.
Hence the error in the HF bandwidth is less than or approximately
equal to $\Delta {\cal E}^{\rm HF}-\Delta {\cal E}^{\rm free}=k_{\rm
  F}(1-2/\pi)$.
So the error in the DMC bandwidth is less than $0.01k_{\rm
  F}(1-2/\pi)\approx 0.007/r_{\rm s}$ for a ferromagnetic HEG and less
than about $0.01k_{\rm F}(1-2/\pi)\approx 0.005/r_{\rm s}$ for a
paramagnetic HEG\@.
Since the bandwidth falls off as $r_{\rm s}^{-2}$, the error is more
significant at large $r_{\rm s}$.
In the worst case (the paramagnetic HEG at $r_{\rm s}=10$) this
argument suggests that DMC overestimates the bandwidth by $\sim 9$\%.
In the next-worse case (paramagnetic, $r_{\rm s}=5$), the bandwidth is
overestimated by $\sim 4$\%.
It is reasonable to assume that DMC underestimates $m^\ast$ by a
similar amount.
The effective mass against system size is plotted in
Fig.\ \ref{fig:qem_v_N}.
The scaling is not the $N^{-1/4}$ predicted by Holzmann \textit{et
  al.}\ near the Fermi surface, \cite{Holzmann_2009} presumably
because we have fitted to the entire band, which shows less severe
finite-size errors.
In fact we observe an ${\cal O}(N^{-1})$ scaling of the finite-size
error.

\begin{figure}[!htbp]
\begin{center}
\includegraphics[width=0.45\textwidth,clip]{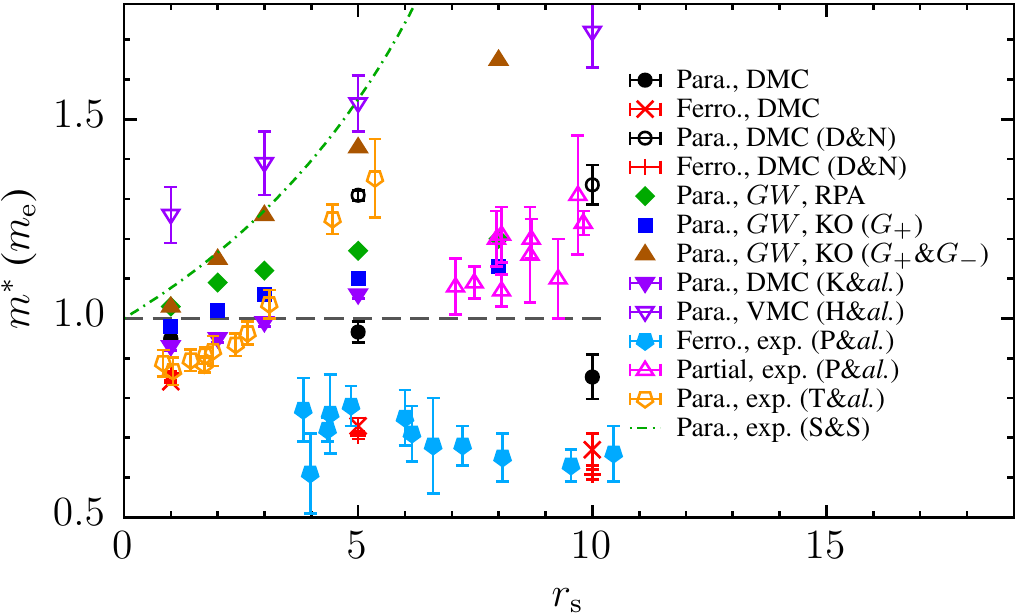}
\caption{Quasiparticle effective mass against density parameter
  $r_{\rm s}$ for paramagnetic and fully ferromagnetic 2D HEGs, from
  various measurements and calculations. \cite{Drummond_2009b,
    Drummond_2013, Giuliani_2005, Kwon_1994, Holzmann_2009,
    Padmanabhan_2008, Tan_2005b, Smith_1972} \label{fig:qem_v_rs}}
\end{center}
\end{figure}

The resulting effective masses of spin-unpolarized and fully
spin-polarized 2D HEGs are plotted in Fig.\ \ref{fig:qem_v_rs}.
In a paramagnetic HEG the effective mass remains close to the bare
electron mass.
In a ferromagnetic HEG $m^\ast$ decreases when the density is lowered.
Our results therefore support the qualitative conclusions of
Ref.\ \onlinecite{Padmanabhan_2008}.
Detailed comparison between theory and experiment is complicated by
finite-well-width effects and the effects of disorder.
Nevertheless, our results suggest that $m^\ast$ in paramagnetic 2D
HEGs does not grow rapidly as the density is reduced.

\subsubsection{Positrons immersed in electron gases}

DMC methods have been used to solve an important quantum impurity
problem, namely that of a single positron immersed in a 3D HEG\@.
The electron-positron correlation energy, known as the relaxation
energy, of a positron in a HEG is required in two-component DFT
calculations for positrons in real
materials. \cite{Boronski_1986,Puska_1994}
In addition, the electron-positron contact pair correlation function
in the HEG allows DFT calculations of positron annihilation rates in
real materials.
Such DFT calculations play a crucial role in the interpretation of
positron-annihilation spectroscopy (PAS) measurements.
In PAS experiments, positrons are injected into materials, where they
rapidly thermalize and settle at sites far from nuclei for a
relatively long time, before annihilating with the production of two
$\gamma$ photons.
The lifetime of the positrons and/or the momentum density of the
annihilation radiation can be measured to obtain information about the
electronic charge density and momentum density, including the Fermi
surface, at the locations at which the positrons
settle. \cite{Krause_1999,Major_2004}
However, positively charged positrons strongly perturb the electronic
structure, and so experimentalists require first-principles
calculations to relate the measurements to underlying electronic
properties.

The ground-state energy of an uncorrelated positron in a homogeneous
system is zero; hence the relaxation energy is equal to the
ground-state energy difference between a positron-in-HEG system and
the HEG without the positron.
DMC calculations of the relaxation energy, pair-correlation function
and annihilating-pair momentum density have been performed in
supercells of 54 electrons. \cite{Drummond_2011}
The resulting data differ appreciably from the parameterized results
of diagrammatic perturbation theory, \cite{Arponen_1979,Boronski_1986}
and when included in two-component DFT calculations lead to
theoretical predictions in improved agreement with
experiment. \cite{Kuriplach_2014,Weber_2015}

\subsection{Van der Waals interactions}

\subsubsection{Binding energies of 2D materials}

2D materials, which are of great interest due to their extreme
mechanical, electronic, and optical properties, consist of atomically
thin layers, often based on a honeycomb motif, with strong covalent
bonds within each layer and weak van der Waals interactions between
layers.
Of particular interest are stacked multilayers and heterostructures of
2D materials, which exhibit interesting physical properties including
the formation of long-range moir\'{e} patterns, natural type-II band
alignment leading to electron-hole bilayer behavior, twist-dependent
superconductivity, etc.
To model the interaction of 2D layers we require an accurate treatment
of van der Waals interactions.
Unfortunately DFT with local exchange-correlation functionals
provides a poor description of van der Waals interactions, which are
nonlocal correlations.
The development of van der Waals correction schemes for DFT depends in
part on the availability of accurate benchmark data for van der Waals
bonded systems.
QMC methods are capable of providing such benchmark data.

QMC studies of 2D materials require the use of supercells subject to
twist-averaged 2D-periodic boundary conditions.
When studying heteromultilayers or twisted homomultilayers, the layers
must be strained to force them to have a common unit cell.
The interlayer binding energy can then be evaluated as the difference
of the bilayer energy and the monolayer energy in the limit of large
supercell size.
Asymptotically, the finite-size errors in the total energies of the
monolayer and bilayer go as $N^{-5/4}$ with the number $N$ of
electrons in the supercell; \cite{Drummond_2008} hence the
twist-averaged monolayer and bilayer total energies can be
extrapolated to infinite system size, allowing the binding energy of
the bilayer to be evaluated in the thermodynamic limit.
Typically one must perform a series of such binding-energy
calculations for the bilayer in which the interlayer distance is
varied, then fit a function of interlayer separation to the results.
From this fitted function one can find the equilibrium separation and
the corresponding equilibrium binding energy, as well as the curvature
about the minimum and hence the breathing-mode phonon frequency.
If one is studying the binding of heterobilayers or twisted
homobilayers, the binding energy per unit cell obtained with
artificially commensurate lattice vectors and a particular local
lattice offset can be regarded as a local contribution to the
interlayer binding energy in a moir\'{e} supercell.
\cite{Szyniszewski_2020}
The relaxation of the structure in the moir\'{e} supercell can then be
described using continuum elasticity theory.

Previous QMC studies have examined the binding energy of layers of
hexagonal boron nitride, \cite{Hsing_2014} layers of graphene in bulk
graphite, \cite{Spanu_2009} and bilayer graphene. \cite{Mostaani_2015}
The latter paper shows that DFT calculations with different
functionals and van der Waals correction schemes give a wide range of
interlayer binding energies and equilibrium separations.
The QMC-calculated breathing-mode frequency is in good agreement with
Raman spectroscopic measurements.
The QMC calculations confirm the experimental observation that Bernal
AB stacking is energetically more favorable than AA stacking, with the
DMC binding energies being $17.7(9)$ and $11.5(9)$ meV per atom,
respectively.

\subsubsection{Van der Waals interactions at surfaces and between molecules}

Zen \textit{et al.}\ have studied a range of molecular crystals,
including various water ices, dry ice (carbon dioxide), ammonia,
benzene, naphthalene, and anthracene, showing that DMC is capable of
achieving chemical accuracy for the lattice-formation energy in each
case, unlike RPA and M{\o}ller-Plesset second-order perturbation
theory (MP2) calculations, which tend to underbind and overbind,
respectively. \cite{Zen_2018}
Other methods capable of achieving high accuracy for the lattice
formation energy, such as RPA with $GW$ single excitations or coupled
cluster with single, double, and perturbative triple excitations
[CCSD(T)] are more expensive than DMC and/or require difficult
fragment-decomposition approaches.

Tsatsoulis \textit{et al.}\ have examined the adsorption energy of a
single water molecule on a $(001)$ LiH surface using quantum chemistry
methods, DFT-based methods, as well as DMC\@. \cite{Tsatsoulis_2017}
They found good agreement between the high-accuracy quantum-chemistry
and DMC methods, which are based on very different methodologies,
indicating that they provide a reliable benchmark.
The MP2 method is found to perform well in this case, whereas van der
Waals-corrected DFT methods either underbind or overbind
significantly.
QMC methods had previously been used to study water molecules on
hexagonal boron nitride \cite{AlHamdani_2015} and graphene,
\cite{Ma_2011} showing that water is weakly physisorbed on either
material, but that the interaction energy is about 15 meV larger in
the former case.
Although van der Waals DFT functionals overbind water to both
materials, this 15 meV difference is consistently reproduced by
different first-principles methods.

DMC methods have also been used to predict the phase diagram of water
ice trapped between two graphene sheets, finding that the hexagonal
and pentagonal phases that are most stable at ambient pressure
transition to a square structure at higher pressure. \cite{Chen_2016}
These results are of relevance to transmission electron microscopy
studies of water confined between graphene sheets.
Again, the DMC results reveal inconsistencies and the need for
improvement of van der Waals-inclusive DFT functionals.

Taken together these results show the useful and important role that
DMC plays in studies of dispersion interactions and in assessing the
performance of other methods for treating surfaces.

\subsection{Solid hydrogen}

Hydrogen is the simplest and most abundant of the elements, yet it
exhibits strikingly rich phase behavior.
At low temperature, hydrogen has been observed to form quantum
crystalline states and orientationally ordered molecular phases.
It has been predicted to exhibit additional exotic behavior,
including a liquid-metal phase at high pressure and low temperature,
a metallic superfluid state at high pressure, and high-temperature
superconductivity.

Experimental X-ray diffraction studies of solid hydrogen have been
carried out over a range of pressures up to 190 GPa, providing limited
structural information. \cite{Akahama_2017}
It has not been possible to determine directly the atomic structures
of high-pressure phases of solid hydrogen, because samples are small
and hydrogen scatters X-rays weakly, making diffraction experiments
challenging.
Our knowledge of the atomic structure of high-pressure crystalline
hydrogen arises in large part from computational modeling in
conjunction with experimental Raman and infrared spectroscopy.

Experiments at around room temperature have demonstrated that at least
four distinct structural states exist at low temperature and at
pressures up to 350 GPa, which are known as phases I, II, III, and
IV\@.
The low-pressure phase I has a hexagonal close-packed structure with
freely rotating molecules at the lattice
sites. \cite{kawamura_X_Ray_2002}
Phase II is a broken-symmetry phase in which molecular rotations are
hindered at low temperature.
As the pressure is increased at low temperature a further phase
transition occurs from phase II to phase III at about 160
GPa. \cite{Goncharov_H_D_200GPa_2011}
A phase IV has been discovered at temperatures above a few hundred K
and pressures above 220 GPa. \cite{Eremets_conductive_H_2011}
A phase IV$'$ has also been reported, which is very similar to phase
IV\@.

\begin{figure}[!htbp]
  \begin{minipage}{0.45\textwidth}
    \centering
    $\vcenter{\hbox{\includegraphics[clip,scale=0.07]{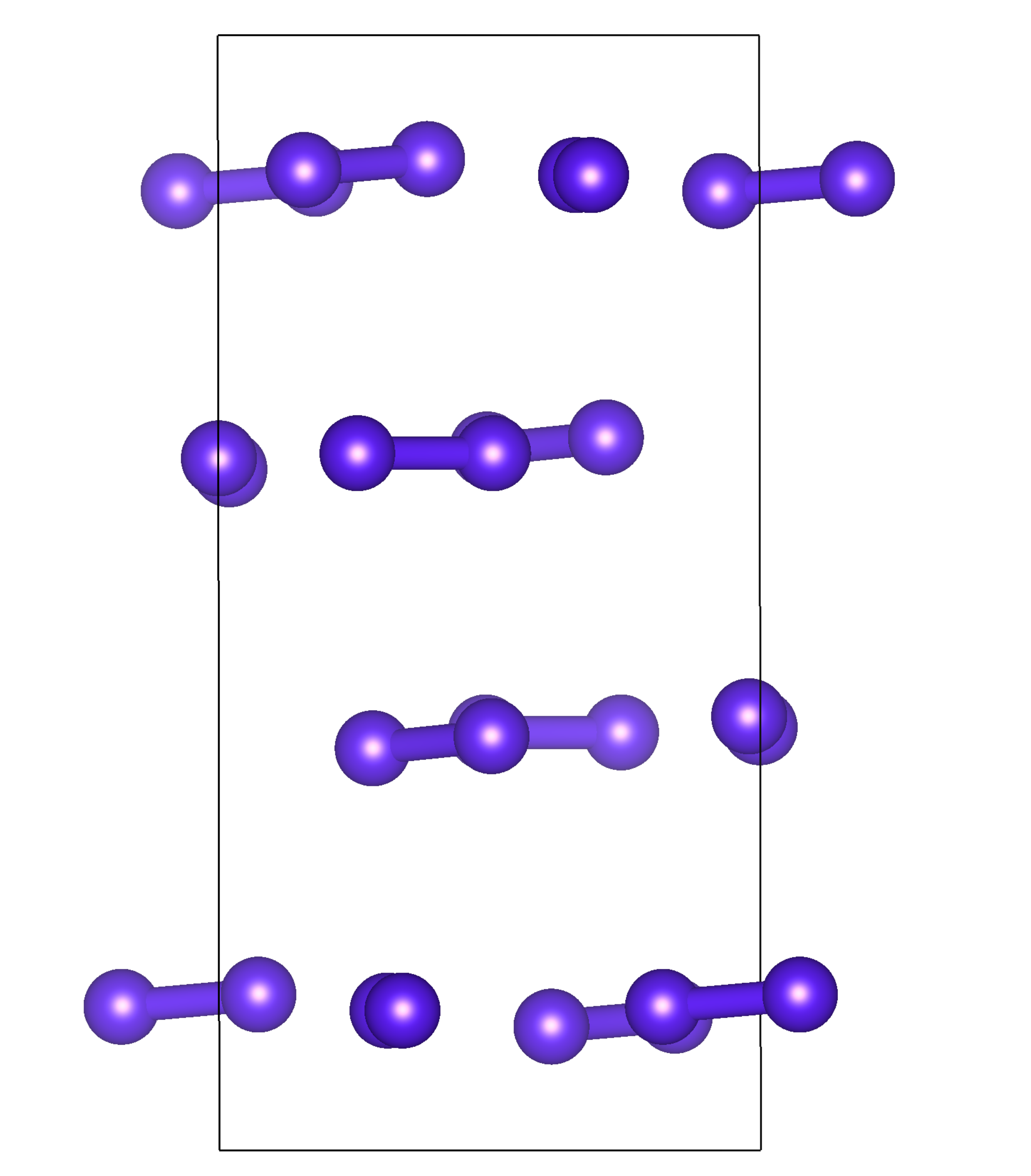}}}$
    ~
    $\vcenter{\hbox{\includegraphics[clip,scale=0.07]{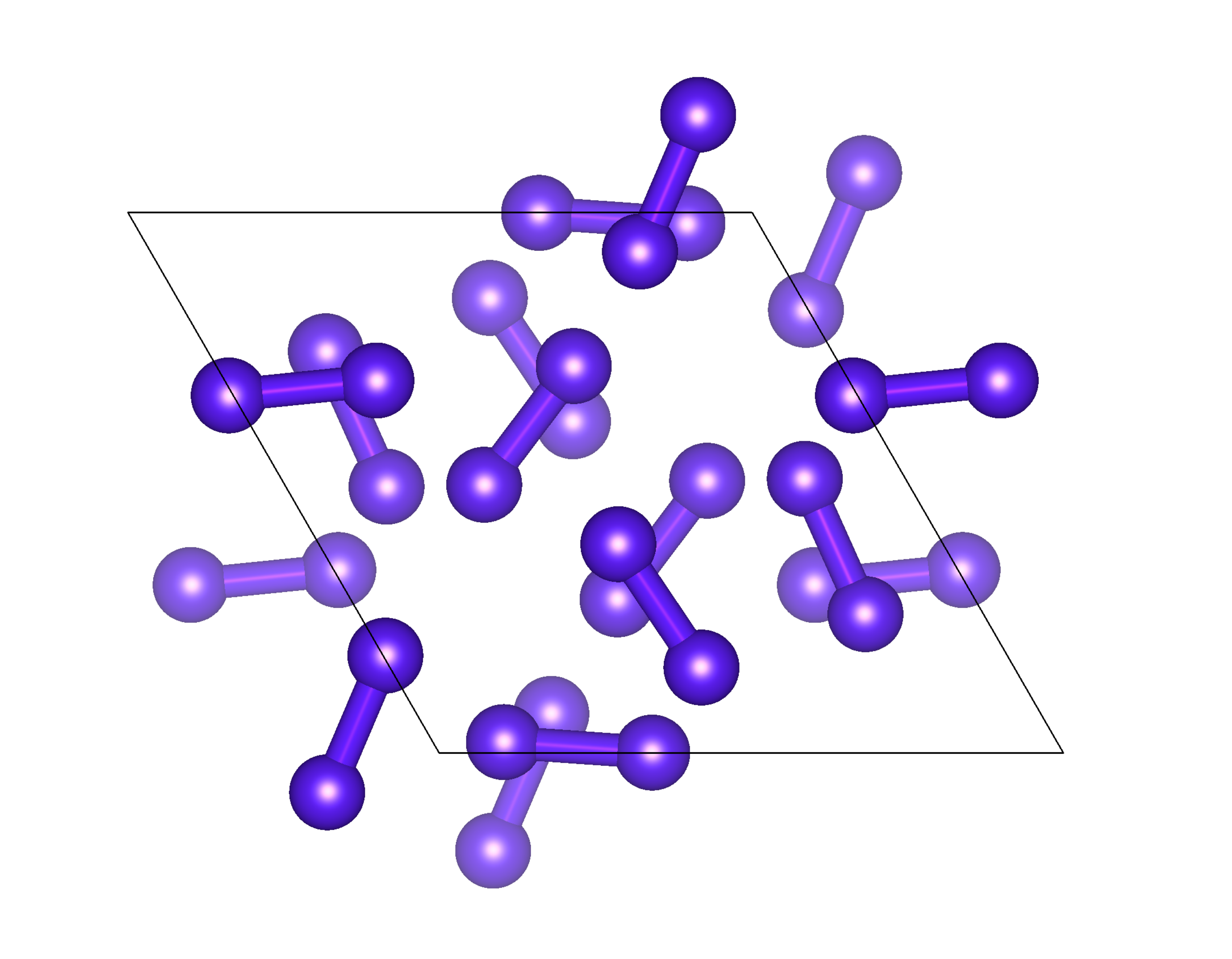}}}$
  \end{minipage}
  \caption{ Side view (left) and top view (right) of the
    \textit{C}2/\textit{c}-24 structure.
    This is essentially a layered structure with small deviations.
    \label{fig:c2c-24}
  }
\end{figure}

\begin{figure}[!htbp]
  \begin{minipage}{0.45\textwidth}
    \centering
    $\vcenter{\hbox{\includegraphics[clip,angle=90,scale=0.07]{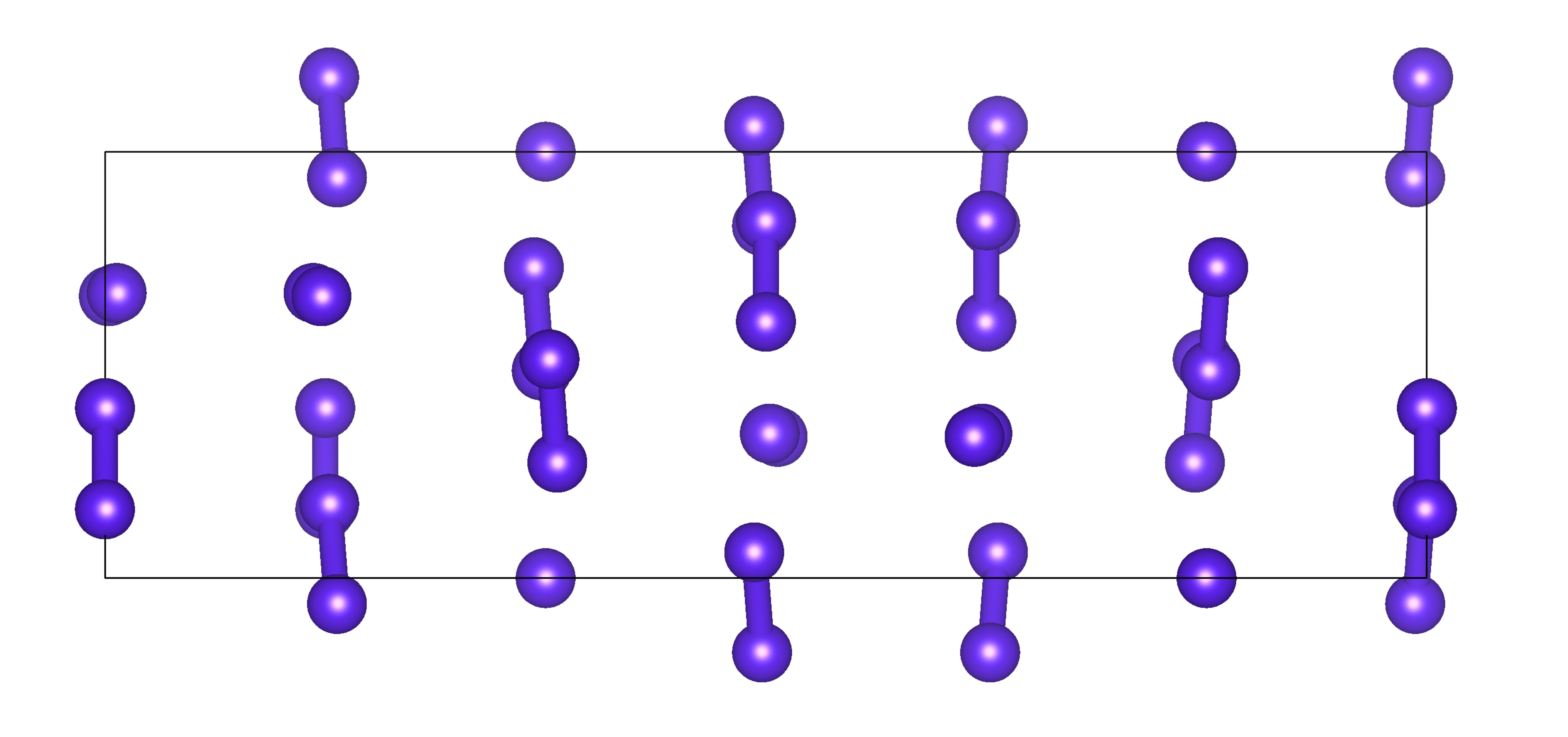}}}$
    ~
    $\vcenter{\hbox{\includegraphics[clip,scale=0.07]{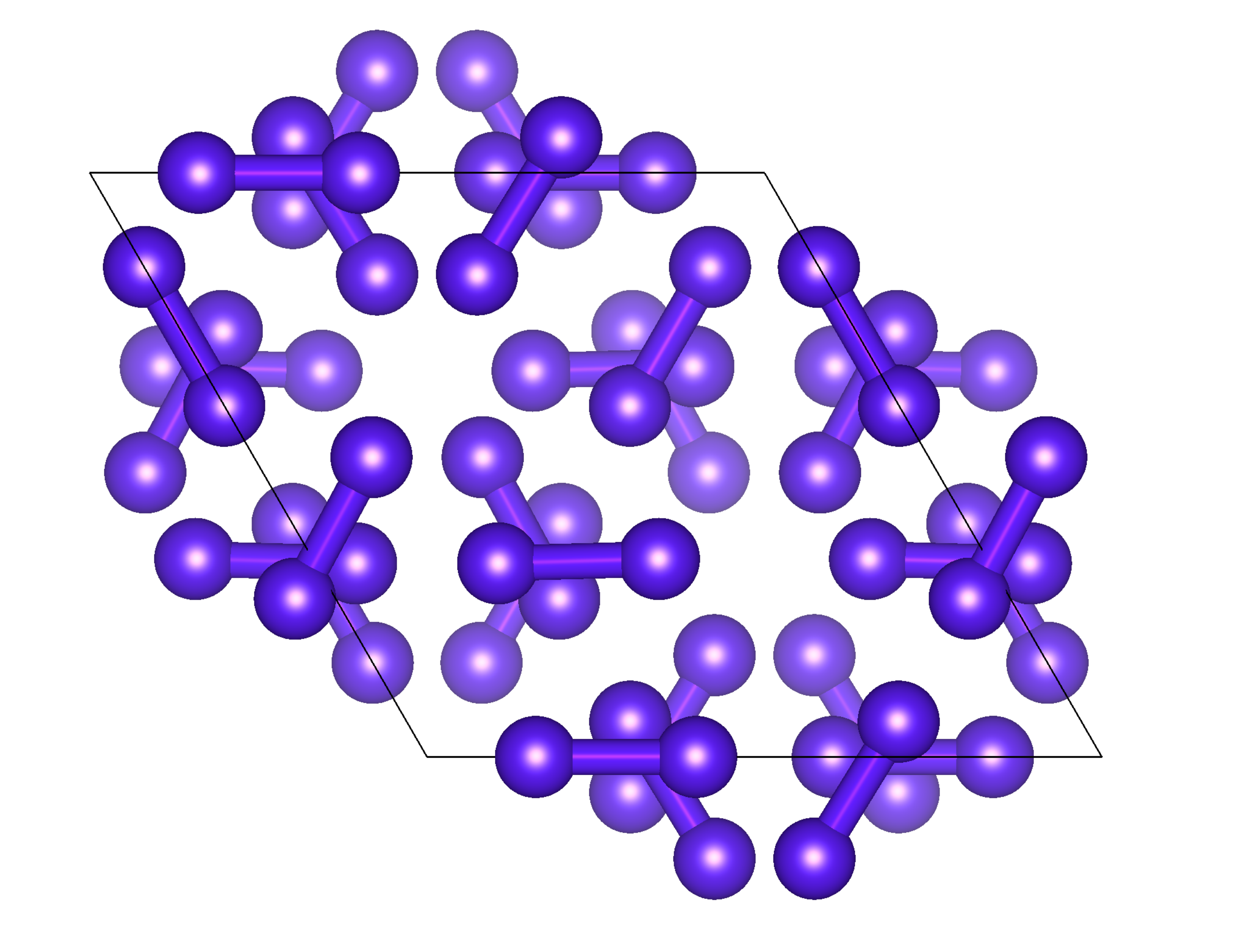}}}$
  \end{minipage}
  \caption{ Side view (left) and top view (right) of the
    \textit{P}6$_1$22 structure.
    \label{fig:p6122}
  }
\end{figure}

\begin{figure}[!htbp]
  \begin{minipage}{0.45\textwidth}
    \centering
    $\vcenter{\hbox{\includegraphics[clip,scale=0.07]{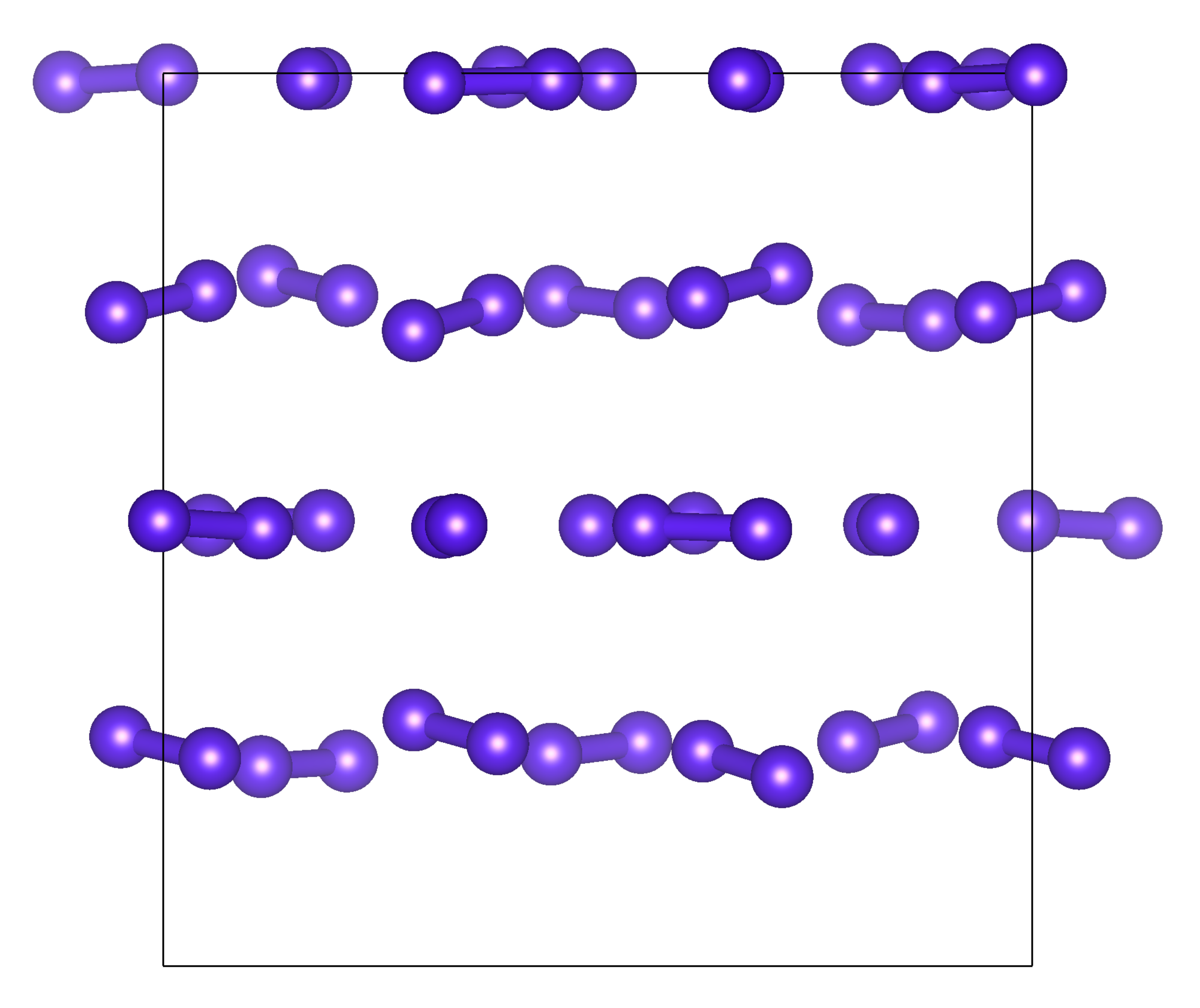}}}$
    ~
    $\vcenter{\hbox{\includegraphics[clip,scale=0.07]{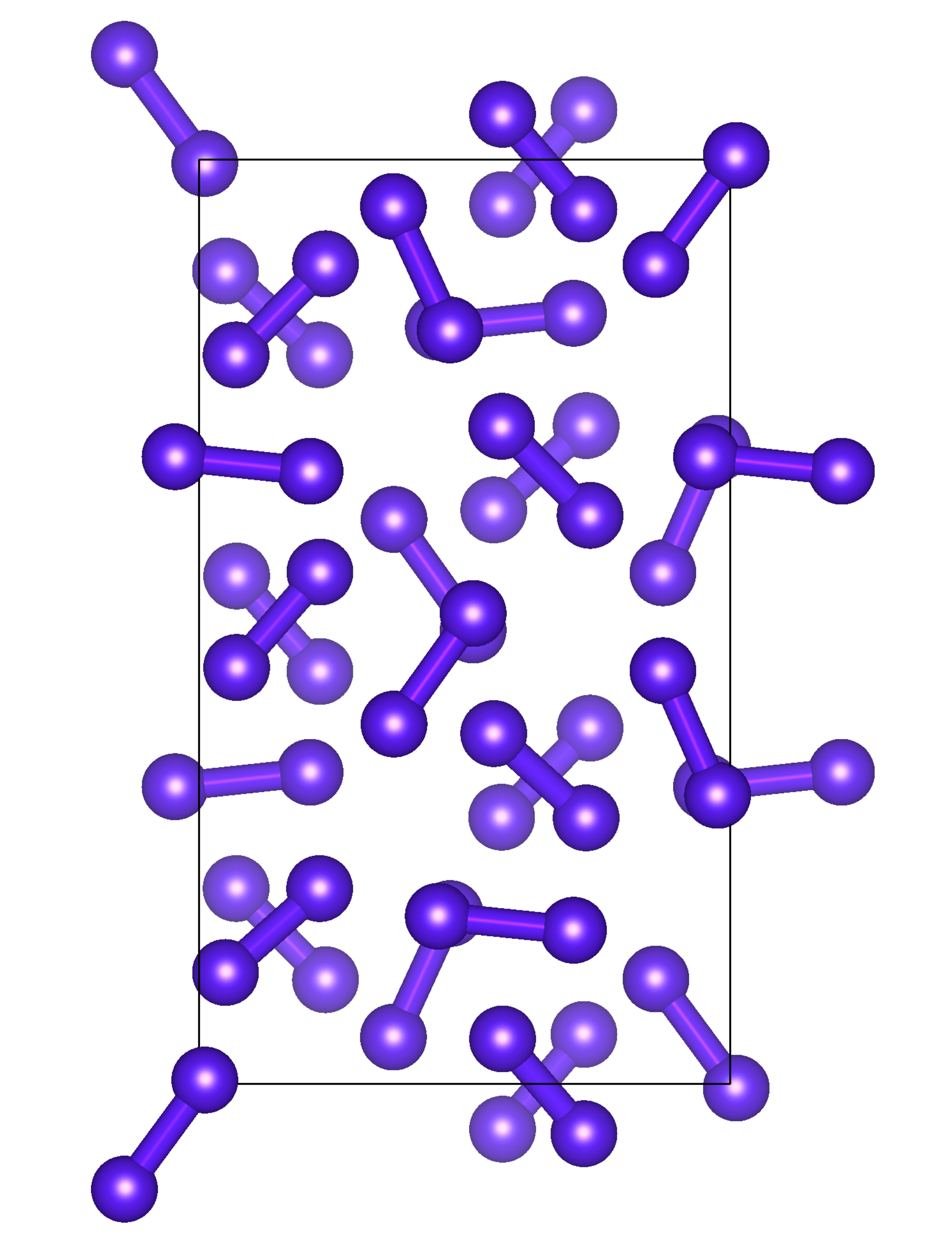}}}$
  \end{minipage}
  \caption{ Side view (left) and top view (right) of the
    \textit{Pc}-48 structure.
    \label{fig:pc-48}
  }
\end{figure}

DFT calculations are relatively inexpensive, and they have been used
to search for low-energy hydrogen structures with vibrational
properties that are in reasonable agreement with experimental Raman
and infrared measurements for phases II, III,
\cite{Pickard_H_phase_III_2007} and IV
\cite{Pickard_H_phase_IV_2012,Pickard_erratum_2012} of hydrogen over a
range of pressures.
The structures were discovered using the ``\textit{ab initio} random
structure searching'' (AIRSS) method. \cite{AIRSS_2011,
  Perspective_APL_Materials_2016}
Phase II may be modeled by a structure of
\textit{P}2$_{\text{1}}/$\textit{c} symmetry.
The first realistic structures for phase III of hydrogen were proposed
by Pickard and Needs. \cite{Pickard_H_phase_III_2007}
They found a structure of \textit{C}2/\textit{c} symmetry with 24
atoms in the primitive unit cell (see Fig.\ \ref{fig:c2c-24}) to be
the most stable.
A very similar structure was subsequently discovered, which exhibits a
hexagonal \textit{P}6$_1$22 symmetry, \cite{Hex_phase_III_2016}
and is depicted in Fig.\ \ref{fig:p6122}.
Two different phase-III-like structures may be formed at high
pressure: hexagonal \textit{P}6$_1$22 below about 200 GPa and
monoclinic \textit{C}2/\textit{c} at higher pressures.
It has been found that quantum nuclear and thermal vibrations play a
central role in stabilizing the \textit{P}6$_1$22 phase.
It is possible that other similar phase-III-like structures of
hydrogen may exist.
A structure of \textit{Pc} symmetry, shown in Fig.\ \ref{fig:pc-48},
was suggested as the best candidate for phase IV at high
pressure.

Although the low-energy structures modeling particular hydrogen
phases can be identified by their vibrational properties, the DFT
phase diagram obtained using those structures does not agree with
experiment.
Pickard \textit{et al.}\ showed that standard DFT functionals predict
metallic structures of hydrogen to be energetically favorable over a
range of pressures up to 400 GPa. \cite{Pickard_erratum_2012}
However, it is known experimentally that, at the pressures of interest
here (below 400 GPa), metallic structures are not in fact favored.
This is an example of the ``band-gap problem,'' in which DFT
electronic band gaps are found to be significantly smaller than those
obtained in experiments.
DMC methods have therefore been used to achieve a consistent theoretical
understanding of high-pressure hydrogen, by calculating the phase
diagram of solid molecular hydrogen using the candidate structures
identified in DFT calculations. \cite{Drummond2015}
Twist-averaged DMC calculations with Slater-Jastrow wave functions
were performed in cells of 96 and 768 atoms.
Finite-size corrections were evaluated using the different methods
described in Sec.\ \ref{sec:fs_lr} as well as extrapolation of the
twist-averaged DMC energies per atom to the thermodynamic limit,
assuming the finite-size error goes as $N^{-1}$.
The fact that different finite-size-correction methods give results in
agreement with each other suggests that finite-size effects are
well-controlled.
Vibrational energy contributions play an important role in hydrogen,
and anharmonicity cannot be neglected on the fine energy scale that
has to be resolved (less than $1$ meV per atom).
DFT anharmonic vibrational free
energies\cite{Monserrat_Anharmonic_2013} were therefore added to the
static-lattice DMC energy data. \cite{Drummond2015}
By comparing the Gibbs free energies of the phases, the
pressure-temperature phase diagram was evaluated and found to be in
qualitative agreement with experiment.
In fact good quantitative agreement with experiment for the
temperature of the transition between phases III and IV was achieved.
The calculated pressure for the transition between phases II and III
is 75 GPa larger than found in experiment, although the isotope
dependence of the II--III transition is
well-reproduced. \cite{Goncharov_H_D_200GPa_2011}
Most importantly, the DMC calculations show that the metallic
structure that is strongly favored in DFT at high pressure is not
energetically competitive, resolving an outstanding disagreement
between theory and experiment.

Experiments have revealed the existence of an additional phase V at
pressures above 325 GPa. \cite{H_Dalladay-Simpson-Gregoryanze_2016}
DFT calculations have been performed to identify the structure of
phase V using the ``saddle-point AIRSS'' approach, which searches for
structures stabilized by anharmonic nuclear motion.
The vibrational results suggest that a \textit{Pca}2$_1$ structure is
a promising model for phase V\@. \cite{Monserrat_2018}
DMC was used to demonstrate that this is indeed one of the most
energetically stable candidate structures.
It is interesting to note that a large fraction of the low-energy
structures of compressed hydrogen at low temperature adopt layered
forms.
This observation may make it easier to determine low-enthalpy
structures of a wider range of hydrogen phases.

The effects of van der Waals forces can be important in systems such
as solid hydrogen, which is one of the reasons why DMC calculations
are more reliable than DFT\@. \cite{Azadi_van_der_waals_2017}
Azadi \textit{et al.}\cite{Azadi_Foulkes_H_2013}\ used DMC methods to
calculate structures of solid hydrogen at high pressure, including
vibrational effects within a self-consistent-field approach.
\cite{Azadi_Dissociation_H_2014}
Azadi and K\"{u}hne performed DMC calculations in which eleven
molecular hydrogen structures with different symmetries were found to
be the most energetically competitive phases within the pressure range
studied of 100--500 GPa, concluding that phase III may be polymorphic.
\cite{Azadi_H_2019}

The electronic structure of high-pressure hydrogen is also of great
interest, with much experimental effort being directed towards the
discovery of metallic hydrogen.
Metalization of solid hydrogen at high pressure occurs via a
structural phase transition rather than band-gap
closure. \cite{H_bonding_Band_Gap_Goncharov_2013}
The difference between the DMC and DFT band gaps in high-pressure
hydrogen is almost independent of system size, and therefore can be
applied as a scissor correction to the infinite-system DFT gap to
obtain the DMC gap in the thermodynamic limit. \cite{Azadi_2017}
Comparisons of static-nucleus DMC energy gaps with the available
experimental data demonstrate the important role played by nuclear
quantum effects in the electronic structure of solid hydrogen.
DMC calculations of quasiparticle and excitonic band gaps have shown
that the exciton binding energies in hydrogen at the high pressures of
interest are smaller than 100 meV\@. \cite{Azadi_2017}

Recent experiments by Eremets \textit{et al.}\ have suggested that
semimetallic molecular structures may occur at pressures above 350
GPa.  \cite{Eremets_semi-metallic_H_2019}
There are still many fascinating questions to address in this field
using QMC methods in conjunction with DFT structure searching and
anharmonic vibrational methods.

\section{Future directions for QMC methods and the CASINO software}
\label{sec:future}

\subsection{Synergy with FCIQMC}

The master equation of FCIQMC can be obtained by substituting the CI
wave function of Eq.\ (\ref{eq:ci_wfn}) with imaginary-time-dependent
coefficients $c_i=c_i(\tau)$ into the imaginary-time Schr\"odinger
equation of Eq.\ (\ref{eq:itse}),
\cite{Booth_FCIQMC_2009}
\begin{equation}
-\frac{\partial c_i}{\partial \tau} =
 (H_{ii}-S) c_i + \sum_{j\neq i} H_{ij} c_j \;,
\end{equation}
where $H_{ij}=\langle D_i \vert \hat H \vert D_j \rangle$ and $S$ is
an adjustable energy shift.
This equation governs the dynamics of discrete walkers that sample
the second-quantized Hilbert space of all Slater determinants that can
be constructed given a fixed orbital basis.
The initiator approximation modifies these dynamics to prevent the
extremely rapid growth in walker number at the start of the
calculation.  \cite{Cleland_initiator_2010, Ghanem_unbias_FCIQMC_2019}
The computational cost of FCIQMC, both without and with the initiator
approximation, formally scales as an exponential of the system size,
but in practice it is possible to study medium-sized systems with
reasonable computational cost.  \cite{Booth_c2_2011,
Cleland_diatomics_2012, LiManni_porphyrins_2016, Veis_TME_2018}

The combination of FCIQMC and DMC is especially appealing in that one
method excels where the other struggles: FCIQMC easily recovers
static correlation but requires very large numbers of walkers to
account for dynamical correlation, whereas DMC correctly captures
dynamical correlation but is incapable of modifying the description of
static correlation provided by the trial wave function.
There are three broad approaches to combining FCIQMC and DMC, detailed
below.

\subsubsection{FCIQMC wave functions in DMC}
\label{sec:future_mdet}

Initiator-FCIQMC is capable of quickly identifying the most important
determinants in the CI wave function and providing their approximate
coefficients.
The most obvious combination of FCIQMC and DMC is simply to build the
trial wave function by selecting a certain number of determinants in
the FCIQMC wave function with the largest weights.

\begin{figure}[!htbp]
  \begin{center}
    \includegraphics[clip,width=0.45\textwidth]{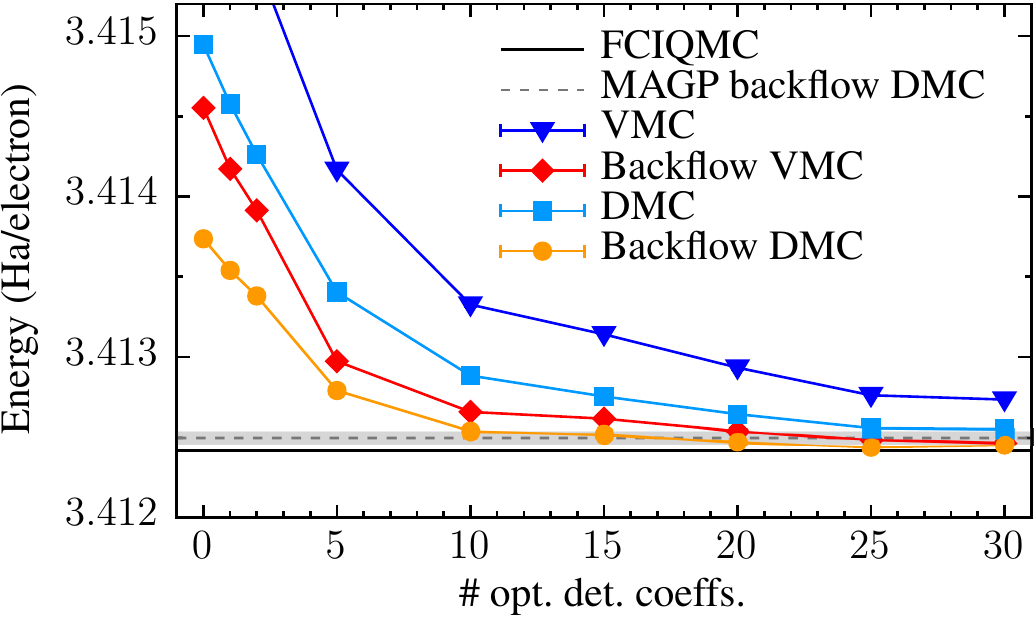}
  \end{center}
  \caption{ VMC and DMC energy of the $r_{\rm s}=0.5$ 3D paramagnetic
    14-electron gas at $\Gamma$ as a function of the size of the
    multideterminant expansion obtained by truncation of the FCIQMC
    wave function.
    Note that the zero of the horizontal axis corresponds to the
    single-determinant wave function.
    The exact (FCIQMC) energy is also shown, along with the MAGP
    backflow DMC result, with its uncertainty represented by the
    shaded area.
    \label{fig:HEG_mdet}
  }
\end{figure}

As an example of the direct application of this principle,
Fig.\ \ref{fig:HEG_mdet} shows the VMC and DMC energies, with and
without backflow, of the 3D paramagnetic 14-electron gas at a density
of $r_{\rm s}=0.5$ at the $\Gamma$ point as a function of the size of
the truncated expansion.
The coefficients of the multideterminant expansion have been
reoptimized in the presence of the Jastrow factor and, where
applicable, of the backflow parameters.
The coefficients of symmetry-equivalent determinants in the expansion
have been constrained to be equal, so, e.g., the expansion with 30
optimizable coefficients contains 1273 determinants, by which point
the backflow DMC energy has converged to within $\sim 35$ ${\upmu}{\rm
  Ha}$ per electron of the exact energy, obtained using FCIQMC\@.
\cite{Neufeld_HEG_2017}
Also shown in Fig.\ \ref{fig:HEG_mdet} is the basis-set limit of the
backflow DMC energy obtained with an MAGP wave function,
\cite{Bugnion_thesis_2014} described in Sec.\ \ref{sec:pairing}.

While these results are very encouraging, total energies for other
systems converge rather more slowly with expansion size, and moderate
fixed-node errors should be expected with medium-sized expansions.
This represents a significant challenge for computing energy
differences, since at fixed expansion size any two systems will likely
incur different fixed-node errors, biasing the result.
This mismatch can be addressed by using different expansion sizes for
each system.
For example, to obtain the binding energy of the C$_2$ molecule to
chemical accuracy one would need to subtract twice the DMC energy of a
carbon atom computed with a single-determinant wave function from the
DMC energy of the carbon dimer obtained with a 60-determinant
expansion.
Criteria to decide how to truncate the CI expansion for each system to
achieve similar fixed-node errors have been proposed.
\cite{Per_mdet_2017, Dash_exstates_2019}
Another approach is to use very large expansions in order to reduce
the fixed-node errors in both systems to the point that the bias in
the difference is negligible.
However this is very expensive and wave-function optimization is
problematic.
A third potential solution is to construct an extrapolation procedure
to obtain the exact energy from DMC energies at various expansion
sizes, which is the subject of current research.

\subsubsection{Jastrow factors and similarity-transformed FCIQMC}
\label{sec:fciqmc_tc}

A recent development in the FCIQMC method is the ability to use
Jastrow factors, \cite{Luo_TC-FCIQMC_2018, Cohen_ST-FCIQMC_2019,
Dobrautz_ST_Hubbard_2019} effectively using a wave function of the
form $e^{\hat J} \vert \Psi_{\rm CI} \rangle$, where $e^{\hat J}$ is
the second-quantized operator associated with the first-quantized
Jastrow factor $e^J$.
Since the FCIQMC method requires the wave function to be expressible
as a CI expansion, the Jastrow factor must be incorporated into the
Hamiltonian.
The effective Hamiltonian in the similarity-transformed FCIQMC
(ST-FCIQMC) method is therefore $e^{-\hat J} \hat H e^{\hat J}$,
which is non-Hermitian.
FCIQMC then serves as a solver for the right eigenvector of this
effective Hamiltonian, which is a CI expansion.

The similarity transformation modifies the Hamiltonian matrix
elements, and the presence of electron-electron Jastrow factor terms
requires the evaluation of six-index integrals in addition to the
four-index integrals used in regular FCIQMC, as well as other methods
in quantum chemistry.
The ST-FCIQMC method has greater memory requirements than FCIQMC, but
the expectation is that this can be offset by the reduction in
computational cost afforded by a more compact representation of the
wave function.

While calculations for first-row atoms appear to suggest that
ST-FCIQMC energies are relatively insensitive to the quality of the
Jastrow factor, \cite{Cohen_ST-FCIQMC_2019} more complex systems can
be expected to benefit from well-optimized Jastrow parameters.
The VMC method provides the ideal framework to optimize these Jastrow
factors.

The presence of the Jastrow factor decreases the importance of those
determinants in the CI wave function that are mainly associated with
dynamical correlation.
These CI expansions should therefore make great trial wave functions
for DMC\@.

\subsubsection{DMC-assisted FCIQMC and benchmarking}
\label{sec:dmc_assist_fciqmc}

The very high accuracy of FCIQMC for small systems and the feasibility
of performing DMC calculations for large systems can also be
exploited.
This can be done by combining the results of independent FCIQMC and
DMC calculations without sharing wave functions between the two
methods.

As described in Sec.\ \ref{sec:app_heg}, it is possible to evaluate
the exact energy of the high-density HEG at small system sizes using
FCIQMC, and to obtain DMC energies at large system sizes that enable a
very reliable extrapolation of the fixed-node energy to infinite
system size.
Modeling the fixed-node error as a slowly-varying function of system
size then allows the estimation of the exact correlation energy of the
infinite system.

This approach has been applied to the ferromagnetic electron gas,
\cite{Ruggeri_heg_2018} and work is underway to produce a similar set
of results for the paramagnetic electron gas.
Besides obtaining the correlation energy itself, the fixed-node error
is produced as a byproduct of this process.
This opens the intriguing possibility of parameterizing a functional of
the density to estimate the fixed-node error incurred in inhomogeneous
systems, which could be used to construct a correction for fixed-node
DMC energies.

Another, more straightforward possibility is to use very accurate
FCIQMC results for small systems as a benchmark against which to gauge
the accuracy of DMC, which is extremely useful in the development of
trial wave functions and other methodological advances.

\subsection{Towards greater efficiency}

\subsubsection{Electron-electron pseudopotentials}

The Coulomb potential energy diverges as $1/r$ when two electrons
approach each other.
Even if the Kato cusp conditions are imposed on the wave function,
this divergence leads to additional variance in the local energy and
hence computational expense in QMC calculations.
To alleviate this problem, Lloyd-Williams \textit{et al.}\ proposed
replacing the electron-electron interaction potential by a local
pseudopotential that reduces to $1/r$ outside a cutoff radius, is
smooth everywhere, and approximately reproduces the scattering states
of the $1/r$ Coulomb potential up to the Fermi wavevector of a HEG of
density parameter $r_\text{s}=2$. \cite{Lloyd_2015}
Their pseudopotential is of polynomial form inside the cutoff radius,
with the polynomial coefficients chosen to minimize the mean squared
difference of the logarithmic derivatives of the exact and pseudo
scattering wave functions at the cutoff radius.
The resulting electron-electron pseudopotentials were shown to give
accurate results for HEGs at a range of densities with a speedup of up
to thirtyfold.
Lloyd-Williams \textit{et al.}\ also showed the effectiveness of the
approach for lithium and beryllium atoms.
This is clearly a promising technique to investigate for QMC studies
of inhomogeneous condensed-matter systems.

\subsubsection{Current developments in computer architectures}

As explained in Sec.\ \ref{sec:parallelisation}, QMC methods are
well-placed to take advantage of the widespread availability of
massively parallel computer architectures.
\textsc{Casino} has also been shown to perform well on many-core
processors such as Intel's Knights Landing processor, although Intel
has now ended the development of this particular class of processor.
It is found to be advantageous to make full use of hyperthreading, due
to the high parallel performance of QMC methods.

Graphics processing units (GPUs) are now a common accelerator
technology on supercomputers.
Unfortunately, allowing QMC methods to make efficient use of GPUs is a
nontrivial task if the CPU and GPU do not share memory.
QMC methods for practical problem sizes involve a large number of
small tasks that must be performed in a particular sequence, dependent
on the outcome of random processes such as accept/reject steps or the
sampling of branching factors.
The steps include small pieces of linear algebra relating to
determinant-updating, evaluation of polynomials and small plane-wave
expansions in the Jastrow factor, and pairwise sums of interaction
potentials.
If these small tasks are devolved to the GPU one by one then they
incur a prohibitive time overhead due to the need to transfer arrays
between the CPU and GPU\@.
We have experimented with a light-touch, compiler-driven GPU
acceleration of \textsc{casino} by means of OpenACC compiler
directives.
Unfortunately, this results in significant slowdowns to
\textsc{casino} due to the cost of transferring data between the CPU
and the GPU\@.
A successful implementation of QMC on a GPU requires a very
significant modification such that a large part of the calculation is
performed on the GPU\@.
Given the comments about fundamental limits on the scaling of QMC
methods in Sec.\ \ref{sec:Nscaling}, it is not clear whether the
benefits of GPUization justify forking \textsc{casino} into GPU and
non-GPU versions.
On the other hand, modern developments in heterogeneous computing
architectures that allow CPUs and GPUs to share memory are expected to
remove many of these issues with GPU computing.

\subsection{Improved accuracy}

\subsubsection{Spin-orbit coupling \label{sec:soc}}

Spin-orbit coupling plays a key role in the electronic structure of
many elements beyond the first row; however, it is not generally
included in QMC calculations because of the additional complication
and expense of using spinor wave functions.
Recent work \cite{Melton_2016} has demonstrated the feasibility of
developing DMC for spin-orbit interactions using a continuous-spin
approach.
An alternative approach would be to use an algorithm similar to the
pseudopotential T-move scheme \cite{Casula_ppot_2006} to propose spin
flips.
Suppose the Hamiltonian contains a spin-dependent term $\hat{V}_{\rm
  SO}$.
Let $\tau$ be the DMC time step, ${\bf X}$ be the
space and spin coordinates of all the electrons, $\Psi$ be the trial
wave function and
\begin{equation} V_{{\bf X},{\bf X}'}=\langle {\bf X}|\hat{V}_{\rm
  SO}|{\bf X}'\rangle \frac{\Psi({\bf X})}{\Psi({\bf X}')}. \end{equation}
Let $V^+_{{\bf
    X},{\bf X}'}=\max\{V_{{\bf X},{\bf X}'},0\}$ and $V^-_{{\bf
    X},{\bf X}'}=\min\{V_{{\bf X},{\bf X}'},0\}$, so that $V_{{\bf
    X},{\bf X}'}=V^-_{{\bf X},{\bf X}'}+V^+_{{\bf X},{\bf X}'}$.
Then the importance-sampled DMC Green's function for the
spin-dependent part of the Hamiltonian is
\begin{widetext}
\begin{equation} \langle {\bf X}|\exp(-\tau \hat{V}_{\rm SO})|{\bf
X}'\rangle \frac{\Psi({\bf X})}{\Psi({\bf
X}')} \approx \left[ \frac{\delta_{{\bf X},{\bf X}'} - \tau V^-_{{\bf
X},{\bf X}'}}{1-\tau\sum_{{\bf X}''} V^-_{{\bf X}'',{\bf
X}'}} \right] \exp\left(-\tau \sum_{{\bf X}''} V_{{\bf X}'',{\bf
X'}}\right) + {\cal O}(\tau^2), \label{eq:spin_tmove} \end{equation}
\end{widetext}
where the negative matrix elements are treated exactly to ${\cal
O}(\tau)$, while the positive matrix elements are localized by
replacing $\langle {\bf X}|\hat{V}_{\rm SO}|{\bf X}'\rangle$ with
$\delta_{{\bf X},{\bf X}'} [\hat{V}_{\rm SO}\Psi({\bf X})]/\Psi({\bf
X})$.
To apply this Green's function to a walker, the first factor on
the right-hand side of Eq.\ (\ref{eq:spin_tmove}) is treated as a
spin-flip transition probability, while the second factor is absorbed
into the weight of the walker.
The contribution to the local energy from $\hat{V}_{\rm SO}$ is
\begin{equation} \frac{\langle\Psi|\hat{V}_{\rm SO}|{\bf
X}'\rangle}{\langle\Psi|{\bf X}'\rangle}=\sum_{{\bf X}''} V_{{\bf
X}'',{\bf X'}}. \end{equation}
This approach leads to a positive error in the DMC energy, which is
second order in the error in the trial wave function.

\subsubsection{Inclusion of vibrational effects in \textit{ab initio}
QMC calculations \label{sec:vib_corr}}

QMC methods provide highly accurate solutions to the electronic
Schr\"{o}dinger equation, but comparison with experiment is always
complicated by the need to include vibrational corrections.
Such corrections are usually evaluated at the DFT level.
For ground-state total-energy calculations the inclusion of
quasiharmonic vibrational Helmholtz free energies is straightforward
and many first-principles DFT codes have the ability to perform such
calculations automatically.
By contrast, codes and scripts for calculating the vibrational
renormalization of band gaps within DFT are not generally available.
However, vibrational effects often alter gaps by a significant
fraction; e.g., the gap of a benzene molecule is reduced by more than
$0.5$ eV due to vibrational effects. \cite{Mostaani_2016}
The vibrational renormalizations of quasiparticle and excitonic gaps
have different physical origins: the total energies that define the
quasiparticle gap should include vibrational Helmholtz free energies,
while the excitonic (optical absorption) gap should be averaged over
the distribution of nuclear positions at temperature $T$ in the
electronic ground state.
For light nuclei (first row), zero-point renormalization is the most
important vibrational effect, with temperature dependence being
relatively weak; on the other hand heavier nuclei behave classically,
leading to negligible zero-point vibrational renormalization, but
significant temperature dependence.

It would be straightforward to implement a QMC
vibrational-renormalization approach based on the Born-Oppenheimer
approximation and the DFT potential-energy landscape, with nuclear
coordinates sampled randomly from a vibrational self-consistent-field
wave function. \cite{Monserrat_Anharmonic_2013}
QMC would be used to evaluate the energy gap at each nuclear
configuration sampled.
The statistical error bar on the QMC gap falls off as the reciprocal
of the square root of the amount of QMC data gathered, irrespective of
whether those data are gathered at different nuclear configurations.
The cost of a vibrationally renormalized QMC gap calculation is not
therefore expected to be significantly larger than the cost of a
static-nucleus gap calculation.

Hydrogen is an important constituent of many compounds, and the
hydrogen nuclei in most materials are more than an order of magnitude
lighter than the other nuclei.
Protons could be treated as distinguishable quantum particles in
\textit{ab initio} QMC calculations, since proton exchange effects are
small.
Including protons in QMC calculations will require (i) bespoke
backflow functions to ensure that the cusps in DFT-generated electron
orbitals occur at the electron-proton coalescence points, even when
the protons have moved; (ii) a bespoke Jastrow factor of Gaussian form
in either the proton positions (Einstein approximation) or the phonon
normal coordinates (from a DFT quasiharmonic phonon calculation with
the nonhydrogen nuclei having infinite mass); and (iii) modifications
to the DMC Green's function near electron-proton coalescence points
similar to those currently used between electrons and fixed nuclei.
\cite{Umrigar_1993}
A novel extension for hydrogen-bearing compounds would be to include
the electrons and protons in the QMC calculations, evaluate the matrix
of force constants for the remaining nuclei using QMC, \cite{Liu_2019}
and then evaluate their zero-point energy within the quasiharmonic
approximation.
We would then have a fully quantum treatment of the protons and a
quasiharmonic treatment of the remaining nuclei.

\subsection{Atomic forces from QMC calculations}

The evaluation of atomic forces in QMC is complicated by statistical
issues.
These affect other expectation values too, such as elements of the
matrix of force constants. \cite{Liu_2019}
Here we focus on VMC forces for simplicity, but the methodology
applies to DMC forces as well.

The force exerted on the $I$th atom of a system along Cartesian
direction $x$ is minus the derivative of the expectation value of the
Hamiltonian with respect to the $x$ component of the position of the
nucleus.
The local force $F$ is the sum of a Hellmann-Feynman term and a Pulay
term,
\begin{equation}
  \begin{split}
  F({\bf R}) = &
    \sum_{i} \frac {-Z_I x_{iI}} {r_{iI}^3} +
    \sum_{J\neq I} \frac {Z_I Z_J x_{IJ}} {r_{IJ}^3} \\ &
   -2
    \Psi^{-1}({\bf R})
    \left[ E({\bf R}) - \langle \hat{H} \rangle \right]
    \frac{\partial \Psi({\bf R})} {\partial x} \;.
  \end{split}
\end{equation}
The local force follows a fat-tailed distribution satisfying $P(F)
\sim |F|^{-5/2}$ as $|F|\to\infty$ due to the error in the nodes of
the trial wave function. \cite{Trail_2008, Badinski_forces_2010,
Badinski_thesis_2008}
This causes statistical problems, since the expectation value of the
force,
\begin{equation}
  \langle F \rangle =
    \int_{-\infty}^{\infty} P(F) F \, {\rm d}F \;,
\end{equation}
is well defined, but its variance,
\begin{equation}
  {\rm Var}[F] = \sigma_F^2 =
    \int_{-\infty}^{\infty}
      P(F) \left(F-\langle F \rangle\right)^2 \, {\rm d}F \;,
\end{equation}
is divergent.
This implies that the standard error is divergent and cannot be used
to provide a confidence interval.
The application of a zero-variance estimator of the force
\cite{Assaraf_ZV_force_2003, Badinski_ZV_2008} and the use of
pseudopotentials, \cite{Badinski_forces_2007, Badinski_forces_2008}
both supported in \textsc{casino}, alleviate the severity of the
problem, but do not completely solve it.

Other methods have been proposed to tackle the fat tails in the
local force distribution by approximating the local observable
\cite{Chiesa_forces_2005} or modifying the VMC sampling distribution.
\cite{Attaccalite_forces_2008}
Recently, L\'opez R\'ios and Conduit \cite{LopezRios_TRE_2019} showed
that this issue can be solved in the statistical analysis stage using
a conceptually simple approach.
(We focus on the right-hand tail for simplicity; the left-hand tail is
treated analogously.)
Let us assume that a random sample of $M$ independent values of the
local force distributed according to $P(F)$ is available, and let
$F^{(m)}$ be the $m$th largest value in the sample.
By definition, the sample quantile $q_m=(m-1/2)/M$ satisfies
\begin{equation}
  \label{eq:quantile_relation}
  \int_{F^{(m)}}^\infty P(F) \, {\rm d}F \approx q_m \;.
\end{equation}
Let $F_{\rm R}$ be a value of the local force such that the
probability distribution can be accurately represented by
\begin{equation}
  \label{eq:tail_form}
  P(F) = \sum_{n=0}^{n_{\rm R}} c_n
           \left|F-F_{\rm c}\right|^{-\frac{5+n}2}
           \,,\quad F>F_{\rm R} \;,
\end{equation}
where $\{c_n\}$ are $n_{\rm R}+1$ unknown coefficients and $F_{\rm c}$
is a parameter which is set to the sample median $F^{(M/2)}$ in
practice.
Substituting Eq.\ (\ref{eq:tail_form}) into Eq.\
(\ref{eq:quantile_relation}) yields
\begin{equation}
  \label{eq:fit_form_raw}
  q_{m} = \sum_{n=0}^{n_{\rm R}} \frac{c_n}{(3+n)/2}
               |F^{(m)}-F_{\rm c}|^{-(n+3)/2} \;.
\end{equation}
Defining
\begin{equation}
  \label{eq:yx_def}
  \begin{split}
    y_m & = q_{m} \left|F^{(m)}-F_{\rm c}\right|^{3/2} \;, \\
    x_m & = |F^{(m)}-F_{\rm c}|^{-1/2} \;,
  \end{split}
\end{equation}
and rearranging, Eq.\ (\ref{eq:fit_form_raw}) reduces to a polynomial,
\begin{equation}
  \label{eq:fit_form}
  y_m=\sum_{n=0}^{n_{\rm R}} \frac{c_n}{(3+n)/2} x_m^n \;.
\end{equation}
By this change of scale, which we refer to as ``$yx$ scale,'' the tail
of the probability distribution function, which takes vanishingly
small values over an infinite range and involves nonlinear
parameters, is transformed into a finite function defined over a
finite range involving linear parameters.

The asymptotic coefficients $\{c_n\}$ can be obtained by using Eq.\
(\ref{eq:fit_form}) as a fit function for the tail data in the sample,
converted according to Eq.\ (\ref{eq:yx_def}).
The use of fit weights
\begin{equation}
  w_m = \left( \ln\frac{q_{M_{\rm R}+1}}{q_m} \right)^{-1}
        |F^{(m)}-F_{\rm c}|^{-3/2} \;,
\end{equation}
where $M_{\rm R}$ is the number of data points in the sample such that
$F>F_{\rm R}$, ensures the asymptotic normality of the regression
coefficients.
This enables the evaluation of
\begin{equation}
  \label{eq:TRE}
  {\cal F} =
    \frac 1 M \sum_{m>M_{\rm R}} F^{(m)} +
    \sum_{n=0}^{n_{\rm R}} c_n \int_{F_{\rm R}}^{\infty}
    \left|F-F_{\rm c}\right|^{-(5+n)/2}
    F \, {\rm d}F \;,
\end{equation}
referred to as the tail-regression estimator (TRE) of $\langle F
\rangle$.
In Eq.\ (\ref{eq:TRE}), the central contribution to the expectation
value is evaluated as a (partial) sample mean, while the integrals in
the tail contribution can be computed analytically.
Crucially, $\cal F$ is asymptotically normally distributed since its
uncertainty arises from that on $\{c_n\}$ and from the finite variance
of the central part, completely bypassing the use of the divergent
variance $\sigma_F^2$.
Parameters $F_{\rm R}$ and $n_{\rm R}$ are chosen so as to minimize
the uncertainty in the resulting estimator.
We note that Ref.\ \onlinecite{LopezRios_TRE_2019} did not provide
closed expressions for the uncertainty in $\cal F$, instead relying on
the bootstrap method \cite{Efron_bootstrap_1986} to compute confidence
intervals.

\begin{figure}[!htbp]
  \begin{center}
    \includegraphics[clip,width=0.45\textwidth]{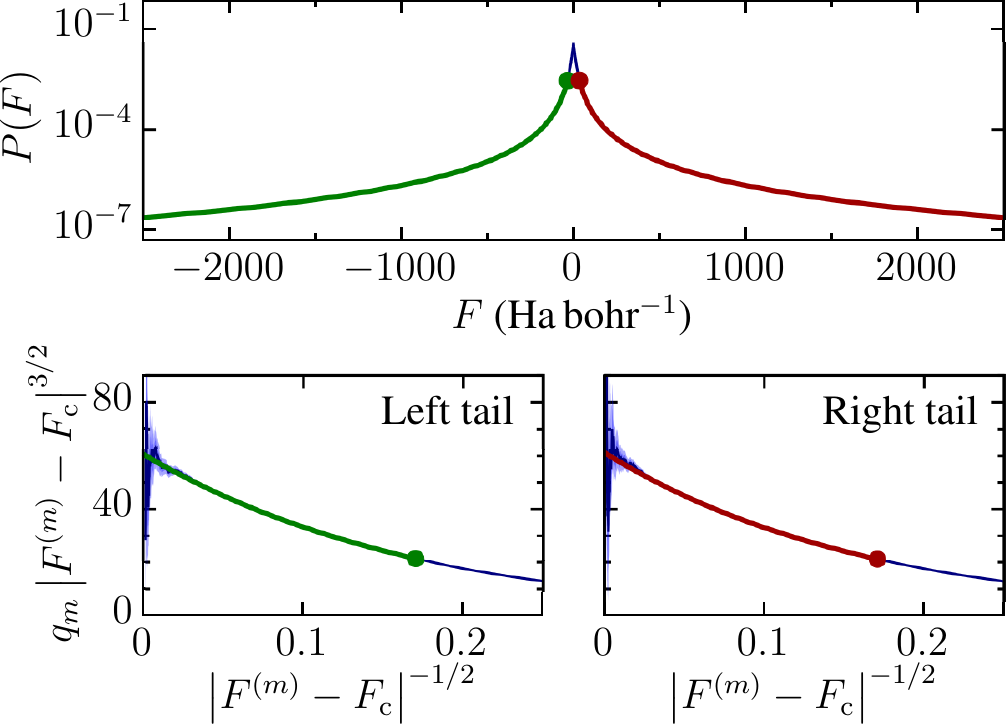}
  \end{center}
  \caption{
    Application of the tail-regression procedure to a sample of $10^7$
    VMC local forces on the lithium atom of the all-electron LiH
    molecule at a compressed bond length of $2$ bohr.
    The thin line in the top panel represents the estimated
    probability distribution (note the logarithmic scale of the
    vertical axis), and the thin lines in the lower panels correspond
    to the tails of the estimated probability distribution in the $yx$
    scale defined in Eq.\ (\ref{eq:yx_def}).
    Fits are shown as thick lines in the three panels, with the left
    and right fit onsets, $F_{\rm L}$ and $F_{\rm R}$, shown as
    circles.
    $68.3\%$ and $95.4\%$ confidence intervals obtained from the
    bootstrap are shown as shaded areas.
    \label{fig:TRE_lih_force}
  }
\end{figure}

We demonstrate the application of the TRE in Fig.\
\ref{fig:TRE_lih_force} using VMC force data for the all-electron
lithium hydride molecule at a compressed bond length of $2$ bohr
using a trial wave function consisting of Slater determinants of
HF orbitals expanded in the cc-pVDZ basis set
\cite{Dunning_cc_basis_1989} multiplied by a DTN Jastrow factor.
The TRE of the sum of the Hellmann-Feynman and Pulay forces on the
lithium atom in the direction of the hydrogen atom is $-0.54(6)$
Ha\,bohr$^{-1}$.
By contrast, the standard estimator of this force is $-0.46$
Ha\,bohr$^{-1}$ with a formally divergent uncertainty (numerically the
apparent uncertainty is $0.18$ Ha\,bohr$^{-1}$).

The TRE is thus an asymptotically normally distributed estimator of
expectation values associated with fat-tailed probability
distributions of known tail exponents, yielding uncertainties
proportional to $M^{-1/2}$, and can be used for distributions with
other tail exponents and for higher moments of the distribution, such
as the variance of the local energy for which standard confidence
intervals are also formally undefined.

The TRE method as described in Ref.\ \onlinecite{LopezRios_TRE_2019}
has the limitation of requiring serially-uncorrelated samples, which
in practice implies spending additional computational time in the
accumulation stage to achieve decorrelation.
Whether this restriction can be worked around in the future or not,
the TRE unequivocally shows that the problem of evaluating expectation
values in QMC associated with fat-tailed probability distributions
of known asymptotic behavior can be solved at the statistical level.

\section{Conclusions}
\label{sec:conclusions}

We have explained the theory underpinning the VMC and fixed-phase DMC
methods, and commented on some of the practical issues affecting QMC
codes such as \textsc{casino}.
We have shown many of the strengths and also limitations of these
methods.
QMC methods are highly successful for studying systems featuring van
der Waals interactions, they give a powerful alternative to many-body
$GW$ methods for calculating accurate quasiparticle and excitonic band
gaps, and they have unique capabilities for accurately solving the
Schr\"{o}dinger equation in model electron(-hole) systems (indeed,
providing exact solutions to the Schr\"{o}dinger equation for
charge-carrier complexes).
On the other hand we have argued that it will prove difficult to
perform useful DMC calculations for systems with several thousand
electrons with current algorithms, irrespective of developments in
supercomputer hardware.
The successful use of DMC in studies of condensed matter therefore
very much depends on the analysis of finite-size effects and/or
extrapolation to the thermodynamic limit.
Furthermore, dealing with systems characterized by strong static
correlation effects remains challenging, because
multideterminant expansions introduce an unwelcome element of
arbitrariness into DMC results.
Despite these limitations, QMC methods continue to play a role of
fundamental importance in the ecosystem of first-principles
computational methods, providing data that expand the accuracy and
capabilities of simpler methods.

\begin{acknowledgments}
The authors would like to thank the Theory of Condensed Matter group
at the University of Cambridge, the Condensed Matter Theory group at
Lancaster University, and the Electronic Structure Theory group at the
Max-Planck Institute for Solid State Research for their support over
the years, and the broader QMC community for its vibrancy and for many
useful discussions.
We would like to thank our long-term collaborators Gareth Conduit,
Bartomeu Monserrat, Matthew Foulkes, Dario Alf\`e, Chris Pickard, and
Ryo Maezono for their contributions to our QMC project.
We also thank Alice Shipley for providing the hydrogen figures shown
in this paper.
Financial support was provided by the Engineering and Physical
Sciences Research Council of the United Kingdom under Grant No.\
EP/P034616/1.
Supporting research data can be freely accessed at
[\onlinecite{opendata}], in compliance with the applicable Open Data
policies.
\end{acknowledgments}

\nocite{*}
\bibliography{casino_jcp}

\end{document}